\documentclass[12pt]{article}

\usepackage[english]{babel}
\usepackage[normalem]{ulem} 
\usepackage{amssymb} 
\usepackage{amsmath}
\usepackage{graphicx} 
\usepackage[pdfpagelabels]{hyperref}
\usepackage{float}
\usepackage{subfig}
\usepackage{slashed}
\usepackage{cite}
\usepackage{simplewick}
\usepackage{mathrsfs}
\usepackage{color}

\numberwithin{equation}{section}

\newcommand{\HRule}{\rule{\linewidth}{0.5mm}}
\newcommand{\sgn}{\operatorname{sgn}}

\begin{document}

\begin{titlepage}
\clearpage\setcounter{page}{0}
\begin{center}


\begin{figure}[htbp]
\begin{center}
\includegraphics[scale=0.12]{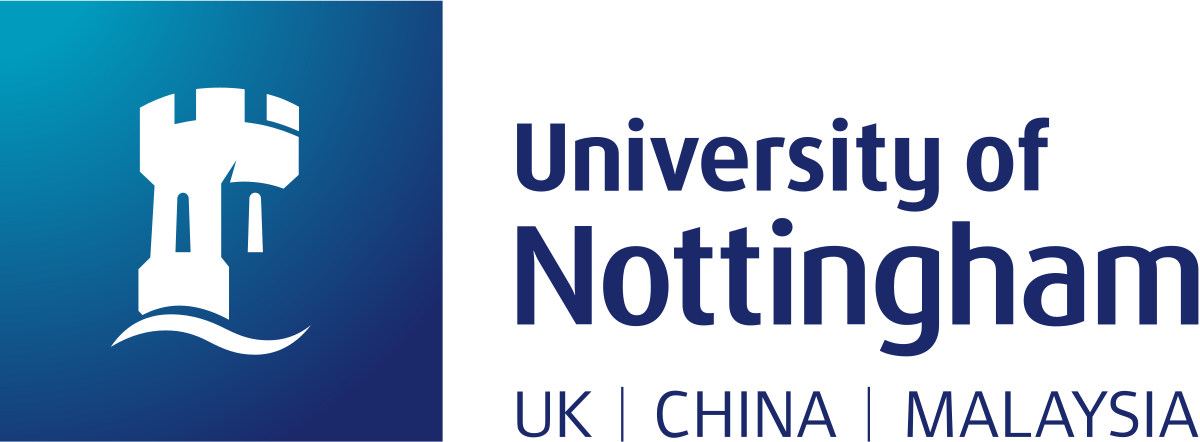}
\end{center}
\end{figure}

\HRule \\[0.4cm]
{ \LARGE \bfseries Astro- and Quantum Physical Tests of Screened Scalar Fields}\\[0.2cm]
\HRule \\[0.5cm]
\textsc{\normalsize By Christian K\"ading, B.Sc., M.Sc.}\\[5.5cm]

 \normalsize
Thesis submitted to the University of Nottingham \\for the degree of Doctor of Philosophy

\hfill
\vfill
Contact: Kkaeding@hse.ru
\end{center}
\end{titlepage}


\pagenumbering{roman}

\thispagestyle{empty}
\clearpage\mbox{}\clearpage

\thispagestyle{empty}
\begin{abstract}
In general, modified gravity theories are modifications or extensions of Einstein's general relativity. 
Some of them give rise to additional scalar degrees of freedom in Nature. 
If these scalar fields exist and are light enough, they should cause a gravity-like fifth force that could, in principle, exceed gravity in its strength. 
However, there are tight constraints on fifth forces from {Solar System}-based tests. 
Screening mechanisms are popular means for avoiding these constraints by suppressing a fifth force in regions of high environmental mass density but allowing for phenomenologically interesting effects in environments of lower densities.
\\
In this thesis, scalar field models with screening mechanisms will be discussed and some astro- and quantum physical tests for their existence presented. 
At first, the impact of disformally coupled symmetrons on gravitational lensing by galaxies will be evaluated. Secondly, it will be shown how fluctuations of a chameleon scalar field induce the open dynamics of a quantum test particle. For this, tools from non-equilibrium quantum field theory will be introduced, developed and applied, and a quantum master equation derived. 
\end{abstract}
\newpage
\thispagestyle{empty}
\renewcommand{\abstractname}{Zusammenfassung}
\begin{abstract}
In der Regel sind modifizierte Gravitationstheorien Modifikationen oder Erweiterungen von Einsteins allgemeiner Relativit\"atstheorie. 
Einige davon bedingen zus\"atzliche skalare Freiheitsgrade in der Natur. 
Falls solche Skalarfelder existieren und sie leicht genug sind, sollten sie eine gravitations\"ahnliche f\"unfte Kraft verursachen, welche sogar st\"arker als die Schwerkraft sein k\"onnte. 
Jedoch schr\"anken im Sonnensystem gemachte Beobachtungen die Existenz einer solchen Kraft stark ein. 
Abschirmmechanismen sind beliebte Methoden, um solche Einschr\"ankungen zu umgehen, indem sie eine f\"unfte Kraft in Regionen hoher Massendichte unterdr\"ucken, aber in Umgebungen niedri\-gerer Dichte ph\"anomenologisch interessante Effekte zulassen.
\\
In dieser Arbeit werden Skalarfeldmodelle mit Abschirmmechanismen diskutiert und ein paar astro- und quantenphysikalische Tests ihrer Existenz pr\"asentiert.
Zun\"achst wird der Einfluss von disform gekoppelten Symmetronen auf den Gravitationslinseneffekt von Galaxien untersucht.
Anschlie{\ss}end wird gezeigt, wie Fluktuationen eines Cha\-m\"a\-le\-on\-skalarfeldes die offene Dynamik eines Quantentestteilchens induzieren. 
Daf\"ur werden Werkzeuge aus der Nicht\-gleich\-ge\-wichts\-quan\-ten\-feld\-theo\-rie eingef\"uhrt, entwickelt und angewandt, sowie damit\\ eine Quantenmastergleichung hergeleitet.
\end{abstract}
\newpage

\thispagestyle{empty}
\tableofcontents
\thispagestyle{empty}
\newpage

\thispagestyle{empty}
\renewcommand{\abstractname}{Acknowledgements}
\begin{abstract}
\noindent 
First of all, I am extremely grateful to Clare Burrage. 
She always supported me in all matters - academically and beyond, and I learned a lot under her supervision.
We worked on some super interesting projects together!
It is very hard to imagine a better PhD supervisor than her.
\\\\
I would also like to express my gratitude to Ji\v{r}\'{i} Min\'{a}\v{r} and Pete Millington for the joyful collaborations we had together and all the things I learned from them. 
Ji\v{r}\'{i} explained to me many things about open quantum dynamics, and Pete taught me much about field theory.
\\\\
Ed Copeland always had an open ear for me, and we had a wonderful collaboration together. 
I am very thankful for that.
\\\\
Another huge thanks goes to Ivette Fuentes.
She made me part of her group during the third year of my PhD, and worked with me on some pretty interesting projects. 
Of her group members I am particularly grateful to Daniel Hartley, Richard Howl, and Ana Luc\'ia B\'aez Camargo Aguilar for the great work we did together. Besides, I would like to thank Paul Juschitz, Jan Kohlrus, Dan Goldwater, and Sir Roger Penrose for the times we spent together with other group members in- and outside the university. 
\\\\
There have been many nice people with whom I have shared my office in Nottingham. 
Of those I am especially grateful to Will Emond and Johannes Dombrowski for being good friends, and making the office hours (and also some after work hours) even more joyful. 
In addition, I would like to thank Ben Thrussell, Mehdi Walji, Toby Wilson, Felipe Maykot, Ben Coltman, Yixuan Li, Dave Stefanyszyn,  Tom Peterken, and all the other people I shared office A103 with. 
\\\\
The last half year of my PhD studies was spent at the TU Wien.
I am very grateful to Anton Rebhan for hosting me in his group there, and to the OeAD for funding my visit. 
Furthermore, I would like to thank David M\"uller and Alex Soloviev for being nice office mates, and Stephan Stetina for helpful conversations.
In addition, I am thanking Mario Pitschmann, Hartmut Abele and Philipp Haslinger for the nice times I spent with them at the Atominstitut.
\\\\
During the second year of my PhD studies I organised the COLLABOR8.1 conference in Nottingham. 
I am very grateful to Katy Clough and Ed Hughes for doing this together with me. 
For help during the organisation of this conference I would also like to thank Wendy Brennan, Shaun Beebe, Mike Merrifield, and Thomas Sotiriou.
\\\\
I did not only organise a conference, I also attended several. 
There I met many lovely people.
Of those I am particularly grateful to Michaela Lawrence, Joakim Str\"omwall, and Ben Elder for spending a lot of time with me during some nice conferences.
\\\\
Last but most importantly I would like to thank my family and friends from Germany and Sweden, who have always supported me throughout the course of my PhD studies.
\end{abstract}
\thispagestyle{empty}
\newpage

\thispagestyle{empty}

\section*{Publications}
The following is a list of articles that this thesis' author released during the course of his PhD studies.
\\\\
\underline{Published:}
\begin{itemize}
\item[\cite{Kading1}] Daniel Hartley, \textbf{Christian K{\"a}ding}, Richard Howl, and Ivette Fuentes
\\
\textit{Quantum simulation of dark energy candidates}
\\
Phys.Rev. D99 (2019) no.10, 105002 
\item[\cite{Kading2}] Clare Burrage, Edmund J. Copeland, \textbf{Christian K{\"a}ding}, and Peter Millington
\\
\textit{Symmetron Scalar Fields: Modified Gravity, Dark Matter or Both?}
\\
Phys.Rev. D99 (2019) no.4, 043539 
\item[\cite{Kading3}] Clare Burrage, \textbf{Christian K{\"a}ding}, Peter Millington, and Ji\v{r}\'{i} Min\'{a}\v{r}
\\
\textit{Open quantum dynamics induced by light scalar fields}
\\
Phys.Rev. D100 (2019) no.7, 076003
\item[\cite{Kading4}] Clare Burrage, \textbf{Christian K{\"a}ding}, Peter Millington, and Ji\v{r}\'{i} Min\'{a}\v{r}
\\
\textit{Influence functionals, decoherence and conformally coupled scalars} 
\\
J.Phys.Conf.Ser. 1275 (2019) no.1, 012041
\end{itemize}
\underline{Preprint:}
\begin{itemize}
\item[\cite{Kading5}] Daniel Hartley, \textbf{Christian K{\"a}ding}, Richard Howl, and Ivette Fuentes
\\
\textit{Quantum-enhanced screened dark energy detection}
\end{itemize}
The focus of this thesis lies on articles \cite{Kading2,Kading3}. Ref.\! \cite{Kading4} is a conference proceedings article which provides a short overview of Ref.\! \cite{Kading3}. 

\newpage
\thispagestyle{empty}
\section*{Conventions}
The following conventions will be applied throughout this work:
\begin{itemize}
\item The $\eta_{\mu\nu} = \text{diag}(-1,+1,+1,+1)$ convention for the Minkowski metric is made use of.

\item Unless appearing explicitly, Planck's constant $\hbar$ and the speed of light in vacuum $c_0$ are set to $1$.

\item The reduced Planck mass $M_P = \sqrt{\frac{\hbar c_0}{8\pi G}} \approx 2.435 \times 10^{27} \,\text{eV}$ is used and only referred to as the Planck mass.

\item A Heaviside theta function in the half-maximum convention is defined as 
\begin{eqnarray*}
\Theta(X) &:=&
\frac{1}{2}[1+\sgn(X)] ~=~
\begin{cases} 
+1 &\mbox{if } X > 0 
\\
1/2 & \mbox{if } X = 0 
\\
0 & \mbox{if } X < 0 ~.
\end{cases} 
\end{eqnarray*}

\end{itemize}

\newpage

\pagenumbering{arabic}

\setcounter{page}{1} 

\section{Introduction}
Modern physics is resting on two main pillars: quantum physics and Einstein's theory of general relativity (GR). 
While quantum physics generally describes physics at very small scales - usually at the atomic and subatomic level, GR finds most of its applications on planetary scales and beyond.
Consequently, in most situations either one of the two theories can be ignored since their respective regimes of importance are so far apart from each other: 
on one hand, quantum physical effects are not considered to be of importance for describing macroscopic objects.
This is because objects containing many quantum particles are coarse grained such that they effectively look like classical entities and therefore do generally not require a probabilistic description. 
On the other hand, the gravitational force described by GR is so weak compared to the other three confirmed forces of Nature that its impact on processes at the quantum level can - usually - safely be ignored.
Nevertheless, some physical objects fall into the regimes of both theories.
Of those, black holes \cite{Schwarzschild,Droste} are some of the most prominent examples.
GR predicts them to be spacetime singularities, i.e.\ as points with an infinitely high mass density.
Certainly, such an infinity can only be of a mathematical nature and must hint at a limitation of the theory\footnote{The problem with an infinitely large dimensionful quantity is that it cannot be measured. Measuring a dimension is fundamentally a process of comparing and counting. For example, for measuring the length of an object the object is compared to a ruler and it is then counted how many times the ruler fits into the object. If the object was infinitely long, it would be practically impossible to confirm this since infinitely many steps of counting (with a finite ruler) would be required. }.
Removing this limitation could be achieved by a UV-completion of GR, which is commonly assumed to be quantum gravity -  a theory that reconciles quantum physics with GR.
There are several candidate theories for quantum gravity that introduce novel fundamental objects or concepts of spacetime (e.g.\ string theory \cite{Polchinski1,Polchinski2}, loop quantum gravity \cite{Rovelli}, causal sets \cite{Dowker}, causal dynamical triangulation \cite{Loll},...), while others  are proposed as classical alternatives to GR which might have properties beneficial for quantisation (e.g.\ shape dynamics \cite{Mercati},...).
It has also been suggested that a quantised version of GR is actually renormalisable and just requires a better understanding of the mathematical structures behind renormalisation \cite{Kreimer2007,Kreimer2008}.
However, to date no complete and fully accepted theory of quantum gravity has been found and the search for it remains one of physics' greatest challenges. 
\\\\
Motivated by this and other challenges concerning gravity, modifications of GR have been studied throughout history. 
One of the earliest was Theodor Kaluza and Oskar Klein's attempt to unify gravity with electromagnetism by considering a fourth spatial but circular dimension \cite{Kaluza,Klein}. 
The principles of the resulting Kaluza-Klein theory provide the basic ideas behind compactifications of modern higher dimensional theories like string theory or supergravity \cite{Freedman}. 
Besides trying to unify gravity with other forces of Nature, implementing Mach's principle\footnote{There are several ideas that are interpreted as Mach's principle (see e.g.\ {Ref.} \cite{BondiMach} for an overview). All of them are based on the experimental observation that inertial frames defined by local physics coincide with the frame in which distant objects, e.g.\ galaxies, are at rest \cite{BondiMach}.} \cite{Mach2012,Mach2013} is another guiding theme of some modified gravity theories, e.g.\ in the aforementioned shape dynamics or in Brans-Dicke theory \cite{Brans}. 
The latter is a so-called scalar-tensor theory of gravitation \cite{Fujii}, comprising a scalar field and the well-known metric tensor from GR that are coupled to each other. 
Such scalar-tensor theories are among the most studied modifications of gravity since, on one hand, adding an extra scalar field is one of the simplest ways of extending GR, while on the other hand some other modifications of gravity can be shown to be equivalent to  a scalar-tensor theory (e.g.\ $f(R)$-gravity \cite{Sotiriou}) or give rise to effective scalar degrees of freedom when higher dimensional spaces are compactified, as is the case for theories following the spirit of Kaluza-Klein theory.
\\\\
Modifications of gravity gained even greater attention after the accelerated expansion of the Universe was discovered \cite{Perlmutter,Riess} and the puzzle of dark energy (DE) - the energy that supposedly drives this expansion - arose.
In this context, some scalar-tensor theories were proposed as possible explanations for the nature of DE.
An overview of such models can, for instance, be found in Refs.\! \cite{Clifton,Joyce}.
Some of these models have already been ruled out by various astrophysical and cosmological observations but others are still viable theories.
An overview of currently allowed and excluded models can be gained from Ref.\! \cite{Ishak}.
Besides being possible explanations for DE, scalar-tensor theories have also been proposed as possible solutions to the question about the nature of dark matter (DM). Some recent ideas on this can be found in Refs.\! \cite{Khoury2014, Copeland2016}.
The scalar-tensor theories that are still viable candidates for successfully explaining DE or DM must be studied further and tested experimentally. 
For this, the observation of gravitational waves from a binary neutron star merger \cite{LIGO2017} provided important constraints from a cosmological point of view \cite{Jain,Langlois2017}, including strong limits on the difference in photon and graviton speeds predicted by some models.
\\\\
In testing and constraining scalar-tensor theories, the involved scalar degrees of freedom play an important role. 
Constraining the model parameters of the associated scalar fields or even finding a new scalar particle in Nature, besides the already known Higgs boson, would also constrain the parameters of the theory or hint at its existence, respectively.
Many theories require some kind of interaction between such a scalar field and ordinary matter, which provides a contact point for different types of experiments. 
One prediction that such an interaction leads to is that of an additional fundamental force - a long ranged fifth force\footnote{It should be stressed that this force is usually not considered to be an explanation of the accelerated expansion of the Universe but rather a byproduct of theories that might be able to explain DE.}.
A problem with such fifth forces is that, to date, they have not been observed in Nature and there are tight constraints on them from {Solar System}-based observations \cite{Adelberger}.
There are several ways of explaining this apparent absence of a fifth force even when the existence of an interaction between a novel scalar field and Standard Model particles is assumed:
the most obvious idea is that the field is very weakly coupled to matter and hence the force is extremely feeble. 
In particular, this force must be much weaker than gravity - the weakest of all currently known fundamental forces \cite{Zee}.
This, however, would make a fifth force very unappealing in the sense that it would be extremely difficult to impossible to observe this force even in astrophysical observations outside the {Solar System}. 
In addition, it would be the result of fine tuning and not very interesting from a theoretical point of view because it would be very challenging to assign a significant role within any kind of phenomenologically interesting physical theory to such a weak force.
\\\\
A popular alternative to a very weak coupling to matter are screening mechanisms. 
Such a mechanism leads to a suppression of the scalar field and its force in environments of high mass density but allows them to act with their full strength in vacuum. 
Scalar field models that have a screening mechanism are known as screened scalar fields.
Screened scalar fields are non-linear theories, which means that their perturbations couple to the background and results in fifth force screening. 
As a consequence of screening, fifth forces sourced by screened scalar fields are strongly suppressed in the {Solar System}, while on larger scales, where the mass density is generally lower, the force can reach phenomenologically interesting strengths that, in principle, can exceed the one of the gravitational force.
\\\\
Over time several types of screening mechanisms have been conceived, which usually either modify the scalar field's kinetics, mass,  matter coupling, or all three of them, in dependence of the surrounding environment.
They resulted in different types of screened scalar field models.
Testing these fields has been of great interest in recent years and a large variety of experiments has been realised, involving both astrophysical observations, and laboratory-based tests. 
A good review of experiments and their conclusions for some popular screened scalar field models can be found in Refs.\! \cite{BurrageSakA,BurrageSakB}.
Even though some of these experiments led to impressive constraints, to date, none of them was able to detect a new type of scalar field or cover all potentially interesting regions of their parameter spaces. 
Therefore, novel ideas are needed in the ongoing hunt for these fields.
\\\\
In this thesis, two new tests of screened scalar fields will be presented - one astrophysical, one quantum physical.
For this, screened scalar fields will be introduced in Section \ref{Sec:ScreenedScalarFields}.
Afterwards, in Section \ref{Sec:Lensing}, it will be discussed whether a particular type of screened scalar fields can cause a deviation from the GR prediction of gravitational lensing by galaxies. 
This test will provide part of the answer to the question whether the fifth force of a symmetron scalar field (see Section \ref{ssec:Symmetrons}) can be an alternative explanation for the galaxy dynamics otherwise attributed to particle DM \cite{Bertone}, as was discussed in Ref.\! \cite{Kading2}.
Next, in Section \ref{Sec:OpenQuantumfromCCSF}, the open dynamics of a quantum test particle induced by conformally coupled scalar fields\footnote{Conformally coupled scalar fields (see Section \ref{ssec:ScalarTensTheo} for more details) are a larger class of fields which comprises screened scalar fields.} will be investigated.
The intention behind this investigation is to find novel ways of constraining screened scalar fields with quantum physical experiments like atom interferometry \cite{Cronin}. 
For this, a quantum master equation will be derived in Section \ref{ssec:MasterEqnDerive} and its predictions discussed in Section \ref{ssec:ExpImplications}.
The derivation of this equation will require the usage of tools from non-equilibrium quantum field theory \cite{CalzettaHuQFT}, some of which have to be newly developed.
These novel tools and the derivation of the master equation are the main results of Refs.\! \cite{Kading3,Kading4} and will therefore be presented here.
Subsequently, closing remarks will be given and an outlook dared in Section \ref{sec:Outlook}. 
Supplementary material can be found in the attached Appendices \ref{app:5Forces} - \ref{app:Contractions}.

\newpage
\clearpage\mbox{}\clearpage
\newpage
\section{Screened scalar fields}\label{Sec:ScreenedScalarFields}

To date there is only one experimentally confirmed elementary scalar field in Nature which, following a broad consensus, is considered to be the Standard Model Higgs boson \cite{Higgs1,Higgs2,Englert,ATLAS,CMS}.
However, there is per se no reason why there should not be more fundamental scalar fields to be found.
In fact, there are several theories that predict the existence of such fields. 
Amongst those theories are higher dimensional ones like string theory \cite{Polchinski1,Polchinski2} and supergravity \cite{Freedman} but also generalisations of GR like $f(R)$-gravity \cite{Sotiriou}. 
\\\\
Even without having a concrete prediction for the existence of novel scalar fields, they can be conceived from a model building standpoint in which they are used for explaining phenomena relevant in different areas of physics.
For example, it is believed that there could be so-called hidden sectors of the Standard Model of particle physics (SM) \cite{Borodulin} in which additional scalar fields reside. 
These extra fields are sometimes considered as possible explanations for DM (see e.g.\ Ref.\! \cite{Battaglieri} for an overview).
The word hidden implies that particles that are housed in this sector do not couple in the usual way to other SM particles, i.e.\ they do not interact via the known gauge bosons, namely photons, gluons, W- and Z-bosons.
Instead they could, for instance, couple via a so-called Higgs portal \cite{Schabinger,Patt}, meaning that they actually interact with the Higgs boson which in turn interacts with the other SM particles.
\\\\
As well as in enlargements of the SM, additional scalar fields commonly appear in the context of theories of modified gravity, i.e.\ in extensions or alterations of GR\footnote{SM extensions and modified gravity theories can actually be intimately connected, see Ref. \cite{HiggsEqui}, but are still often considered separately.}.
Some of these modified gravity theories are so-called scalar-tensor theories \cite{Fujii} in which the scalar degree of freedom is in some way coupled to the gravitational metric tensor.
A common way of realising such {a} coupling between scalar field $\Phi$ and metric $g_{\mu\nu}$ is via a conformal factor, e.g.\ $A(\Phi)g_{\mu\nu}$.  
Recently it has been shown \cite{HiggsEqui} that there is an equivalence between Higgs portal scalar extensions of the SM and conformally coupled scalar-tensor modifications of gravity.
This means that there is actually an intimate relationship between these two areas of physics in which additional scalar fields could appear.
\\\\
In general, the interplay between scalar field and metric in gravitational scalar-tensor theories, of which Brans-Dicke theory \cite{Brans} was one of the earliest, leads to a coupling between the scalar and the trace of the energy-momentum tensor $T_{\mu\nu}$ of other matter (see Appendix \ref{app:CouplingBetwe}). 
This type of interaction gives rise to a gravity-like fifth force of Nature.
A problem with theories that predict fifth forces is that their prediction has so far not been confirmed in any experiment or observation.
In fact, there are very tight constraints on gravity-like fifth forces from {Solar System}-based tests \cite{Adelberger}.
Besides assuming that theories predicting such fifth forces are plainly not accurate descriptions of Nature, there are several ways of solving this problem.
For example, the force carrying scalar could be very heavy which would result in a short ranged force, effectively rendering it to be weak.
It would also be possible to take the coupling between scalar and matter to be always extremely weak.
Both ideas would be sensible reasons for the absence of an additional force of Nature.
However, such fine tunings do not provide any deeper explanation for why this would be the case.
\\\\
Another option, that also leads to a more interesting phenomenology, is given by so-called screening mechanisms.
These are different methods of suppressing a force in environments of high mass or energy density.
They arise from the non-linear nature of the scalar theories they appear in and give a fifth force {an} environmental dependence{, i.e.\ on the energy momentum tensor $T_{\mu\nu}$ or the curvature}.
More precisely, if $T^\mu_{~\mu}$ is sufficiently large, the force resulting from its coupling to the scalar field is weak or even entirely vanishing due to a screening mechanism.
In situations where the trace of the energy-momentum tensor is smaller, the force is less suppressed and can be strong enough to be of phenomenological interest. 
For non-relativistic matter the assumption $T^\mu_{~\mu} = -\rho$ (see e.g.\ Ref.\! \cite{Joyce}), with $\rho$ being a mass density, is made.
Assuming that $\rho$ in the {Solar System} is large enough, a screening mechanism makes it possible to circumvent the constraints on fifth forces by {Solar System}-based tests and provides the associated scalar fields with a rich phenomenology that can be tested in experiments and observations{\footnote{This of course requires that the screening model parameters are such that the densities of the Sun and the Solar System planets are large enough to suppress a fifth  force sufficiently much.}}.  
Such tests usually take place on beyond {Solar System} scales or, for some models (see e.g.\ chameleons in Section \ref{ssec:Chameleons}), under special laboratory conditions, e.g.\ in an artificial vacuum within a vacuum chamber. 
Scalar fields that have a screening mechanism are called screened scalar fields.
\\\\
Screened scalar fields are the main subject of this thesis. 
Therefore, they will be more extensively discussed in the current section.
At first, the reader will be reminded about scalar fields in general in Section \ref{ssec:ScalarFields}.
Then some aspects of scalar-tensor theories of gravity will be discussed in Section \ref{ssec:ScalarTensTheo}, more specifically, conformal and disformal couplings will be introduced, and Einstein and Jordan frames discussed.
Subsequently, in Section \ref{ssec:fR}, it will be shown that a generalisation of GR named $f(R)$-gravity is equivalent to a scalar-tensor theory.
Finally, three of the most prominent screened scalar field models will be introduced, namely galileons, symmetrons, and chameleons {(confer Section \ref{ssec:ScreenedScalarModels})}.    

\newpage
\subsection{Scalar fields}\label{ssec:ScalarFields}

Scalar fields, here denoted by $\phi$, are the most simple objects in classical and quantum field theory (QFT).
They represent Lorentz invariant\footnote{Meaning that $\phi(x)$ and $\phi(\Lambda^{-1}x)$, with $\Lambda$ as an element of the Lorentz group \cite{Costa}, both fulfil the same equation of motion, which is reflected in their actions being equal.} scalar degrees of freedom\footnote{Here the term scalar means that this object is represented by a single complex number. However, all scalar fields considered here are assumed to be real.} with, in $4$-dimensional spacetime, the dimension of an energy.
Here some relevant facts about scalar fields will be provided, and the notation that will be used in this thesis will be introduced. 
{For this, the $(-,+,+,+)$-metric convention will be used, and $c_0 = \hbar = 1$ set.}
\\\\
A free scalar field fulfils its equation of motion, the Schr\"odinger-Gordon or Klein-Gordon wave equation \cite{Wentzel}
\begin{eqnarray}\label{eqn:KG}
(\Box - m_\phi^2)\phi = 0,
\end{eqnarray} 
where $\Box$ is the d'Alembert operator
\begin{eqnarray}
\Box = -\frac{\partial^2}{\partial t^2} + \vec{\nabla}, 
\end{eqnarray}
and $m_\phi$ the mass associated to the scalar field $\phi$. 
The Lagrangian density\footnote{Later a Lagrangian density will often be just referred to as a Lagrangian.} of such a free field is given by
\begin{eqnarray}\label{eqn:FreeLagrangian}
\mathcal{L}_0 &=& -\frac{1}{2}(\partial\phi)^2 -\frac{1}{2}m_\phi^2\phi^2,
\end{eqnarray}
and the resulting action is 
\begin{eqnarray}
{S_0} &=& \int d^4x \mathcal{L}_0.
\end{eqnarray}
An interacting theory describing interactions of the scalar with itself and other fields, represented by $\chi$, is obtained by subtracting an interaction potential density $V(\phi,\chi)$ from the free Lagrangian in Eqn.\! (\ref{eqn:FreeLagrangian}), which yields
\begin{eqnarray}\label{eqn:InterLagrangian}
\mathcal{L} &=& -\frac{1}{2}(\partial\phi)^2 -\frac{1}{2}m_\phi^2\phi^2 - V(\phi,\chi),
\end{eqnarray}
and an equation of motion
\begin{eqnarray}
(\Box - m_\phi^2)\phi = \frac{\partial}{\partial\phi}V(\phi,\chi),
\end{eqnarray}
following from the Euler-Lagrange equations \cite{Zee}.
\\\\
A QFT of scalars is usually obtained in either one of two ways: by canonical quantisation \cite{Srednicki} or by Feynman's path integral formalism \cite{FeynmanPaths}.
\\\\
For the former a conjugate momentum density \cite{PeskinSchroeder}
\begin{eqnarray}
\pi(x) &:=& \frac{\partial\mathcal{L}}{\partial \dot{\phi}(x)} 
\end{eqnarray}
is defined.
Both, $\phi$ and $\pi$, are then promoted to operators
\begin{eqnarray}
\phi \to \hat{\phi},
\\
\pi \to \hat{\pi},
\end{eqnarray}
which have to obey the equal time commutation relations (in the Heisenberg picture \cite{BHall}) \cite{PeskinSchroeder}
\begin{eqnarray}
[\hat{\phi}(t,\vec{x}),\hat{\pi}(t,\vec{y})] &=& i \delta^{(3)}(\vec{x}-\vec{y}), 
\end{eqnarray}
\begin{eqnarray}
[\hat{\phi}(t,\vec{x}),\hat{\phi}(t,\vec{y})] &=& [\hat{\pi}(t,\vec{x}),\hat{\pi}(t,\vec{y})] ~=~ 0,
\end{eqnarray}
where $[\cdot,\cdot]$ is a commutator.
\\\\
The scalar field operator can be decomposed in plane waves via \cite{PeskinSchroeder}
\begin{eqnarray}
\hat{\phi}(x) &=& \int d\Pi_{\vec{k}} \left( \hat{a} e^{-iE_{\vec{k}}t + i\vec{k}\cdot\vec{x}}  + \hat{a}^\dagger e^{iE_{\vec{k}}t -i\vec{k}\cdot\vec{x}} \right)
\end{eqnarray}
where $\hat{a}^\dagger$ and $\hat{a}$ are creation and annihilation operators, respectively, and
\begin{eqnarray}\label{eqn:ScFiNota1}
\int d\Pi_{\vec{k}} &:=& \int_{\vec{k}} \frac{1}{2E_{\vec{k}}},
\end{eqnarray}
where the energy 
\begin{eqnarray}
E_{\vec{k}} &:=& \sqrt{\vec{k}^2 + m_\phi^2 + \partial_\phi^2 V(\phi,\chi)|_{\phi_0 \in \{ \varphi:~\partial_\phi \mathcal{L}|_{\varphi} \,=\,0\}}  }\,,
\end{eqnarray}
and 
\begin{eqnarray}\label{eqn:ScFiNota2}
\int_{\vec{k}} &:=& \int \frac{d^3k}{(2\pi)^3}
\end{eqnarray}
are included.
\\\\
The creation and annihilation operators fulfil \cite{PeskinSchroeder}
\begin{eqnarray}
[\hat{a}_{\vec{p}},\hat{a}_{\vec{k}}^\dagger] &=&(2\pi)^3 2E_{\vec{p}} \delta^{(3)}(\vec{p}-\vec{k}),
\end{eqnarray}
\begin{eqnarray}
[\hat{a}_{\vec{p}},\hat{a}_{\vec{k}}] &=& [\hat{a}_{\vec{p}}^\dagger,\hat{a}_{\vec{k}}^\dagger] ~=~ 0.
\end{eqnarray}
With the help of the annihilation operator a vacuum state $|0\rangle$ fulfilling
\begin{eqnarray}
{\hat{a}}_{\vec{k}}|0\rangle &=& 0
\end{eqnarray}
is defined.
Computing the expectation value of two scalar fields in this state, 
\\
$\langle 0 | {\hat{\phi}(x) \hat{\phi}}(y) | 0 \rangle$, leads to the vacuum transition amplitude for a particle to go from spacetime point $y$ to spacetime point $x$ given by the {free} positive frequency Wightman propagator \cite{PeskinSchroeder,CalzettaHuQFT}
\begin{eqnarray}\label{eqn:PosWightmanPropy}
\langle 0 | {\hat{\phi}(x) \hat{\phi}}(y) | 0 \rangle ~=~ \Delta^>_{xy} &=&  \int d\Pi_{\vec{k}} e^{-iE_{\vec{k}}(x^0 - y^0) + i\vec{k}\cdot(\vec{x}-\vec{y})}.
\end{eqnarray}
Complex conjugating Eqn.\! (\ref{eqn:PosWightmanPropy}) or swapping $x$ and $y$ in its arguments leads to the {free} negative frequency Wightman propagator $\Delta^<_{xy}$.
Applying the time ordering operator ${\hat{T}}$, given by
\begin{eqnarray}
{\hat{T}}{\hat{\phi}(x) \hat{\phi}}(y) &:=& \Theta(x^0 - y^0){\hat{\phi}(x) \hat{\phi}}(y) + \Theta(y^0 - x^0){\hat{\phi}(y) \hat{\phi}}(x),
\end{eqnarray}
leads to the well-known {free} Feynman propagator in terms of the Wightman propagators \cite{CalzettaHuQFT}
\begin{eqnarray}\label{eqn:FeynProp}
\Delta^{\rm F}_{xy} &=& \langle 0 | {\hat{T}} {\hat{\phi}(x) \hat{\phi}}(y) | 0 \rangle 
\\
&=& \Theta(x^0 - y^0)\Delta^>_{xy} +\Theta(y^0 - x^0)\Delta^<_{xy}
\\
&=& -i \int \frac{d^4k}{(2\pi)^4}\frac{e^{ik(x-y)}}{k^2 + m_\phi - i\epsilon},
\end{eqnarray}
where $\epsilon \to 0$.
The complex conjugated version of Eqn.\! (\ref{eqn:FeynProp}) is the {free} Dyson propagator $\Delta^{\rm D}_{xy}$.
More information on some useful properties of those propagators can be found in Appendix \ref{app:PropoScaPro}.
\\\\
An alternative to canonical quantisation is Feynman's path integral formalism.
In this formalism an integral is taken over all spacetime paths that a classical particle could take, which effectively describes the behaviour expected of a quantised object.
Both, canonical quantisation and the path integrals, lead to the same physical predictions as they are equivalent \cite{GreinerQuantization}.
For example, the vacuum Feynman propagator in Eqn.\! (\ref{eqn:FeynProp}) can be obtained from path integrals 
as \cite{PeskinSchroeder}
\begin{eqnarray}
\Delta^{\rm F}_{xy} &=& \langle 0 | {\hat{T}} {\hat{\phi}(x) \hat{\phi}}(y) | 0 \rangle ~=~ \frac{\int\mathcal{D}\phi e^{iS_0}\phi(x)\phi(y)}{\int\mathcal{D}\phi e^{iS_0}},
\end{eqnarray}
where the denominator is a normalisation factor, $S_0$ is the free action of $\phi$, and 
\begin{eqnarray}
\mathcal{D}\phi &=& \prod\limits_i d\phi(x_i)
\end{eqnarray}
is an expression for all possible integration paths of $\phi$.
Since such path integrals describe integrations over functions, they are also known as functional integrals \cite{Srednicki}.
\\\\
Higher order correlation functions in general states are given by{\footnote{From now on and for the rest of this thesis the hat notation for operators in correlation functions will be dropped for convenience.}} \cite{Rammer}
\begin{eqnarray}\label{eqn:HigherOrderCorre}
\langle \phi_1(x_1) ... \phi_N(x_N) \rangle &=& \frac{\int\mathcal{D}\phi e^{iS}\phi_1(x_1)...\phi_N(x_N)}{\int\mathcal{D}\phi e^{iS}} {.}
\end{eqnarray}
{If they are given by time-ordered products of fields in Gaussian states, then they can} be reduced to products of two-point functions with Wick's theorem \cite{Wick} (see also e.g.\ Ref.\! \cite{PeskinSchroeder}).
\\\\
Before ending this short overview of scalar fields, a useful feature shall be discussed:
it is often demanded that scalar fields and all their derivatives vanish at infinity, which leads to
\begin{eqnarray}\label{eqn:Schwartz}
\int\limits^\infty_{-\infty} d^4x \partial_\mu \left( \partial_\alpha\phi(x) f(x) \right) = \partial_\alpha\phi(x) f(x)|^{+\infty}_{-\infty} =0,
\end{eqnarray}
where $\alpha\in \mathbb{N}_0^n$ \footnote{Here the subscript $0$ indicates that this set includes the number $0$. $n$ denotes that $\alpha$ is a multi-index which here indicates derivatives with respect to different variables, e.g.\ $x^0$, $x^1,...$ .} and $f$ is a function of $x$. 
\\\\
This is, for example, physically required in order to ensure that the total energy of the field is conserved.
According to Ref.\! \cite{Schwichtenberg}, a variation of the free scalar field Lagrangian leads to the continuity equation
\begin{eqnarray}\label{eqn:Continuity}
\partial^\mu T^\phi_{\mu\nu} &=& 0
\end{eqnarray}
of the scalar field energy-momentum tensor \cite{Blencowe}
\begin{eqnarray}
T^\phi_{\mu\nu} &=& \partial_\mu \phi \partial_\nu \phi - \frac{1}{2} \eta_{\mu\nu} \partial_\rho \phi \partial^\rho \phi - \frac{1}{2}\eta_{\mu\nu} m_\phi^2 \phi^2,
\end{eqnarray} 
whose $T^\phi_{00}$ component corresponds to the spatial energy density of the field.
Energy conservation can then be expressed as 
\begin{eqnarray}
\partial_t E_\phi &=& 0,
\end{eqnarray}
where $E_\phi$ is the total energy of $\phi$,
meaning \cite{Schwichtenberg}
\begin{eqnarray}\label{eqn:EnergyConservation}
\partial_t \int_{\text{vol}} d^3x T^\phi_{00} &=&  \int_{\text{vol}} d^3x \vec{\nabla} \vec{T}^\phi
\nonumber
\\
&=& \int_{\delta\text{vol}} d^2x \vec{T}^\phi {\cdot\vec{n}}
\nonumber
\\
&=& 0
\end{eqnarray}
where $\text{vol}$ denotes an infinite volume, $\delta\text{vol}$ its surface, {$\vec{n}$ a normal unit vector on $\delta\text{vol}$,} and $\vec{T}^\phi$ the spatial part of $T^\phi_{\mu\nu}$.
In the first line of Eqn.\! (\ref{eqn:EnergyConservation}), Eqn.\! (\ref{eqn:Continuity}) was made use of, and in the second line the divergence theorem \cite{Bronshtein} was applied.
Since surface terms at infinity are required to vanish, energy is conserved, resulting in the third line.
\\\\
One way of mathematically implementing that scalar fields vanish at infinity is given by interpreting them as operator-valued distributions acting on Schwartz functions (see e.g.\ Refs.\! \cite{Zeidler,Schottenloher}).
Besides giving a formal justification for this particular behaviour of scalar fields, this interpretation also provides a more realistic description of them since in experiments the field strength is never measured at a single spacetime point but instead in a particular region of space and within a finite time interval \cite{Schottenloher}. 
\\\\
Now it will be explained how this interpretation is actually implemented.
In general, operator-valued distributions are maps from a function space $\mathscr{F}$ to a set of operators $\mathscr{O}$, which fulfil some additional requirements (see e.g.\ Ref.\! \cite{Schottenloher}).
A function $\xi(x)$ of the real Schwartz space $\mathscr{S}(\mathbb{R}^n)$ is simply speaking a rapidly decaying function fulfilling \cite{Iske}
\begin{eqnarray}
\lim_{|x|\to\infty} \xi(x) p(x) \to 0
\end{eqnarray}
for any polynomial $p(x)$, and all its derivatives $\partial_\alpha \xi(x) \in \mathscr{S}(\mathbb{R}^n)$.
Scalar field operators can then be expressed as \cite{Grange}
\begin{eqnarray}
\hat{\phi}(x) &=& {\hat{\mathcal{T}}_x \hat{\Phi}}(\xi), 
\end{eqnarray} 
where ${\hat{\mathcal{T}}}_x$ is a translation operator in coordinate space and ${\hat{\Phi}}$ is an operator-valued distribution acting on the Schwartz function $\xi$, such that
\begin{eqnarray}
{\hat{\mathcal{T}}_x \hat{\Phi}}(\xi) &=& \int d^4y \hat{\chi}(y) \xi(x-y).
\end{eqnarray}
The operator $\hat{\chi}$ is here the ``naive" scalar field operator that locally depends on one spacetime point.
Due to the convolution with the Schwartz function, the scalar field is smeared out, in accordance with Heisenberg's uncertainty principle.
In addition, the scalar field satisfies
\begin{eqnarray}
\forall n \in \mathbb{N}:~\int dx \partial^n_x\hat{\phi}(x) &\sim & \int dx \partial^n_x\xi(x-y), 
\end{eqnarray}
which has to vanish at $\pm \infty$, as desired.

\newpage
\subsection{Scalar-tensor theories of gravity}\label{ssec:ScalarTensTheo}

In general, a scalar-tensor theory comprises at least one scalar degree of freedom and at least one tensorial object, which are coupled to each other in some way. 
In the case of a scalar-tensor theory of gravity \cite{Fujii}, the scalar degrees of freedom are interacting with the gravitational metric tensor $g_{\mu\nu}$. 
Such a theory provides a potential alternative to GR and is therefore considered to be a modified gravity theory.
Early examples include Jordan's theory \cite{Jordan} and Brans-Dicke theory \cite{Brans}. 
A motivation for modifying gravity in such a way can be found in string theory and other higher dimensional theories like braneworld models. 
There it is usually predicted that the metric tensor of 4-dimensional gravity is accompanied by additional scalar fields effectively resulting from compactifications of the higher dimensions \cite{Fujii}.
Besides, a generalisation of GR like $f(R)$-gravity \cite{Sotiriou} can be shown to be equivalent to a scalar-tensor theory, as will be elaborated on in Section \ref{ssec:fR}. 
In addition, scalar-tensor theories are used for attempts to solve mysteries in modern cosmology, e.g.\ the question about the nature of DE. 
For this particular example, quintessence models \cite{Steinhardt2003} have been of great interest.
Alternatively, part of the origin of the cosmological constant could be explained by a scalar field with a non-vanishing vacuum expectation value leading to constant terms in its own Lagrangian that can be interpreted as contributions to the cosmological constant and consequently DE. 
\\\\
The most common way of coupling the scalar $\Phi$ to the metric tensor is via a conformal factor $A(\Phi)$. 
In fact, scalar-tensor theories of gravity are only defined up to a conformal transformation leading from one so-called conformal frame to another \cite{Fujii}.
These conformal frames are merely different mathematical formulations - some calculations are maybe easier to perform in one frame than in another.
Of course the theoretical prediction for a physical measurement cannot be altered due to a change in the mathematical formulation.
However, the interpretation of the physics can differ, as will become more apparent when considering particular examples:
the Einstein and the Jordan frame which are the two most prominent conformal frames. 
\\\\
Considering a total action as a sum of a gravitational action and an action that describes the dynamics of SM matter, then a theory  of gravity in the Jordan frame with metric $\tilde{g}$ and scalar field $\Phi$ may look schematically like
\begin{eqnarray}\label{eqn:JordanFrame}
S &=& \frac{M_P}{2} \int d^4x \sqrt{-\tilde{g}}\Phi \tilde{R} + S_m(\tilde{g}_{\mu\nu},\psi) ,
\end{eqnarray}
where $\tilde{R}$ is the Jordan frame Ricci scalar in the modified Einstein-Hilbert action \cite{Ortin}, and $S_m$ is the matter action with $\psi$ as a representative of any SM matter field. 
For simplicity, even though they are also subject to conformal transformations, the kinetic and potential terms of $\Phi$ were not considered here since they are not relevant for the current discussion.
See Section \ref{ssec:fR} for an example in which all terms are included. 
\\\\
It is clear that the scalar $\Phi$ in Eqn.\! (\ref{eqn:JordanFrame}) only couples to the gravitational sector but not to the SM particles.
In contrast, the Einstein frame formulation recovers the canonical Einstein-Hilbert action but has a coupling between scalar field and ordinary matter:
\begin{eqnarray}\label{eqn:EinsteinFrame}
S &=& \frac{M_P}{2} \int d^4x \sqrt{-g} R + S_m(\Phi^{-1}g_{\mu\nu},\psi).
\end{eqnarray}
Here quantities without tilde $\tilde{~}$ belong to the Einstein frame.
It can be seen that Jordan frame action in Eqn.\! (\ref{eqn:JordanFrame}) and Einstein frame action in Eqn.\! (\ref{eqn:EinsteinFrame}) lead to two distinct physical interpretations: a test particle in the Jordan frame moves on a geodesic of $\tilde{g}$ that, as a results of the  presence of the scalar field in the gravitational sector, differs from the one predicted by GR. 
However, the same test particle in the Einstein frame would, if there was only gravity, move on a GR geodesic.
Though, its actual trajectory deviates from this geodesic since the particle interacts with a scalar field which leads it to the experience of a fifth force.
In the former case gravity was modified, in the latter case gravity was unchanged but a novel interaction introduced.
Nevertheless, in both cases the experimentally measured trajectory of the test particle will be the same since a conformal frame is  only a mathematical concept and not part of an observable reality\footnote{Whether this is also true at the quantum level is still a subject to debate. Some references \cite{Faraoniquant,Banerjee1} claim that there is no equivalence between Jordan frame and Einstein frame in quantised theories. However, more recent work, see Ref. \cite{Pandey1}, shows that at least for Brans-Dicke theory \cite{Brans}, both frames are equivalent even at the quantum level.}. 
\\\\
In order to go from the Einstein to the Jordan frame, a conformal transformation of the form
\begin{eqnarray}\label{eqn:Conformal}
\tilde{g}_{\mu\nu} &=& A(\Phi)g_{\mu\nu}
\end{eqnarray}
is applied. 
Extending this, there is a more general concept that also includes a so-called disformal coupling. 
It was proposed by J. Bekenstein in Ref.\! \cite{Bekenstein}.
A disformal transformation, including also a conformal coupling, is given by \cite{Minamitsuji}
\begin{eqnarray}\label{eqn:Disformal}
\tilde{g}_{\mu\nu} &=& A(X_\Phi,\Phi)g_{\mu\nu} + B(X_\Phi,\Phi) \Phi,_{\mu}  \Phi,_{\nu},
\end{eqnarray} 
where the last term is called the disformal coupling, $X_\Phi :=g^{\mu\nu}\Phi,_\mu\Phi,_\nu$, $\Phi,_{\mu} := \nabla_\mu\Phi$, and $\nabla_\mu$ is a covariant derivative.
Eqn.\ (\ref{eqn:Disformal}) is the most general metric transformation that involves one scalar field, and preserves causality and locality.
Clearly, calculations with Eqn.\ (\ref{eqn:Disformal}) are generally more involved than those with Eqn.\! (\ref{eqn:Conformal}) but disformal couplings have the advantage that they allow the scalar field to couple to massless fields like photons.
Massless fields are conformally invariant and can therefore not interact with the scalar via a conformal coupling.
Disformal couplings on the other hand allow for such interactions.
This makes them ideal candidates for modified gravity theories that aim to predict deviations from the gravitational lensing effects predicted by GR.
Such deviations will be discussed in Section \ref{Sec:Lensing}.
\\\\
Before ending this subsection it shall be noted that there is a most general{\footnote{In fact, there are actually even more general models called beyond Horndeski theories, see e.g.\ Ref. \cite{Gleyzes:2014dya}, but they will not be discussed here.}} and healthy \cite{Kobayashi} scalar-tensor theory of gravity, where healthy means that it avoids Ostrogradsky instabilities \cite{Ostrogradsky,Woodard}. 
It is called Horndeski theory \cite{Horndeski} and defined by the Lagrangian \cite{Deffayet2011,Kobayashi}
\begin{eqnarray}\label{eqn:Horndeski}
\mathcal{L} &=& G_2(\phi,X_\phi) -G_3(\phi,X_\phi)\Box\phi + G_4(\phi,X_\phi)R 
\nonumber
\\
&\phantom{=}&
+  G_4,_{{X_\phi}}(\phi,X_\phi)[(\Box\phi)^2 - \phi,^{\mu\nu}\phi,_{\mu\nu}]
+ G_5(\phi,X_\phi)G^{\mu\nu}\phi,_{\mu\nu} 
\nonumber
\\
&\phantom{=}&
- \frac{1}{6}G_5,_{X_\phi}(\phi,X_\phi)[(\Box\phi)^3 -3\Box\phi\phi,^{\mu\nu}\phi,_{\mu\nu} +2\phi,_{\mu\nu}\phi,^{\nu\lambda}\phi,^\mu_\lambda ],
\end{eqnarray}
where $R$ is the Ricci scalar, $G^{\mu\nu}$ the Einstein tensor \cite{Nakahara}, and $G_2,...,G_5$ are arbitrary functions of $\phi$ and $X_\phi$. 
It is straightforward to see that the Lagrangian in Eqn.\! (\ref{eqn:InterLagrangian}) for a scalar field with canonical kinetic term, mass $m$, and potential $V$ is a special case of Eqn.\! (\ref{eqn:Horndeski}) and obtained from setting
\begin{eqnarray}
G_3(\phi,X_\phi) &=& G_4(\phi,X_\phi) ~=~ G_5(\phi,X_\phi) ~=~ 0,
\end{eqnarray}
and
\begin{eqnarray}
G_2(\phi,X_\phi) &=& -\frac{1}{2}X_\phi - \frac{1}{2}m^2\phi^2 - V(\phi).
\end{eqnarray}
In order to have a scalar-tensor theory, however, $G_4$ must be non-vanishing, such that the tensor field has a kinetic term.
{For this, the minimal choice is 
\begin{eqnarray}
G_4(\phi,X_\phi) &=& \frac{M_P}{2}\phi.
\end{eqnarray}
}
\newpage
\subsection{Equivalence of $f(R)$ gravity and scalar-tensor theories}\label{ssec:fR}

As mentioned before, there are several reasons for motivating gravitational scalar-tensor theories.
One of them is $f(R)$-gravity which will now be shown to be equivalent to such a theory.
$f(R)$-gravity was initially proposed in Ref.\! \cite{Buchdahl} as an attempt to avoid the occurrence of empty or singular states in the evolution of the Universe.
It is a modified gravity theory that introduces a change to the usual Einstein-Hilbert action \cite{Ortin}, such that a total action containing a gravitational action and an action that describes the dynamics of SM matter is given by 
\begin{eqnarray}\label{eqn:fRaction}
S &=& \frac{M_P^2}{2}\int d^4x \sqrt{-\tilde{g}}f(\tilde{R}) + S_m(\tilde{g}_{\mu\nu},\psi),
\end{eqnarray}
where $f$ is an arbitrary analytic function of the Ricci scalar $\tilde{R}$ \footnote{The action in Eqn.\! (\ref{eqn:fRaction}) is in the Jordan frame and therefore all quantities derived from the metric $\tilde{g}_{\mu\nu}$ that it contains are labelled with a tilde $\tilde{~}$.}.
Einstein's theory of gravity is recovered by simply setting $f(\tilde{R}) =\tilde{R}$.
Using an arbitrary function $f(\tilde{R})$ instead of only $\tilde{R}$ in the gravitational action can be motivated by the idea that the Ricci scalar in the original Einstein-Hilbert action is only the first term in an expansion of the analytic function $f$.
The additional terms of this expansion provide possible extensions of GR that could potentially be tested in astrophysics and cosmology. 
An overview and more details of $f(R)$-gravity can be gained from the review in Ref.\! \cite{Sotiriou}.   
The following discussion is also partially based on this article.
\\\\
It shall now be shown that Eqn.\! (\ref{eqn:fRaction}) is equivalent to the action of a scalar-tensor theory.
For this, a new scalar field $\chi$ is introduced, and with it an action defined:
\begin{eqnarray} \label{eqn:alternativeAction}
S &=& \frac{M_P^2}{2}\int  d^4x \sqrt{-\tilde{g}}(f(\chi) + f'(\chi)(\tilde{R}-\chi)) + S_m(\tilde{g}_{\mu\nu},\psi)  .
\end{eqnarray}
Next, the aim is to show that Eqns.\! (\ref{eqn:fRaction}) and (\ref{eqn:alternativeAction}) are dynamically equivalent, meaning that the equation of motion resulting from varying one action should lead to a condition that recovers the other one.
Indeed, varying the action in Eqn.\! (\ref{eqn:alternativeAction}) with respect to $\chi$ and requiring this variation to vanish, i.e.\ $\delta S/\delta \chi =0$, yields
\begin{eqnarray}
f'(\chi) + f''(\chi) (\tilde{R}-\chi) -f'(\chi) &=& 0
\nonumber
\\
\Leftrightarrow f''(\chi) (\tilde{R}-\chi) &=& 0.
\end{eqnarray}
This means that $\tilde{R} = \chi$ if $f''(\chi) \neq 0$, which then leads back to Eqn.\! (\ref{eqn:fRaction}).
Subsequently, after defining $\Phi := f'(\chi)$ and $\mathcal{V}(\Phi) := \Phi\chi - f(\chi)$, Eqn.\! (\ref{eqn:alternativeAction}) becomes
\begin{eqnarray}\label{eqn:alteredAction}
S &=& \frac{M_P^2}{2}\int  d^4x \sqrt{-\tilde{g}}(\Phi \tilde{R} - \mathcal{V}(\Phi)) + S_m(\tilde{g}_{\mu\nu},\psi).  
\end{eqnarray}
In addition, the identification $\chi = \tilde{R}$ leads to $\tilde{R} = V'(\Phi)$ and $\Phi = f'(\tilde{R})$.
Next, following Ref.\! \cite{Pearson}, a conformal transformation from the Jordan to the Einstein frame with
\begin{eqnarray}
\tilde{g}_{\mu\nu} = \Phi^{-1} g_{\mu\nu}, 
\end{eqnarray}
where
\begin{eqnarray}
\sqrt{-\tilde{g}} &=& \Phi^{-2}\sqrt{-g}, 
\end{eqnarray}
and 
\begin{eqnarray}
\tilde{R}  &=& \Phi\left( R + 3\Big( \frac{\Box\Phi}{\Phi} - \frac{3}{2} \Big( \frac{\nabla\Phi}{\Phi} \Big)^2 \Big) \right),
\end{eqnarray}
is applied to Eqn.\! (\ref{eqn:alteredAction}):
\begin{eqnarray}\label{eqn:EinsteinfrLagr}
S &=& \frac{M_P^2}{2}\int  d^4x\sqrt{-g} \left( R + 3\frac{\Box\Phi}{\Phi} - \frac{9}{2}(\nabla\ln\Phi)^2- \frac{\mathcal{V}(\Phi)}{\Phi^2} \right) 
\nonumber
\\
&\phantom{=}&
+ S_m(\Phi^{-1}g_{\mu\nu},\psi) ,
\end{eqnarray}
where $\nabla\Phi/\Phi = \nabla\ln\Phi$ was used.
\\\\
Substituting $\Phi = e^{2\beta\phi/M_p}$ with $\beta = 1/\sqrt{6}$ and defining $V(\phi) := M_P^2\mathcal{V}(\Phi)/2\Phi^2$ changes Eqn.\! (\ref{eqn:EinsteinfrLagr}) into
\begin{eqnarray}\label{eqn:fRfinal}
S &=& \int d^4x \sqrt{-g}\left( \frac{M_P^2}{2}R - \frac{1}{2}(\partial\phi)^2 - V(\phi) \right) + S_m\left( e^{-{2}\beta\phi/M_p} g_{\mu\nu},\psi \right), 
\nonumber
\\
\end{eqnarray}
where a term proportional to $\Box\phi$ was dropped since it would vanish after integrating over $x$, as was explained in Section \ref{ssec:ScalarFields}.
The action in Eqn.\! (\ref{eqn:fRfinal}) is now in a nice form and represents a Einstein frame scalar-tensor theory. 
By this result it was shown that $f(R)$-gravity can be expressed in terms of a scalar-tensor theory, which means that {it} suffers from the same problems with being compatible with experimental tests that were mentioned earlier. 
\newpage
\subsection{Screened scalar field models}\label{ssec:ScreenedScalarModels}

Screening mechanisms are means to circumvent {Solar System} constraints on scalar fifth forces.
They arise from the non-linearity of their respective scalar field theories and lead to a suppression of a fifth force in environments of high mass density.
Over the years there have been several different types of scalar field models emerging that address the problem of strongly constrained fifth forces via a screening mechanism. 
The most prominent models can be assigned to one of the following categories of screening mechanisms:
\begin{itemize}
\item kinetic screening and Vainshtein mechanism - the kinetic term is modified (or additional kinetic terms are added) and depends on the {environmental mass density or curvature, such that the kinetics of the field is suppressed in dense or strongly curved regions} and the resulting force has a short range;
\item varying coupling - the coupling to matter varies with the environmental density, such that it becomes very weak in dense regions;
\item varying mass - the effective mass of the field varies with the environmental density, such that it becomes large in dense regions and causes the fifth force to be short-ranged.
\end{itemize}
In what follows, an important example model for each of these three categories will be given\footnote{For each example there are partial text overlaps with Ref. \cite{Kading1}.}.
Some of them will be relevant in subsequent sections and introduced in as much detail as necessary for their later purpose.
\\\\
Each model will be presented with the assumption of non-relativistic matter as a source, meaning that the trace of the energy-momentum tensor fulfils $\rho = - T^\mu_{~\mu}$, where $\rho$ is the mass density of the source. 
This assumption is made such that field profiles and forces around explicit example sources can be given.
Furthermore, the conformal factor $A(\varphi)$ that was introduced in Section \ref{ssec:ScalarTensTheo} is assumed to be of the form
\begin{eqnarray}\label{eqn:Sometrafo}
A(\varphi) &=& e^{a\varphi^\alpha/\mathcal{M}^\alpha},
\end{eqnarray}
where $\mathcal{M}$ is a constant with the dimension of a mass, and for all models presented in this section $(a,\alpha) =(2,1)$ or $(a,\alpha) =(1,2)$.
\\\\
Assuming $\mathcal{M}^\alpha\gg\varphi^\alpha$, Eqn.\! (\ref{eqn:Sometrafo}) can be expanded:
\begin{eqnarray}
A(\varphi) &=& 1 + a\frac{\varphi^\alpha}{\mathcal{M}^\alpha} + \mathcal{O}(\varphi^{2\alpha}/\mathcal{M}^{2\alpha}).
\end{eqnarray}
A fifth force experienced by a test particle with mass $m$ at position $\vec{x}$ is given by
\begin{eqnarray}\label{eqn:FifthForce}
\vec{F}_\varphi(\vec{x}) &=& - \frac{m}{2}\vec{\nabla}_x  A(\varphi(\vec{x})).
\end{eqnarray}
How this expression for a fifth force is derived from a conformal transformation of the metric tensor is explained in Appendix \ref{app:DerivOf5Forces}.
For $(a,\alpha) =(2,1)$ Eqn.\! \ref{eqn:FifthForce} becomes 
\begin{eqnarray}\label{eqn:FifthForcen11}
\vec{F}_\varphi(\vec{x}) &\approx& -\frac{m}{\mathcal{M}}\vec{\nabla}_x \varphi(\vec{x}),
\end{eqnarray}
and for $(a,\alpha) =(1,2)$
\begin{eqnarray}
\vec{F}_\varphi(\vec{x}) &\approx& -\frac{m}{\mathcal{M}^2}\varphi(\vec{x})\vec{\nabla}_x \varphi(\vec{x}),
\end{eqnarray}
which can both often be found in the literature (see e.g.\ Refs.\! \cite{Kading2,BurrageForce}).


\subsubsection{Galileons}\label{ssec:Galileons}

Galileons are a well-known example of scalar fields with a screening mechanism that falls into the first category, i.e.\ they are screened by the Vainshtein mechanism \cite{Vainshtein1972}, which will be explained later in this {section}.
They were first described in the context of Dvali-Gabadadze-Porrati (DGP) braneworld models \cite{Dvali2000}, and are higher derivative field theories with second order equations of motion.
In a conformal coupling a galileon $\varphi$ appears as
\begin{eqnarray}
\tilde{g}_{\mu\nu} &=& e^{2\varphi/\mathcal{M}}g_{\mu\nu},
\end{eqnarray}
and, following Ref.\! \cite{Nicolis2008}, the most general galileon Lagrangian in flat space is obtained by substituting
\begin{eqnarray}
G_2(\varphi,X_\varphi) &=& -\frac{1}{2}X_\varphi \label{eqn:GalG2}
\\
G_3(\varphi,X_\varphi) &=& \frac{1}{2\Lambda^3}X_\varphi
\\
G_4(\varphi,X_\varphi) &=& -\frac{\lambda_4}{4\Lambda^6}X_\varphi^2
\\
G_5(\varphi,X_\varphi) &=& \frac{3}{2}\frac{\lambda_5}{\Lambda^9}X_\varphi^2 \label{eqn:GalG5}
\end{eqnarray}
into the Horndeski Lagrangian in Eqn.\! (\ref{eqn:Horndeski}), where $R=0$ and $G^{\mu\nu}=0$ due to the flat space metric.
Here $\Lambda$ is a coupling constant with {a} dimension of an energy, and $\lambda_4$ and $\lambda_5$ are dimensionless couplings.
In addition, there is a matter coupling term $-\varphi\rho/\mathcal{M}$, coming from the coupling between scalar field and SM matter in the Einstein frame matter Lagrangian (see Eqn.\! (\ref{eqn:EinsteinFrame})), added to Eqn.\! (\ref{eqn:Horndeski}).
The coupling constant $\mathcal{M}$ has mass dimension $1$.
\\\\
The Lagrangian obtained from adding Eqns.\! (\ref{eqn:GalG2})-(\ref{eqn:GalG5}) into Eqn.\! (\ref{eqn:Horndeski}) is invariant under the galilean shift
\begin{eqnarray}
\varphi(x) \mapsto \varphi(x) + c +b_\mu x^\mu,
\end{eqnarray} 
where $c$ and $b_\mu$ are some constants.  
The matter coupling term $-\varphi\rho/\mathcal{M}$ breaks this symmetry but if $\mathcal{M}\gg \Lambda$, then this breaking is mild since in this case the matter coupling term is relatively small compared to the non-canonical kinetic terms.     
\\\\
The screening of the galileon works by the Vainshtein mechanism \cite{Vainshtein1972}.
In short, within a certain radius - the Vainshtein radius $R_v$ - around a massive object, the non-linear terms of the galileon Lagrangian dominate over and consequently suppress the canonical kinetic term. 
Since, in this case, the galileon is strongly self-coupled \cite{Ali}, the resulting force is much weaker than gravity even if the unscreened force would be at least equally strong due to the assumption $\mathcal{M} \leq M_P$.
In order to illustrate the effect of this screening, the Lagrangian
\begin{eqnarray}\label{eqn:CubicLagrangian}
\mathcal{L} &=& -\frac{1}{2}(\partial\varphi)^2 - \frac{\Box\varphi}{2\Lambda^3}(\partial\varphi)^2 - \frac{\varphi}{\mathcal{M}}\rho
\end{eqnarray}
will be investigated.
It describes the so-called cubic galileon for which $\lambda_4 = \lambda_5 =0$ is assumed.
As an example situation, a static point source with mass density $\rho = M\delta^{(3)}(\vec{x})$ will be considered.
In this spherically symmetric scenario, the idea behind the Vainshtein mechanism becomes more apparent:
if the radial coordinate $r$ fulfils $r\ll R_v$, then the field must satisfy $\Box\varphi/\Lambda^3\gg 1$, such that the canonical kinetic term can be neglected. 
When, on the other hand, $r\gg R_v$, then the field must satisfy $\Box\varphi/\Lambda^3\ll 1$, such that the non-canonical kinetic term can be neglected.
Following Ref.\! \cite{Gabadadze2012}, Eqn.\ (\ref{eqn:CubicLagrangian}) leads to an equation of motion
\begin{eqnarray}
\Box\varphi -\frac{1}{\Lambda^3}\left[(\varphi,_{\mu\nu})^2 -(\Box\varphi)^2\right] &=& \frac{\rho}{\mathcal{M}},
\end{eqnarray}
which in the considered scenario can be rewritten as \cite{Gabadadze2012}
\begin{eqnarray}\label{eqn:RadCubic}
\frac{\varphi'}{r} + \frac{2}{\Lambda^3}\frac{\varphi'^2}{r^2} &=& \frac{M}{\mathcal{M}}\frac{1}{4\pi r^3},
\end{eqnarray}
where a prime $'$ indicates a derivative with respect to the radius $r$.
In- and outside the Vainshtein radius, Eqn.\! (\ref{eqn:RadCubic}) can be approximately solved for $\varphi'$ by neglecting either one of the two terms on the left-hand side, and yields
\begin{eqnarray}\label{eqn:GaliDeriv}
\varphi'(r) &=&
	\begin{cases} \sqrt{\frac{\Lambda^3 M}{8\pi\mathcal{M}r}} &\mbox{if } r \ll R_v \\
				 \frac{M}{\mathcal{M}}\frac{1}{4\pi r^2} &\mbox{if } r\gg R_v ,
	\end{cases}
\end{eqnarray}
where the Vainshtein radius
\begin{eqnarray}
R_v &=& \left(\frac{M}{2\pi\mathcal{M}}\right)^{1/3}\frac{1}{\Lambda}
\end{eqnarray}
is the radius at which both solutions in Eqn.\! (\ref{eqn:GaliDeriv}) coincide.
\\\\
Looking at the resulting screened fifth force described in Eqn.\! (\ref{eqn:FifthForce}), and comparing its magnitude $F_\varphi$ to the magnitude of the gravitational force on a test mass $m$, \cite{Bloomfield2014} $F_G = mM/8\pi M_P^2 r^2$, leads to
\begin{eqnarray}
\frac{F_\varphi}{F_G} &=& 
	\begin{cases} \sqrt{8\pi} \left(\frac{M_P}{\mathcal{M}}\right)^2\left(\frac{r}{R_v}\right)^{3/2} &\mbox{if } r \ll R_v \\
				 2\left(\frac{M_P}{\mathcal{M}}\right)^2 &\mbox{if } r\gg R_v .
	\end{cases}
\end{eqnarray}
Clearly, within in the Vainshtein radius the fifth force is strongly suppressed, while outside this region it can be even stronger than gravity.
This illustrates the essence of the Vainshtein mechanism.
\\\\
The consistency of the assumptions 
\begin{align}
\Box\varphi/\Lambda^3\gg 1 ~~~&\mbox{if } r \ll R_v
\end{align}
and
\begin{align}
\Box\varphi/\Lambda^3\ll 1 ~~~&\mbox{if } r \gg R_v
\end{align}
can be checked by using Eqn.\! (\ref{eqn:GaliDeriv}) in order to find
\begin{eqnarray}
\Box\varphi/\Lambda^3 &=& \frac{1}{r^2}\partial_r (r^2 \varphi')/\Lambda^3
\nonumber
\\
&=&
	\begin{cases} \frac{3}{2\sqrt{8\pi}} \left(\frac{R_v}{r} \right)^{3/2}  &\mbox{if } r \ll R_v \\
				 0 &\mbox{if } r\gg R_v .
	\end{cases}
\end{eqnarray}
This is in agreement with the assumptions of the Vainshtein mechanism.
\\\\
Solving the equation of motion for galileons in more generality, i.e.\ without distinguishing between solutions in- and outside the Vainshtein radius is an intricate enterprise.
Considering the seemingly simple case of a cubic galileon around a static spherically symmetric source with constant density $\rho$ and radius $R$, already leads to a solution for the field profile around the source that is of a rather complicated form \cite{Bloomfield2014}:
\begin{eqnarray}\label{eqn:CubicGaliProf}
\varphi(r) &=& \frac{\Lambda^3}{8}\bigg(r^2 \bigg[\sqrt{1 + \frac{R_v^3}{r^3}} -1 \bigg] 
+ 3\sqrt{R_v^3 R} \bigg[\sqrt{\frac{r}{R}} ~{}_2F_1\bigg(\frac{1}{6},\frac{1}{2};\frac{7}{6};-\frac{r^3}{R_v^3} \bigg)
\nonumber
\\
&\phantom{=}&
\phantom{\frac{\Lambda^3}{8}\bigg(r^2 \bigg[\sqrt{1 + \frac{R_v^3}{r^3}} -1 \bigg] + 3}
-{}_2F_1\bigg(\frac{1}{6},\frac{1}{2};\frac{7}{6};-\frac{R^3}{R_v^3} \bigg)\bigg]\bigg),
\end{eqnarray}
where ${}_2F_1$ denotes a Gaussian hypergeometric function. 
However, if the fifth force is of interest, only $\varphi'$ is required (see Eqn.\! (\ref{eqn:FifthForcen11})), which has a comparably simple solution
\begin{eqnarray}
\varphi'(r) &=& \frac{\Lambda^3}{4} r \left( \sqrt{1+ \frac{R_v^3}{r^3}} -1\right).
\end{eqnarray}
Further solutions with different geometries can also be found in Ref.\! \cite{Bloomfield2014}.


\subsubsection{Symmetrons}\label{ssec:Symmetrons}

\begin{figure}[htbp]
\begin{center}
\includegraphics[scale=0.1]{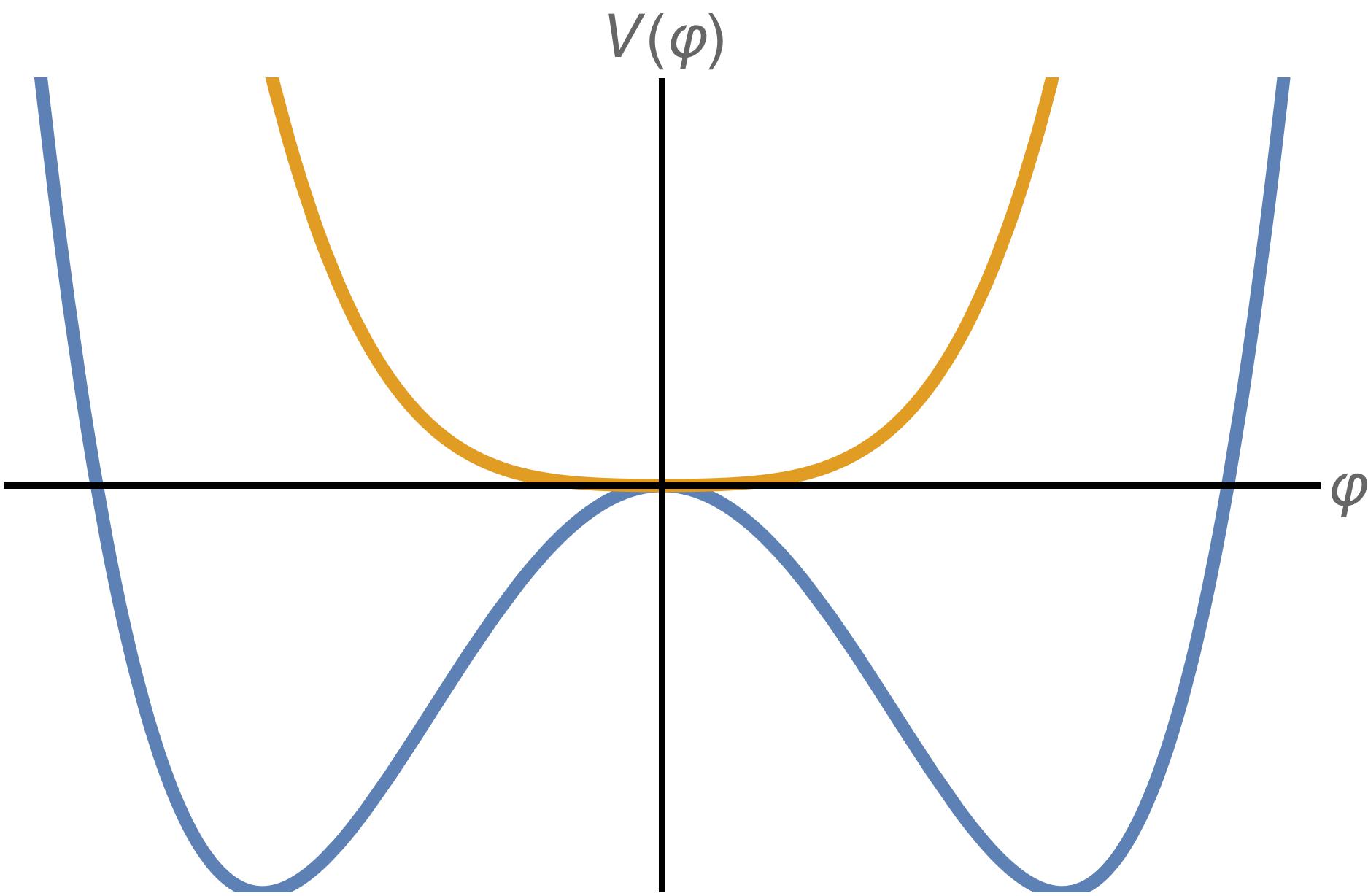}
\caption{(Cf. Ref. \cite{Kading1}) At low values of the density $\rho$, the potential allows for spontaneous breaking of the $\mathbb{Z}_2$ symmetry and therefore gives a non-vanishing vacuum expectation 
value to the symmetron (blue line). However, for $\rho \gg \mu^2 M^2$, the symmetron can only have a vanishing vacuum expectation value and is therefore screened (orange line). }
\label{Fig:SymmetronPotential}
\end{center}
\end{figure}

The symmetron is another commonly studied screened scalar field model.
It was first mentioned in Refs.\! \cite{Dehnen1992, Gessner1992, Damour1994, Pietroni2005, Olive2008, Brax2010}, described with its current name in Refs.\! \cite{Hinterbichler2010,Hinterbichler2011}, and initially introduced as a DE candidate.
Its fifth force has recently been considered as an explanation for the dynamics of galaxies which is otherwise attributed to particle DM \cite{Copeland2016,OHare2018,Kading2}.
The symmetron screening mechanism belongs to the varying coupling category, meaning that the field's coupling to matter is dependent on the environmental density and changes accordingly.
\\\\
The Lagrangian of a symmetron denoted by $\varphi$ is 
\begin{eqnarray}\label{eqn:SymmetronLagrangian}
\mathcal{L}_\varphi &=& -\frac{1}{2}\left(\partial\varphi\right)^2 -\frac{1}{2}\left( \frac{\rho}{\mathcal{M}^2} - \mu^2 \right)\varphi^2 - \frac{\lambda}{4}\varphi^4,
\end{eqnarray}
where $\lambda$ and $\mathcal{M}$ determine the strength of self-interaction and symmetron-matter coupling respectively{\footnote{There is in principle no restriction on the size of the coupling constant $\lambda$. However, for perturbative treatments it has to be smaller than $1$. There are actually constraints on the symmetron model that even consider $\lambda < 10^{-70}$ (see e.g.\ Ref. \cite{BurrageSakB}).}}. 
$\mu$ has the dimension of a mass. 
The effective potential\footnote{In this thesis, the term effective potential refers to a sum of single potentials. For example, the symmetron effective potential in Eqn.\! (\ref{eqn:EffecPotSymme}) is the sum of all potential terms appearing in the Lagrangian in Eqn.\! (\ref{eqn:SymmetronLagrangian}).} in this Lagrangian 
\begin{eqnarray}\label{eqn:EffecPotSymme}
V_\text{eff.} &=& \frac{1}{2}\left( \frac{\rho}{\mathcal{M}^2} - \mu^2 \right)\varphi^2 + \frac{\lambda}{4}\varphi^4
\end{eqnarray}
has a $\mathbb{Z}_2$ symmetry (see Figure \ref{Fig:SymmetronPotential}) which can be spontaneously broken in environments of low mass density, i.e.\ where $\rho < \mu^2\mathcal{M}^2$, such that the symmetron obtains a non-vanishing vacuum expectation value 
\begin{eqnarray}\label{eqn:SymmetronBackgroundProf}
\varphi_0 &=& \pm \sqrt{\frac{2}{\lambda}}\left( \mu^2 - \frac{\rho}{\mathcal{M}^2} \right) ,
\end{eqnarray}
which leads to a mass
\begin{eqnarray}\label{eqn:SymmetronMass}
m^2 &=& 2 \left(\mu^2 - \frac{\rho}{\mathcal{M}^2}\right).
\end{eqnarray}
However, in regions of high density, i.e.\ where $\rho \gg \mu^2 \mathcal{M}^2$, the symmetry is restored and $\varphi$ can only take on a vanishing vacuum expectation value, resulting in a mass
\begin{eqnarray}
m^2 &=&\frac{\rho}{\mathcal{M}^2}-\mu^2.
\end{eqnarray}
It is common practice to split a scalar field into a background value $\varphi_0$ and a small fluctuation $\delta\varphi$, such that $\varphi = \varphi_0 + \delta\varphi$, where $\delta\varphi$ is the actual carrier of the fifth force. 
At first order its interaction to matter is approximately proportional to $\rho\varphi_0\delta\varphi$.
This produces a force $F_\varphi \sim \varphi_0 \nabla\delta\varphi$. 
Consequently, the interaction to matter is turned off and the force is suppressed when the $\mathbb{Z}_2$ symmetry is restored in environments of large mass density, since $\varphi_0 = 0$ there. 
\\\\
In order to ensure the $\mathbb{Z}_2$ symmetry, the conformal factor for a transformation between Einstein and Jordan frame involving the symmetron is given by
\begin{eqnarray}
\tilde{g}_{\mu\nu} &=& e^{\varphi^2/\mathcal{M}^2}g_{\mu\nu}.
\end{eqnarray}
The possible impact of symmetrons on gravitational lensing and the consequences for attempts of explaining some effects otherwise attributed to particle DM will be discussed in Section \ref{Sec:Lensing}.
For this, it will prove useful to have the field profile of the symmetron condensate around a static spherically symmetric source with constant density $\rho$ and radius $R$. 
Therefore, the equation of motion is needed which is derived from Eqn.\! (\ref{eqn:SymmetronLagrangian}) and reads
\begin{eqnarray}
\frac{1}{r}\frac{\partial^2}{\partial r^2}(r\varphi) &=& \left( \frac{\rho}{\mathcal{M}^2} - \mu^2 +\lambda\varphi^2\right)\varphi,
\end{eqnarray}
which has as an obvious solution $\varphi(r)=0$.
The other solution for the symmetron profile outside the source is given by Ref.\! \cite{Ben}
\begin{eqnarray}\label{eqn:SymmetronProfile}
\varphi(r) &=& \varphi_{0,\text{out}} 
\nonumber
\\
&\phantom{=}&
-
(\varphi_{0,\text{out}} - \varphi_{0,\text{in}})\frac{R}{r}e^{m_{\text{out}}(R-r)}\frac{R m_{\text{in}}-
\tanh(m_{\text{in}} R)}{R m_{\text{in}}+R m_{\text{out}}\tanh(m_{\text{in}} R)},
\nonumber
\\
\end{eqnarray}
where ``in'' and ``out'' denote quantities depending on the density in- and outside the sphere, respectively.
\\\\
Besides the symmetron, there is another, very similar scalar field whose screening mechanism works with a vanishing coupling in dense environments.
It is the so-called environmentally dependent dilaton which is commonly appearing in discussions of string theory compactifications.
In this context it was first elaborated on in Ref.\!{\cite{Damour1994}}, but a more modern discussion with implications for cosmology can be found in Ref.\! \cite{Dilaton}.   
Following Ref.\! \cite{Joyce}, the dilaton Lagrangian can be written as
\begin{eqnarray}\label{eqn:DilatonLagrangian}
\mathcal{L} &=& -\frac{1}{2}(\partial\varphi)^2 - V_0 e^{-\varphi/M_P}
- \frac{(\varphi - \varphi_*)^2}{2\mathcal{M}^2}\rho,
\end{eqnarray}
where $V_0$ is a constant of mass dimension $4$,
$\mathcal{M}$ is the coupling to the matter density $\rho$
and $\varphi_*$ is a critical constant field value for which the dilaton decouples. 
This decoupling at $\varphi \approx \varphi_*$ is the essence of the dilaton screening mechanism. 
Under the assumption $\varphi \ll {M_P}$ it can be shown that the dilaton profile has the same functional form as the symmetron in Eqn.\! (\ref{eqn:SymmetronProfile}).


\subsubsection{Chameleons}\label{ssec:Chameleons}

\begin{figure}[H]
\centering
\includegraphics[scale=0.1]{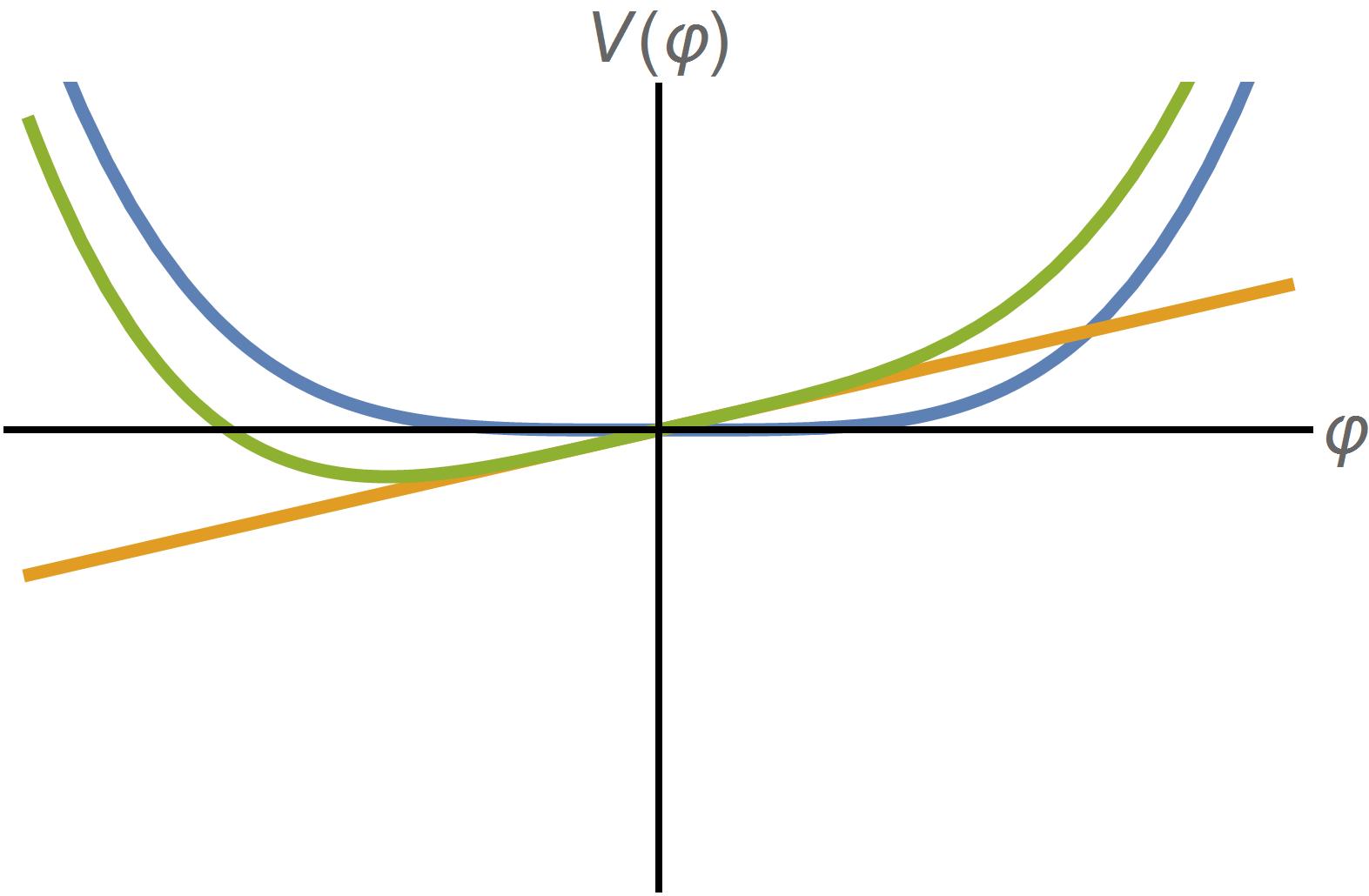}
\includegraphics[scale=0.1]{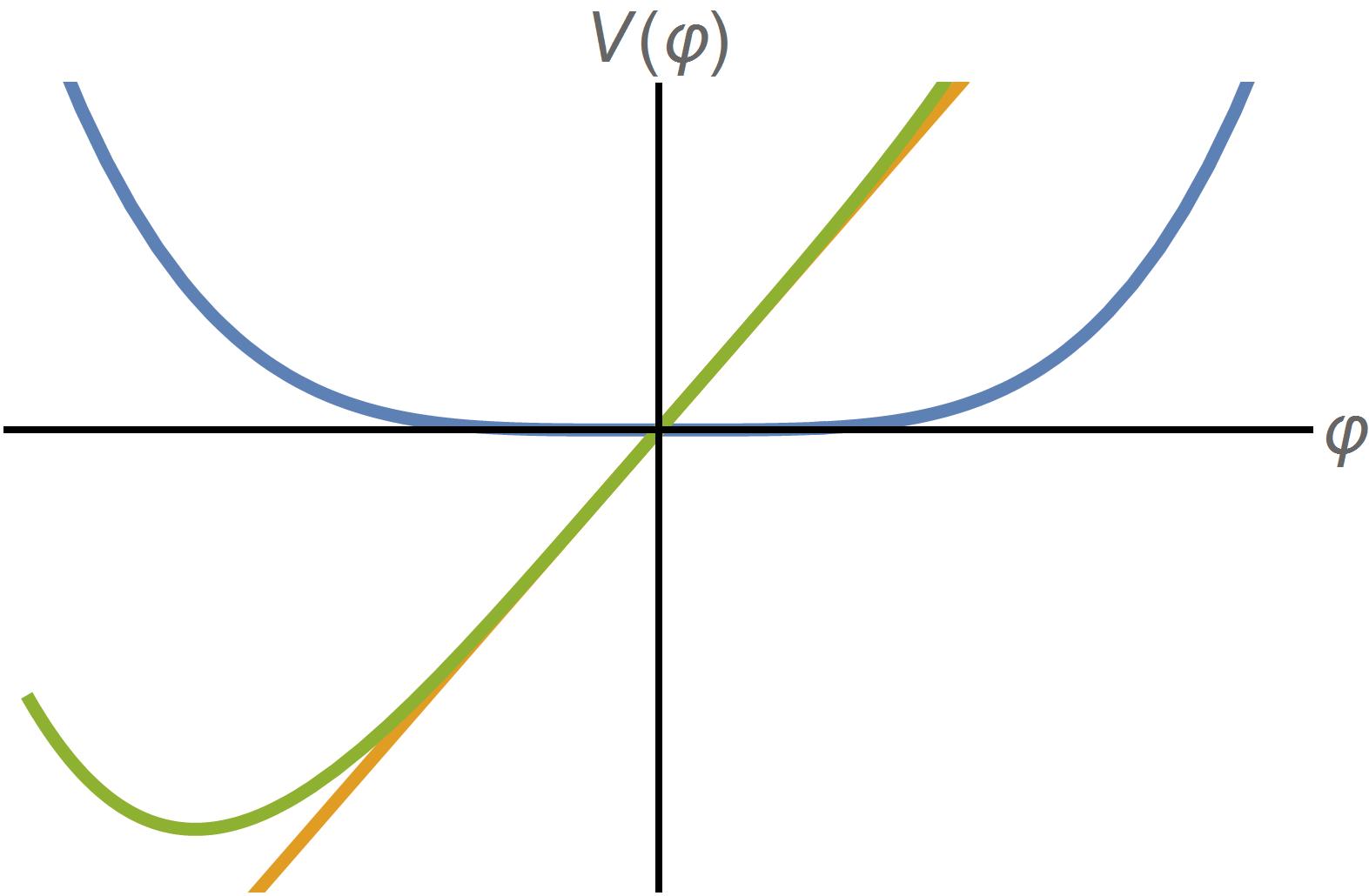}
\caption{The blue line represents the self-interaction potential of the chameleon for $n=-4$, the orange line depicts the interaction potential of the chameleon with matter, and the green line 
is the effective potential given by the sum of those two. In contrast to the two potentials alone, the effective potential has a non-vanishing minimum which allows the chameleon to have a non-vanishing mass. 
The left (right) figure represents the case of a low (high) environmental mass density.}
\label{Fig:ChameleonPotential}
\end{figure}

\begin{figure}[H]
\begin{center}
\includegraphics[scale=2.00]{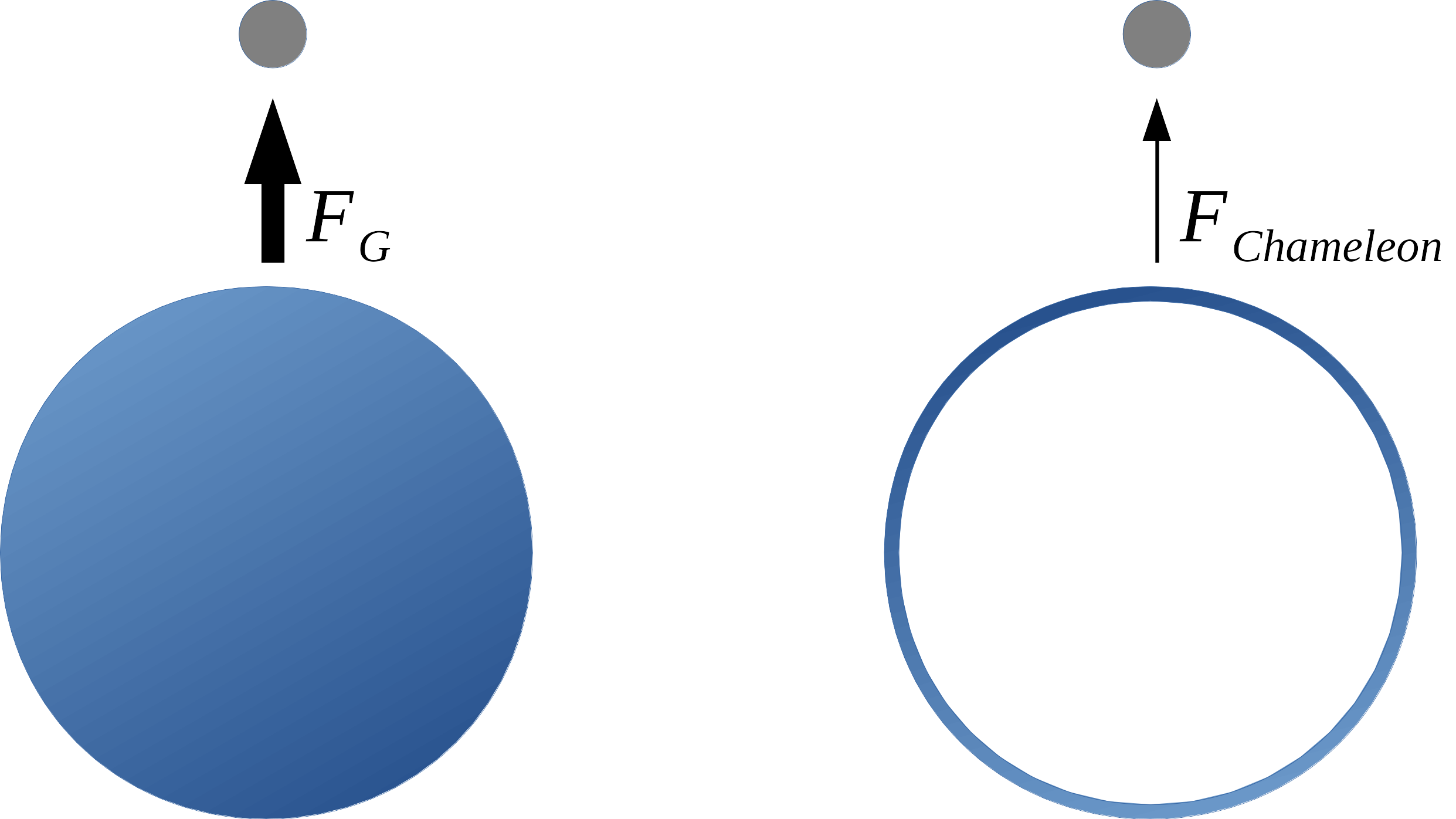}
\caption{Comparison of the gravitational force $F_G$ and the chameleon fifth force $F_\text{Chameleon}$. A test particle (in grey) experiences the full strength of the gravitational force sourced by all mass charges within the sphere, while, under the assumption that the sphere is sufficiently large, it only experiences the fifth force being sourced by a thin shell of chameleon charges around the object. {Charges within this shell are less screened than those deeper inside the source mass, and therefore cause a fifth force that has a longer range. Consequently, even if some charges underneath the thin shell are spatially closer to the test particle, the particle is effectively only affected by the force coming from all charges in the thin shell.} The blue colour indicates the charges that effectively contribute to each force.}
\label{fig:ComparisonGravityChameleonForce}
\end{center}
\end{figure}

The chameleon scalar field model was first introduced in Refs.\! \cite{Khoury20032,Khoury2003} and deals with a screened scalar field $\varphi$ whose non-vanishing effective mass {$m$} is dependent on the environmental density.
As its animal counterpart is adaptive to the colour of its surrounding, the chameleon field adapts its mass to the environment - a denser environment leads to a heavier chameleon mass.
\\\\
In a conformal coupling a chameleon appears as
\begin{eqnarray}
\tilde{g}_{\mu\nu} &=& e^{2\varphi/\mathcal{M}} g_{\mu\nu},
\end{eqnarray}
and is described by the Lagrangian density \cite{Khoury2003}
\begin{eqnarray}\label{eqn:ChamLagrangian}
\mathcal{L}_\varphi&=&-\frac{1}{2}\left(\partial\varphi\right)^2-\frac{\Lambda^{4+n}}{\varphi^n}-\frac{\varphi}{\mathcal{M}}\rho,
\end{eqnarray}
where $n \in \mathbb{Z}^+  \cup 2\mathbb{Z}^-\backslash \left\{-2\right\}$ distinguishes between different chameleon models and $\rho$ is the density of matter.
$\Lambda$ and $\mathcal{M}$ determine the strength of the self-interaction and the chameleon-matter coupling, respectively.
The models $n=-2$ and $n \in 2\mathbb{Z}^- - 1$ are not valid chameleons since the former case has only a constant mass ${m} \sim \Lambda$ and the latter case leads to imaginary vacuum expectation values.  
\\\\
In contrast to each of its summands, the effective potential
\begin{eqnarray}\label{eqn:ChamEffPot}
V_\text{eff.} &=& \frac{\Lambda^{4+n}}{\varphi^n}+\frac{\varphi}{\mathcal{M}}\rho
\end{eqnarray}
has a minimum (see Figure \ref{Fig:ChameleonPotential}), which means that the chameleon can take on a non-vanishing vacuum expectation value
\begin{eqnarray}
\varphi_0 &=& \left(n \Lambda^{4+n} \frac{\mathcal{M}}{\rho} \right)^{\frac{1}{n+1}},
\end{eqnarray}
and consequently a non-vanishing mass
\begin{eqnarray}
m^2 &=& \frac{n(n+1)\Lambda^{4+n}}{\varphi_0^{n+2}},
\end{eqnarray}
which has a dependence on the environmental density $\rho$ due to the form of $\varphi_0$.
It can easily be seen that, independent of the specific chameleon model, $m$ increases with $\rho$.
This is the essential quality of a chameleon scalar field model.
\noindent
For gaining a better understanding of how the chameleon screening mechanism works, it is useful to look at the example of a field profile around a static and homogeneous sphere with density $\rho$ and radius $R$.
The corresponding equation of motion is derived from Eqn.\! (\ref{eqn:ChamLagrangian}):
\begin{eqnarray}
\frac{1}{r}\frac{\partial^2}{\partial r^2}(r\varphi) &=& \frac{\rho}{\mathcal{M}} - n \frac{\Lambda^{4+n}}{\varphi^{n+1}}.
\end{eqnarray}
It is solved by \cite{Khoury2003}
\begin{eqnarray}
\varphi\left(r\right) &=& -\left(\varphi_\text{out}-\varphi_\text{in}\right)\frac{R}{r}e^{-m_{\text{out}}r}+\varphi_\text{out},
\end{eqnarray}
with ``in'' and ``out'' denoting quantities depending on the density in- and outside the sphere, respectively. 
If the source is screened, $\varphi_\text{in}$ is actually the minimum of the chameleon within the source. 
From this, the fifth force on a test particle according to Eqn.\! (\ref{eqn:FifthForce}) reads
\begin{eqnarray}
F_\varphi ~\sim ~  \partial_r \varphi(r) ~\sim ~ \exp\left[-m r \right]/r,
\end{eqnarray}
which means that this force is suppressed in the same manner as a Yukawa potential \cite{Yukawa1935} (see also e.g.\ Ref.\! \cite{PeskinSchroeder}).
As a consequence, a heavy chameleon carries a shorter ranged force than a light one, which effectively renders this force weaker.
\\\\
This particular behaviour of the fifth force gives rise to the thin-shell effect which allows only the thin outermost layer of mass of a large object to effectively contribute to the chameleon force.
More precisely, the chameleon charges deep within the object also source the force but since the force is so strongly suppressed and decays quickly in dense environments, the force caused by these deep charges is very weak in comparison with the one coming from the thin-shell and therefore plays a negligible role.
A comparison between gravity and a chameleon fifth force is depicted in Figure \ref{fig:ComparisonGravityChameleonForce}.  
\\\\
Due to the thin-shell effect, the chameleon fifth force coming from objects in the {Solar System} is screened.
Nevertheless, there have been successful attempts to constrain the chameleon parameter space with Earth-based experiments since it is possible to create situations in which the field reaches its unscreened regime in the vacuum of a vacuum chamber.
An overview of the variety of experiments that have been conducted may be gained from Refs.\! \cite{BurrageSakA,BurrageSakB}.
This is also made possible by the thin-shell effect.
More precisely, if the walls of a vacuum chamber are sufficiently thick, anything outside the chamber will not contribute to the fifth force which allows it to only be screened by the chamber itself and the objects within it\footnote{A perfectly unscreened situation could only be reached in an infinitely large vacuum chamber.}. 
This could be pictured as the chameleon becoming so heavy in the chamber walls that it gets stuck there and cannot communicate with the outside world.
\\\\
In Section \ref{Sec:OpenQuantumfromCCSF} the $n=-4$ chameleon model will be used as an example for a field that can induce open quantum dynamical effects on test particles in a vacuum chamber.

\newpage
\section{Lensing with symmetron fields}\label{Sec:Lensing}

Actually detecting screened scalar fields would equal a scientific breakthrough. 
This is because the finding of additional scalar fields in Nature would hint at physics beyond GR and the SM, and therefore start a new exciting era in physics research - both in theory and experiments.
In order to find or at least constrain screened scalar fields, there have been plenty of different types of tests conceived.
Amongst those, astrophysical and cosmological observations are perfect testing grounds for modified gravity theories and in particular the phenomenology of screend fields.
On one hand, this is due to the fact that it is very challenging to detect these fields on {Solar System} scales and therefore larger distance scales with lower mass densities have to be considered, which are subjects of astrophysics and cosmology.
On the other hand, some of the most interesting predictions for screened scalar fields, for example, that they could at least partially explain the origin of DE (see e.g.\ Refs.\! \cite{Copeland2006,SaksteinPhD,JainOthers}) or DM (e.g.\ Refs.\! \cite{Copeland2016,KhouryAlter}), concern phenomena which are most important on such large scales.
\\\\
Rotation curve measurements \cite{Rubin1,Roberts,Rubin2,Bosma} were the first to strongly hint at the existence of a significant amount of non-baryonic matter in the Universe \cite{Lisanti}.
This was recently confirmed by measurements of the PLANCK collaboration \cite{PLANCK2015}.
Nevertheless, to date, no DM particle has been directly detected, which motivates the search for alternative explanations to particle DM.
In Ref.\! \cite{Copeland2016} it was shown that certain effects on galaxy dynamics usually attributed to particle DM could actually be explained by the presence of a symmetron scalar field, as introduced in Section \ref{ssec:Symmetrons}.
More precisely, it turned out that its fifth force may be able to explain the radial acceleration for rotating galaxies \cite{Sanders1990,Janz,McGaugh}, and the energy stored in the symmetron field could be sufficient to lead to a stabilisation of disk galaxies \cite{Ostriker}.
All this would work without the presence of particle DM.
Further studies regarding the influence of symmetron fields on local stars in the Milky Way were done in Ref.\! \cite{OHare2018} and supported the idea of using light scalar fields as alternatives to the established cold DM models.
\\\\
An important piece in answering whether a symmetron fifth force could be a suitable replacement for particle DM is gravitational lensing.
This effect describes the deflection that light experiences when it passes a gravitational source, and is important for indirect observations of DM \cite{Massey} and other objects that emit only little to no light, e.g.\ black holes \cite{Bozza}.
If a symmetron fifth force is supposed to entirely replace particle DM, then it must be able to explain the gravitational lensing that is observed due to the apparent presence of DM. 
For this, as will be illustrated later in this section, it is absolutely crucial that the symmetron not only couples conformally but also disformally to matter.
Ref. \cite{Dai} provides first constraints on the strength of the coupling between photons and an additional scalar field from modified gravity theories, which is responsible for galactic dynamics in the absence of particle DM.
\\\\
In this section it will be discussed whether disformally coupling symmetron fields can explain the discrepancy between the observed baryonic mass of large astrophysical objects, like galaxies, and the measured lensing mass, which would otherwise be explained by the presence of particle DM.
For this, a more detailed description of gravitational lensing will be given in Section \ref{ssec:GravLens} before it will be explained why it is necessary for scalar fields to not only be conformally but also disformally coupled in order to see any effect (see Section \ref{ssec:ConfDisfLens}). 
Afterwards, the magnitude of the effect of a disformally coupling symmetron field will be investigated in Section \ref{ssec:LensSingleSymm}.
Subsequently, the study will be repeated but with an additional scalar field of potential cosmological relevance coupling to the symmetron \ref{ssec:CosmoScalar}. 
\\\\
Sections \ref{ssec:LensSingleSymm} and \ref{ssec:CosmoScalar} present the computational contributions of this thesis' author to article \cite{Kading2}. 

\newpage
\subsection{Gravitational lensing}\label{ssec:GravLens}

\begin{figure}[htbp]
\begin{center}
\includegraphics[scale=1.00]{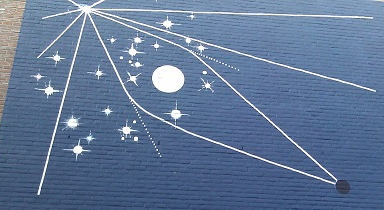}
\caption{Wall of a house in Leiden, the Netherlands, depicting the schematics of gravitational lensing. Picture taken by the author during the `Dark Energy in the Laboratory' workshop at Lorentz Center, Leiden, in November 2017.}
\label{fig:Leiden}
\end{center}
\end{figure}

\begin{figure}[htbp]
\begin{center}
\includegraphics[scale=0.3]{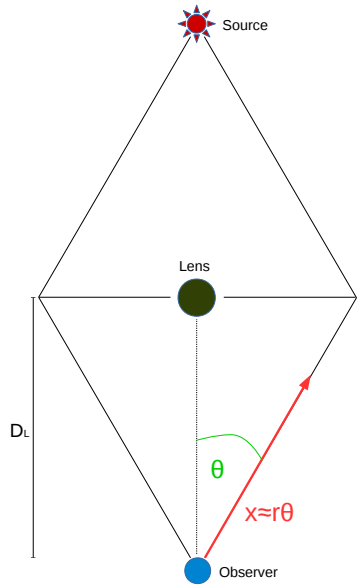}
\caption{Schematics of gravitational lensing in two dimensions with thin-lens approximation. The source sends out light which is distorted by the lens and then received by an observer. $D_L$ denotes the distance between the observer and the lens plane in which the lens is situated. The radial coordinate $r$ equals $0$ at the position of the observer and is orthogonal to the lens plane. $\theta$ denotes the angle between $r$ and the light ray reaching the observer {(here it represents either $\theta^1$ or $\theta^2$)}. The coordinate $x$ is approximated by $r\theta$ since $\theta$ is assumed to be very small.}
\label{fig:LensingScheme}
\end{center}
\end{figure}

Although Einstein's theory of gravity was already formulated more than 100 years ago \cite{Einstein}, some of its most prominent predictions are still subjects of modern research.
For example, gravitational waves have just been directly detected for the first time in 2015 \cite{Abbott2016}, and the first exemplar of a black hole has only recently been photographed \cite{Akiyama}.
However, the first test that GR successfully passed was that of the predicted bending of light due to the gravitational field of the sun \cite{Eddington}.
This first successful observation was actually an example of gravitational lensing - an effect where a massive body deflects light  with its gravitational field.
Since gravity is a rather weak force, the object that acts as a lens needs to have a large mass in order to cause any observable effect.
Consequently, lensing by gravity is used in astrophysical contexts, e.g.\ in the search for exoplanets \cite{ExoPlanet} or black holes \cite{Bozza}, but is not an effect that humans encounter in their daily lives.
\\\\
Figure \ref{fig:Leiden} shows the schematics of the gravitational lensing effect. 
The large white dot in the centre of the sketch is a very massive object, for example, a black hole or a galaxy cluster, and called the lens.
Light coming from a star, the white dot in the top left of the figure where the light rays start, is subject to the gravitational field of the massive object. 
As a consequence, the trajectory of the light, depicted by the white lines, is changed and bent towards the lens.
An observer on a planet, depicted by the black dot, sees the light arriving from a direction different from the one he would see if the the starlight was travelling in a straight line towards him.   
The angle under which the observed trajectory of the light is changed due to the gravitational lensing is determined by the mass of the lens, and the distances between light source, lens and observer (see Figure \ref{fig:LensingScheme}).
\\\\
In the following it will now be shown how this deflection angle is derived in GR in a Friedmann-Lema\^itre-Robertson-Walker (FLRW) background, which describes an homogeneous, isotropic and expanding universe \cite{AmendolaTsujikawa}.
For this, if not otherwise stated, the arguments of the discussion in Ref.\! \cite{AmendolaTsujikawa} will be made use of.
The FLRW metric without the presence of any matter is given by
\begin{eqnarray}\label{eqn:FLRWmetric}
ds^2 &=& a^2(\tau)(-d\tau^2 + \delta_{ij}dx^idx^j), 
\end{eqnarray}
where $a$ is the scale factor that describes the expansion of the universe, and $\tau$ is the conformal time given by
\begin{eqnarray}
d\tau &=& \frac{dt}{a(t)}.
\end{eqnarray}
When there is actually a mass present that perturbs this background metric, then the metric reads
\begin{eqnarray}\label{eqn:PertubedFLRW}
ds^2 &=& a^2(\tau)[-(1+2\Psi)d\tau^2 + (1+2\Phi)\delta_{ij}dx^idx^j] , 
\end{eqnarray}
where $\Phi$ and $\Psi$ are Newtonian potentials that fulfil $\Phi,\Psi \ll 1 $
and 
\begin{eqnarray}\label{eqn:NoSlip}
\Phi &=& - \Psi{,}
\end{eqnarray}
{which can be derived from the perturbed FLRW Einstein equations.}
\\\\
The geodesic equation \cite{Straumann}
\begin{eqnarray}\label{eqn:Geodesic}
\frac{d^2x^\mu}{d\lambda^2} + \Gamma^\mu_{\alpha\beta}\frac{dx^\alpha}{d\lambda}\frac{dx^\beta}{d\lambda} &=& 0
\end{eqnarray}
is the equation of motion of an uncharged particle in a gravitational field, with $\Gamma$ being a Christoffel symbol, and $\lambda$ an affine parametrisation of the particle's path.
Christoffel symbols resulting from Eqn.\! (\ref{eqn:PertubedFLRW}) can be found in Appendix \ref{app:PertubredFLRW}.
After defining the photon momentum
\begin{eqnarray}\label{eqn:PhotonMomentum}
k^\mu := \frac{dx^\mu}{d\lambda},
\end{eqnarray}
Eqn.\! (\ref{eqn:Geodesic}) becomes
\begin{eqnarray}\label{eqn:ModGeodesicEqn}
\frac{dk^\mu}{d\lambda} + \Gamma^\mu_{\alpha\beta}k^\alpha k^\beta &=& 0.
\end{eqnarray}
It will be assumed that the observer and the lens are sufficiently far away from each other in order to apply the so-called thin-lens approximation. 
In this approximation the lens is assumed to have no thickness and lie within a two-dimensional plane, the lens plane, perpendicular to the line between light source and observer. 
This is valid if the thickness of the lens is much smaller than the distance between lens and observer, and the two angles parametrising the deflection of the light with respect to the lens plane are very small. 
The resulting coordinate system that will be used here is then given by
\begin{eqnarray}\label{eqn:Coordiates}
\{ \tau, r, x^1 \approx r\theta^1, x^2 \approx r\theta^2 \},
\end{eqnarray}
where $r$ is a radial coordinate, and $\theta^{1,2}$ are the deflection angles.
This setup is visualised in Figure \ref{fig:LensingScheme}.
\\\\
The momentum vector $k^\mu$ can be split up into a background vector $\hat{k}^\mu$ and its perturbation $\delta k^{\mu}$ arising from the presence of the lens mass:
\begin{eqnarray}
k^\mu &=& \hat{k}^\mu + \delta k^{\mu}.
\end{eqnarray}
When considering only the background, then the trajectory of photons can only depend on $\tau$ and $r$ since there is no lensing mass that could cause deflections of the light.
Consequently, $d\tau = dr$ is fulfilled, and the $x^{1,2}$ components of the background momentum have to vanish:
\begin{eqnarray}\label{eqn:X12vanish}
\hat{k}^{x^{1,2}}&=&0.
\end{eqnarray}
Using this together with the on-shell mass condition $k^\mu k_\mu =0$ leads to
\begin{eqnarray}\label{eqn:reuqals0}
\hat{k}^r &=& \hat{k}^0.
\end{eqnarray}
In order to find the deflection angles $\theta^{1,2}$, Eqn.\! (\ref{eqn:Geodesic}) will now be considered for the $x^{1,2}$ components. 
When using the Christoffel symbols from Appendix \ref{app:PertubredFLRW} and ignoring all terms higher than first order in the momentum perturbations and the Newtonian potentials, this yields for $x^1$
\begin{eqnarray}\label{eqn:FirstOrderEqn}
\frac{dk^{x^1}}{d\lambda}  +  2\mathcal{H}\hat{k}^0\delta k^{x^1} -\Phi,_{x^1}(\hat{k}^r)^2+\Psi,_{x^1}(\hat{k}^0)^2 &=& 0,
\end{eqnarray}
where $\mathcal{H}$ is the conformal Hubble parameter $\mathcal{H} := a'/a$, with $'$ denoting a derivative with respect to the conformal time $\tau$, and $\Phi,_{x} := \partial\Phi/\partial x$.
\\\\
Next, using Eqns.\! (\ref{eqn:PhotonMomentum}) and (\ref{eqn:reuqals0}), while taking into account Eqn.\ (\ref{eqn:X12vanish}), leads from Eqn.\! (\ref{eqn:FirstOrderEqn}) to the deflection equation
\begin{eqnarray}
\label{eqn:deflection}
\frac{d^2{x^1}}{d\lambda^2}  +  2\mathcal{H}\frac{d\tau}{d\lambda}\frac{dx^1}{d\lambda} -(\Phi,_{x^1}-\Psi,_{x^1})\left(\frac{d\tau}{d\lambda}\right)^2 &=& 0,
\end{eqnarray}
of which there is an identical expression for $x^2$.
\\\\
Replacing the first term on the left-hand side of Eqn.\! (\ref{eqn:deflection}) with
\begin{eqnarray}\label{eqn:Replacementofx22} 
\frac{d^2x^1}{d\lambda^2} &=& \frac{d}{d\lambda}\left(\frac{dr}{d\lambda}\frac{dx^1}{dr}\right) ~=~ \frac{d\hat{k}^r}{d\lambda}\frac{dx^1}{dr}+\hat{k}^r\cdot\hat{k}^r \frac{d^2x^1}{dr^2},
\end{eqnarray}
and making use of
\begin{eqnarray}\label{eqn:GeodesicResults}
\frac{d\hat{k}^0}{d\lambda} &=& -2 \mathcal{H}(\hat{k}^0)^2,
\end{eqnarray}
which was derived from the geodesic equation (\ref{eqn:Geodesic}), leads to 
\begin{eqnarray}
- 2\mathcal{H}\hat{k}^0\frac{dx^1}{d\lambda} +(\Phi,_{x^1}-\Psi,_{x^1})(\hat{k}^0)^2 &=& -2\mathcal{H}(\hat{k}^0)^2\frac{dx^1}{dr} + (\hat{k}^r)^2 \frac{d^2x^1}{dr^2},
\nonumber
\\
\end{eqnarray}
which can be rewritten as the force equation
\begin{eqnarray} \label{eqn:force1}
\frac{d^2x^1}{dr^2} &=& \Phi,_{x^1}-\Psi,_{x^1}
\end{eqnarray}
by applying Eqn.\! (\ref{eqn:reuqals0}). 
Substituting $x^1\approx r\theta^1$ into Eqn.\! (\ref{eqn:force1}), and assuming that the photons {were} emitted by a star at $r=0$ under an original angle $\theta^1_0$, then gives
\begin{eqnarray}
\theta^1 &=& \theta_0^1 + \frac1r\int_0^rdr''\int_0^{r''} dr' (\Phi,_{x^1}-\Psi,_{x^1}),
\end{eqnarray}
which is the deflection angle in terms of the Newtonian potentials and the distance between light source and observer.
Obviously, $\theta^2$ follows an equation of the same form.

\newpage

\subsection{Effect of conformal and disformal couplings on lensing}\label{ssec:ConfDisfLens}

Before quantitatively discussing the effect of disformally coupled fields it shall now be explained why it is actually necessary to consider disformal instead of conformal couplings when investigating effects of scalar fields on the lensing of light.
Clearly, a conformal coupling to photons has to vanish since they are conformally invariant \cite{Bambi}, meaning that the conformally coupled scalar field cannot influence the dynamics of light\footnote{There is, however, the possibility that the energy stored in the conformally coupling field causes a small perturbation to the gravitational potential of a massive object and therefore indirectly influences the lensing effect.}.
In addition, the trace of the electromagnetic energy-momentum tensor \cite{Straumann}
\begin{eqnarray}
T^{\mu\nu} &=& \frac{1}{4\pi}\left( F^{\mu\alpha}F^{\nu}_{~\alpha} - \frac{1}{4}g^{\mu\nu}F_{\alpha\beta}F^{\alpha\beta} \right)
\end{eqnarray}
is nil, meaning that a term $A(\Phi)T^\mu_{~\mu}$ appearing in a Lagrangian must vanish\footnote{The Lagrangians given in Section \ref{Sec:ScreenedScalarFields} are only considered for non-relativistic matter with $T^\mu_{~\mu} = -\rho$. The replacement $-\rho \leftrightarrow T^\mu_{~\mu}$ gives $A(\Phi)T^\mu_{~\mu}$ which is valid also for relativistic matter.}.
\\\\
Besides such arguments, it can also explicitly be shown that a conformally coupled scalar field does not change the geodesic equation (\ref{eqn:Geodesic}) and can therefore not modify gravitational lensing.
For this, the following transformation between Jordan and Einstein frame is considered:
\begin{eqnarray}\label{eqn:ConfTrafoLens}
\tilde{g}_{\mu\nu} & = & A(\varphi)g_{\mu\nu}{,} 
\\
\tilde{g}^{\mu\nu} & = & A^{-1}(\varphi)g^{\mu\nu},
\end{eqnarray}
and the following arguments are based on those presented in Section \ref{ssec:GravLens}.
A deviation of geodesics from those predicted in GR would be visible in the Jordan frame due to a change in the Christoffel symbols, i.e.\ if the Jordan frame Christoffel symbols do not equal those from GR.
The modified Christoffel symbols given in terms of the Einstein frame metric $g_{\mu\nu}$ and the scalar field $\varphi$ are\begin{eqnarray}\label{eqn:ModChristofferConf}
\tilde{\Gamma}^{\alpha}_{\beta\gamma} & = & \frac12 \tilde{g}^{\alpha\delta}\left(\tilde{g}_{\beta\delta},_{\gamma}+\tilde{g}_{\gamma\delta},_{\beta}-\tilde{g}_{\beta\gamma},_{\delta}\right)\nonumber \\
 & = & \frac12 A(\varphi)^{-1}g^{\alpha\delta}\left(\frac{\partial}{\partial x^\gamma}(A(\varphi) g_{\beta\delta})+\frac{\partial}{\partial x^\beta}(A(\varphi) g_{\gamma\delta})-\frac{\partial}{\partial x^\delta}(A(\varphi) g_{\beta\gamma})\right)\nonumber \\
 & = & \Gamma^{\alpha}_{\beta\gamma} + \frac12 A(\varphi)^{-1}g^{\alpha\delta}\left(A,_{\gamma}g_{\beta\delta} + A,_{\beta}g_{\gamma\delta} -A,_{\delta}g_{\beta\gamma}\right)\nonumber \\
  & = & \Gamma^{\alpha}_{\beta\gamma} + \frac12 A(\varphi)^{-1}\left(A,_{\gamma}g^{\alpha}_{\beta} + A,_{\beta}g^{\alpha}_{\delta} -A,_{\delta}g^{\alpha\delta}g_{\beta\gamma}\right) \,.
\end{eqnarray}
Next, a coupling $A(\varphi) = 1 + \varphi/\mathcal{M} + \mathcal{O}(\varphi^2/\mathcal{M}^2)$ with $\varphi \ll \mathcal{M}$ is assumed\footnote{In order to give a simplified presentation, a linear coupling is assumed here. Non-linear conformal couplings lead to the same conclusion as linear ones, i.e.\ they do not affect gravitational lensing. For the discussions that actually involve the symmetron in Sections \ref{ssec:LensSingleSymm} and \ref{ssec:CosmoScalar} the usual quadratic coupling will be used.}, 
\begin{eqnarray}
\Delta\Gamma^{\alpha}_{\beta\gamma} &{:=}& \tilde{\Gamma}^{\alpha}_{\beta\gamma} - \Gamma^{\alpha}_{\beta\gamma} 
\end{eqnarray}
defined, and only an $x$-component (either $x^1$ or $x^2$ as given in Eqn.\! (\ref{eqn:Coordiates})) is considered since only it is relevant for the deflection equation (compare with the derivation of Eqn.\! (\ref{eqn:deflection})).
Doing this, yields
\begin{eqnarray}
\Delta\Gamma^{x}_{\alpha\beta}  &=& \frac{1}{2\mathcal{M}}\left(\varphi,_{\beta}g^{x}_{\alpha} + \varphi,_{\alpha}g^{x}_{\beta} -\varphi,_{\delta}g^{x\delta}g_{\alpha\beta} \right).
\end{eqnarray}
This expression is then contracted with momentum vectors $k^\alpha$ and $k^\beta$, in order to reproduce the second term on the left-hand side of the geodesic equation (\ref{eqn:ModGeodesicEqn}), giving 
\begin{eqnarray}\label{eqn:ModChristoffel1}
\Delta\Gamma^{x}_{\alpha\beta}k^\alpha k^\beta  & = & \frac{1}{2\mathcal{M}}\left(\varphi,_{\beta}k^{x}k^{\beta} + \varphi,_{\alpha}k^{x}k^{\alpha} -\varphi,^{x}g_{\alpha\beta}k^\alpha k^\beta \right) .
\end{eqnarray}
The first two terms in Eqn.\! (\ref{eqn:ModChristoffel1}) are dropped since they must be of second order due to $k^x = \delta k^x$.
This only leaves 
\begin{eqnarray}
\Delta\Gamma^{x}_{\alpha\beta}k^\alpha k^\beta  & = & -\frac{1}{2\mathcal{M}} \varphi,^{x}g_{\alpha\beta}k^\alpha k^\beta , 
\end{eqnarray}
which has to vanish due to the on-shell condition for photons.
Consequently, it is 
\begin{eqnarray}
\Delta\Gamma^{x}_{\alpha\beta}k^\alpha k^\beta  & = & 0 , 
\end{eqnarray}
and hence there is no change to the geodesic equation {(\ref{eqn:ModGeodesicEqn})} due to the presence of a conformally coupled scalar field.
\\\\
Next, it will be shown that this {does not happen} for disformal couplings, i.e.\ disformal couplings actually lead to a modification of the geodesic equation.
For this, a disformal transformation of the form 
\begin{eqnarray}\label{eqn:DisformalCoupLens}
\bar{g}_{\mu\nu} & = & A(\varphi)g_{\mu\nu} +B\varphi,_{\mu}\varphi,_{\nu}~,
\\
\bar{g}^{\mu\nu} & = & A^{-1}(\varphi)\left( g^{\mu\nu}-\frac{B}{C} \varphi,^{\mu}\varphi,^{\nu} \right) 
\end{eqnarray}
will be considered, where $B$ is a constant of mass dimension $-4$ and 
\\$C:=A+B(\partial\varphi)^2$.
The Jordan frame metric is here denoted by a bar $\bar{~}$ in order to distinguish it from the Jordan frame metric obtained from a conformal transformation in Eqn.\! (\ref{eqn:ConfTrafoLens}).
From Eqn.\! (\ref{eqn:DisformalCoupLens}) modified Christoffel symbols are derived:
\begin{eqnarray}
\bar{\Gamma}^{\alpha}_{\beta\gamma} & = & \frac{1}{2}\bar{g}^{\alpha\delta}(2\bar{g}_{\delta(\beta},_{\gamma)}-\bar{g}_{\beta\gamma},_\delta)  \nonumber \\
& = & A^{-1}\left( g^{\alpha\delta}-\frac{B}{C} \varphi,^\alpha\varphi,^\delta \right)\bigg[\partial_{(\gamma}(A(\varphi)g_{\beta)\delta} +B\varphi,_{\beta)}\varphi,_\delta)
\nonumber
\\
&\phantom{=}&
\phantom{A^{-1}\left( g^{\alpha\delta}-\frac{B}{C} \phi,^\alpha\phi,^\delta \right)}
-\frac{1}{2}\partial_{\delta}(A(\varphi)g_{\beta\gamma} +B\varphi,_{\beta}\varphi,_\gamma)\bigg] 
\nonumber 
\\
& = & \tilde{\Gamma}^{\alpha}_{\beta\gamma} + \frac{B}{C} \varphi,^\alpha\bigg[\varphi,_{\gamma\beta} - \varphi,^\delta g_{\delta(\beta},_{\gamma)} + \frac{1}{2}\varphi,^\delta g_{\beta\gamma},_\delta
\nonumber
\\
&\phantom{=}&
\phantom{\tilde{\Gamma}^{\alpha}_{\beta\gamma} + \frac{B}{C} \varphi,^\alpha}
+\frac{1}{2A}\varphi,^\delta A,_\delta g_{\beta\gamma}-A^{-1} A,_{(\gamma} \varphi,_{\beta)}\bigg] ,
\end{eqnarray}
where $A^{-1}-A^{-1}(\partial\varphi)^2\frac{B}{C} = \frac{1}{C}$ was used, and the symmetrisation 
\begin{eqnarray}
X_{(\alpha}X_{\beta)} := \frac{1}{2}(X_{\alpha}X_{\beta}+X_{\beta}X_{\alpha})
\end{eqnarray}
introduced.
\\\\
Again assuming $A(\varphi) = 1 + \varphi/\mathcal{M} + \mathcal{O}(\varphi^2/\mathcal{M}^2)$, and only working with the $x$ Christoffel symbols leads to 
\begin{eqnarray}
 \bar{\Gamma}^x_{\beta\gamma} & = &  \frac{B}{\mathcal{C} } \varphi,^x\bigg[\varphi,_{\gamma\beta} - \varphi,^\delta g_{\delta (\beta},_{\gamma)} + \frac{1}{2}\varphi,^{\delta}g_{\beta\gamma},_\delta + \frac{1}{2\mathcal{M} }(\partial\varphi)^2 g_{\beta\gamma} - \frac{1}{\mathcal{M} }\varphi,_\gamma\varphi,_\beta\bigg]
\nonumber
\\
&\phantom{=}&
+\tilde{\Gamma}^x_{\beta\gamma},
\end{eqnarray}
where $\mathcal{C} := 1+\varphi/\mathcal{M} +B(\partial\varphi)^2$, and $\tilde{\Gamma}^x_{\beta\gamma}$ is the Christoffel symbol modified by the conformal coupling as given in Eqn.\! (\ref{eqn:ModChristofferConf}).
\\\\
Now defining 
\begin{eqnarray}
\Delta\bar{\Gamma}^x_{\beta\gamma} &:= & \bar{\Gamma}^x_{\beta\gamma} - \Gamma^x_{\beta\gamma}, 
\end{eqnarray}
and then contracting with $k^\beta$ and $k^\gamma$ makes 
\begin{eqnarray}\label{eqn:DisfoModificatio}
\Delta\bar{\Gamma}^x_{\beta\gamma}k^\beta k^\gamma & = & \frac{B}{\mathcal{C}}\varphi,^x \bigg[\varphi,_{\gamma\beta} - \frac{1}{\mathcal{M}}\varphi,_\gamma\varphi,_\beta - \frac{1}{2} \varphi,_\rho g^{\rho\delta}(2g_{\delta(\beta},_{\gamma)}-g_{\beta\gamma},_\delta)\bigg]k^\beta k^\gamma \nonumber \\
& = & \frac{B}{\mathcal{C}}\varphi,^x \bigg[\varphi,_{\gamma\beta} - \frac{1}{\mathcal{M}}\varphi,_\gamma\varphi,_\beta -\varphi,_\rho \Gamma^\rho_{\beta\gamma}\bigg] k^\beta k^\gamma,
\end{eqnarray}
which is a non-vanishing modification of the geodesic equation (\ref{eqn:ModGeodesicEqn}) and therefore implies an influence of the scalar field on the lensing of light.
\\\\
With this, a force law in the spirit of Eqn.\! (\ref{eqn:force1}) shall now be derived. 
When again using the FLRW metric given in Eqn.\! (\ref{eqn:FLRWmetric}), the corresponding Einstein frame Christoffel symbols can be found in Appendix \ref{app:PertubredFLRW}. 
Substituting Eqn.\! (\ref{eqn:DisfoModificatio}) into Eqn.\! (\ref{eqn:ModGeodesicEqn}) gives
\begin{eqnarray} \label{eqn:disgeodesic}
\frac{dk^{x^1}}{d\lambda_s} &+& (\Gamma^{x^1}_{\alpha\beta} + \Delta\bar{\Gamma}^{x^1}_{\alpha\beta})k^\alpha k^\beta = 0 
\nonumber 
\\
\frac{dk^{x^1}}{d\lambda_s} &+& \Gamma^{x^1}_{\alpha\beta}k^\alpha k^\beta + \frac{B}{\mathcal{C}}\varphi,_{x^1}\bigg[\varphi,_{\alpha\beta} - \frac{1}{\mathcal{M}}\varphi,_\alpha\varphi,_\beta - \varphi'\Gamma^0_{\alpha\beta} - \varphi,_z\Gamma^z_{\alpha\beta}\bigg]k^\alpha k^\beta = 0 ,
\nonumber
\\
\end{eqnarray}
where $z \in \{r,x^1,x^2\}$. 
\\\\
The last two terms in the square brackets of Eqn.\! (\ref{eqn:disgeodesic}) will be evaluated separately.
Since the Christoffel symbols of $x^1$,$x^2$ and $r$ differ only by the respective index, it is sufficient to calculate only one addend of the last term in order to find an expression for the whole sum.
\\\\
Starting with the $\tau$ term gives 
\begin{eqnarray}\label{eqn:TauTerm}
\varphi'\Gamma^0_{\alpha\beta}k^\alpha k^\beta & = & \varphi' [(\mathcal{H} + \Psi')(k^0)^2 + 2\Psi,_z k^z k^0 + (\mathcal{H} + \Phi' +2\mathcal{H}(\Phi-\Psi))(k^z)^2] \nonumber \\ 
& = & \varphi'[(k^0)^2(2\mathcal{H} + (\Phi' - \Psi') +2\mathcal{H}(\Phi-\Psi)) + 2\Psi,_\alpha k^\alpha k^0] ,
\nonumber
\\
\end{eqnarray}
and, as a representative example, the $x^1$ term becomes
\begin{eqnarray}
\varphi,_{x^1}\Gamma^{x^1}_{\alpha\beta}k^\alpha k^\beta &=&  \varphi,_{x^1} [\Psi,_{x^1}(k^0)^2 + 2(\mathcal{H} +\Phi')k^{x^1}k^0 + \Phi,_{x^1}(k^{x^1})^2  
\nonumber 
\\
&\phantom{=}& \phantom{\phi,_{x^1}} 
+ 2 \Phi,_{x^2}k^{x^1}k^{x^2} + 2 \Phi,_r k^{x^1}k^{r} - \Phi_{x^1}(k^{x^2})^2 - \Phi_{x^1}(k^r)^2] 
\nonumber 
\\
&=& \varphi,_{x^1} [2\mathcal{H}k^{x^1}k^0 - (\Phi,_{x^1} - \Psi,_{x^1})(k^0)^2 + 2\Phi,_\alpha k^\alpha k^{x^1}] .
\nonumber
\\
\end{eqnarray}
Consequently, the sum over $z$ adds up to
\begin{eqnarray}\label{eqn:ZSum}
 \varphi,_z\Gamma^z_{\alpha\beta}k^\alpha k^\beta & = & 
 2\varphi,_z k^z(\mathcal{H}k^0 + \Phi,_\alpha k^\alpha)- a^2(k^0)^2 \varphi,_{z}(\Phi-\Psi),^{z}.
\end{eqnarray}
Combining Eqns.\! (\ref{eqn:TauTerm}) and (\ref{eqn:ZSum}) yields
\begin{eqnarray}\label{eqn:Sumofbothterms}
(\varphi'\Gamma^0_{\alpha\beta} + \varphi,_z\Gamma^z_{\alpha\beta})k^\alpha k^\beta & = & 2\mathcal{H} \varphi,_\alpha k^\alpha k^0 + 2\mathcal{H} \varphi'(\Phi - \Psi)(k^0)^2 + 2\varphi' k^0 \Psi_\alpha k^\alpha  \nonumber \\
&\phantom{=}& + 2 \varphi,_z k^z \Phi,_\alpha k^\alpha - a^2 (k^0)^2 \varphi_\alpha (\Phi-\Psi),^\alpha ,
\end{eqnarray}
where 
\begin{eqnarray}
a^2 \varphi,_\alpha (\Phi-\Psi),^\alpha = -\varphi'(\Phi - \Psi)' + \sum\limits_z \varphi,_{z}(\Phi-\Psi),_{z} 
\end{eqnarray}
was used.
\\\\
Next, substituting the result Eqn.\ (\ref{eqn:Sumofbothterms}) into Eqn.\ (\ref{eqn:disgeodesic}), expanding $k^\mu$ wherever it is reasonable, identifying $\hat{k^0} = \hat{k^r}$, and using $k^\mu = dx^\mu/d\lambda$ leads to the deflection equation:
\begin{eqnarray}
\label{eqn:deflect}
0 &=& \frac{d^2 x^1}{d\lambda^2} + 2\mathcal{H}\frac{d\tau}{d\lambda}\frac{dx^1}{d\lambda} - (\Phi -\Psi),_{x^1} \left( \frac{d\tau}{d\lambda} \right)^2 
\nonumber 
\\ 
&\phantom{=}& 
+ \frac{B}{\mathcal{C}}\varphi,^{x^1}\bigg\{ \varphi,_{\alpha\beta} \frac{dx^\alpha}{d\lambda} \frac{dx^\beta}{d\lambda}  - \frac{1}{\mathcal{M}} \varphi,_\alpha \varphi,_\beta \frac{dx^\alpha}{d\lambda} \frac{dx^\beta}{d\lambda} 
\nonumber 
\\ 
&\phantom{=}&
+ \left( \frac{d\tau}{d\lambda} \right)^2 [ a^2 \varphi,_\alpha (\Phi-\Psi),^\alpha - 2\varphi' (\Psi' + \Psi,_r) - 2\varphi,_r(\Phi' + \Phi,_r)  
\nonumber 
\\
&\phantom{=}&
\phantom{}
-2\mathcal{H}\varphi'(\Phi-\Psi)] + 2\frac{d\tau}{d\lambda}\mathcal{H} \varphi,_\alpha \frac{dx^\alpha}{d\lambda} \bigg\}.
\end{eqnarray}
The force law is finally obtained by substituting Eqns.\! (\ref{eqn:Replacementofx22}) and (\ref{eqn:GeodesicResults}) into Eqn.\! (\ref{eqn:deflect}), and dividing the result by $(d\tau/d\lambda)^2$:
\begin{eqnarray}\label{eqn:DisfoForceLaw}
 \frac{d^2 x^1}{dr^2} &=& (\Phi - \Psi),_{x^1} 
\nonumber
\\ 
&\phantom{=}&
 - \frac{B}{\mathcal{C}}\varphi,^{x^1} \bigg[ \bigg(\varphi,_{\alpha\beta} - \frac{1}{\mathcal{M}} \varphi,_\alpha \varphi,_\beta\bigg)\frac{dx^\alpha}{dr} \frac{dx^\beta}{dr} + a^2 \varphi,_\alpha (\Phi-\Psi),^\alpha 
\nonumber 
\\
&\phantom{=}&
-2\varphi'(\Psi'+\Psi,_r) -2\varphi,_r (\Phi' + \Phi,_r) - 2\mathcal{H}\varphi' (\Phi-\Psi) +2\mathcal{H}\varphi,_\alpha\frac{dx^\alpha}{dr}\bigg] .
\nonumber
\\
\end{eqnarray}
This result differs from Eqn.\! (\ref{eqn:force1}) by an additional term of second order in $\varphi$, which causes a modification of the gravitational lensing of light. 
\\\\
Only considering the case of a static lens reduces Eqn.\! (\ref{eqn:DisfoForceLaw}) to
\begin{eqnarray}
\frac{d^2 x^1}{dr^2} &=& (\Phi - \Psi),_{x^1} - \frac{B}{\mathcal{C}}\varphi,^{x^1} \bigg[\bigg(\varphi,_{zy} - \frac{1}{\mathcal{M}} \varphi,_z \varphi,_y\bigg)\frac{dx^z}{dr} \frac{dx^y}{dr} 
\nonumber 
\\
&\phantom{=}&
\phantom{(\Phi - \Psi),_{x^1} - \frac{B}{\mathcal{C}}\varphi,^{x^1} }
+a^2 \varphi,_z (\Phi -\Psi),^z - 2\varphi,_r \Phi,_r + 2\mathcal{H}\varphi,_z\frac{dx^z}{dr}\bigg] ,
\nonumber
\\
\end{eqnarray}
{where $y,z \in \{r,x^1,x^2\}$.}

\newpage

\subsection{Disformally coupled symmetron}\label{ssec:LensSingleSymm}

\begin{figure}[H]
\centering
\includegraphics[scale=0.6]{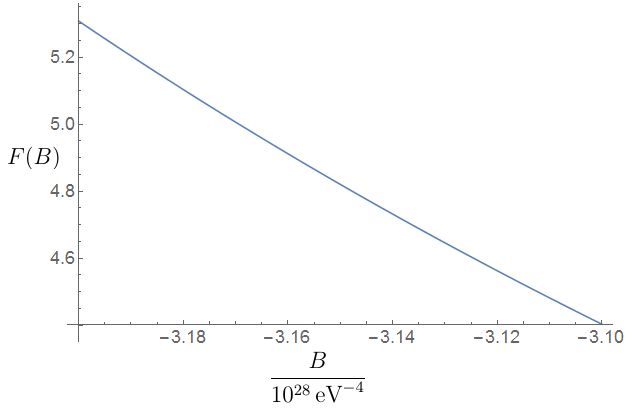} 
\caption{$F(B)$ is the lensing contribution by the disformal coupling relative to the Newtonian potential. The plot shows the range of $B$ for which $F(B)\approx 5$, such that the symmetron fifth force could be an explanation for the modification of gravitational lensing otherwise attributed to particle DM.}
\label{fig:SingleSymmetron}
\end{figure}

The authors of Ref.\! \cite{Copeland2016} showed that the radial acceleration for rotating galaxies \cite{Sanders1990,Janz,McGaugh} and the stabilisation of disk galaxies \cite{Ostriker} could be explained by the presence of a symmetron scalar field (see Section \ref{ssec:Symmetrons}).
In Ref.\! \cite{Kading2} it was then investigated whether symmetrons could also explain the enhancement of gravitational lensing of galaxies that is usually attributed to particle DM.
Since conformally coupled fields cannot influence gravitational lensing (as shown in Section \ref{ssec:ConfDisfLens}), one considered idea was that a disformal coupling of the type 
\begin{eqnarray}
\tilde{g}_{\mu\nu} &=& A^2(\varphi)g_{\mu\nu} + B\varphi,_\mu\varphi,_\nu~, \label{eqn:DisfomSymLensing}
\\
\tilde{g}^{\mu\nu} &=& A^{-2} (\varphi)\left(g^{\mu\nu} - \frac{B}{C}\varphi,^\mu\varphi,^\nu \right),
\end{eqnarray}
with $B$ being constant at leading order,
\begin{eqnarray}\label{eqn:CoCoSiSy}
A(\varphi) &=& 1+ \frac{\varphi^2}{2\mathcal{M}^2} + \mathcal{O}\left( \frac{\varphi^4}{\mathcal{M}^4} \right) ,
\end{eqnarray}
and 
\begin{eqnarray}\label{eqn:Cee}
C &=& 1 + \frac{\varphi^2}{\mathcal{M}^2} + B(\partial\varphi)^2,
\end{eqnarray}
could be sufficient to explain the discrepancy between observed baryonic mass and the lensing mass of galaxies.
In fact, observations indicate that the ratio between baryonic mass and DM in galaxies is roughly $1:5$ \cite{Weinberg}.
\\\\
Here it will be checked whether a disformally coupling symmetron is indeed a sufficient explanation for the observed DM content of galaxies.
For this, an analysis analogous to Section \ref{ssec:ConfDisfLens} shall be performed in order to study the ratio between the GR contributions to the lensing force equation and its corrections due to the disformal coupling.
Following the treatment in Section \ref{ssec:ConfDisfLens}, a force law in $x^i$-direction for a static lens under consideration of the transformation in Eqn.\! (\ref{eqn:DisfomSymLensing}) is obtained:
\begin{eqnarray}
\frac{d^2x^i}{dr^2} &=& (\Phi - \Psi),_{x^i} - \frac{B}{C} \varphi,^{x^i}\left[ \bigg(\varphi,_{zy} - \frac{2}{\mathcal{M}^2}\varphi\varphi,_z \varphi,_y\bigg)\frac{dx^z}{dr} \frac{dx^y}{dr}
 \right. 
\nonumber
\\
&\phantom{=}&
\phantom{(\Phi - \Psi),_{x^i} - \frac{B}{C} \varphi,^{x^i}}
\left. +a^2 \varphi,_z  (\Phi - \Psi),^z -2\varphi,_r \Phi,_r + 2\mathcal{H} \varphi,_z \frac{dx^z}{dr}  \right], 
\nonumber
\\
\end{eqnarray}
where $z,y \in\{ r, x^1, x^2 \}$.
\\\\
This can be further simplified by assuming the $a_\text{today}=1$, using the no-slip condition in Eqn.\! (\ref{eqn:NoSlip}), and only considering lensing in the lens plane ($r=D_L$) where derivatives $dx^i/dr$ are expected to vanish\footnote{Without the thin-lens approximation the lens plane is defined by the plane that goes through the centre of the lens, and is orthogonal to the line between the centres of lens and observer. The path of light does then not have a kink as depicted in Figure \ref{fig:LensingScheme} but is instead differentiable in the lens plane. In this case, the function $x^i(r)$ has an extremum at $r=D_L$, which requires $dx^i/dr|_{D_L} =0$.}.
As a result, this gives
\begin{eqnarray}\label{eqn:23ForceLAw}
\left. \frac{d^2x^i}{dr^2} \right|_{D_L} &=& \bigg[ 2\Phi,_{x^i} 
\left. 
- \frac{B}{C}\varphi,^{x^i} \bigg( \varphi,_{rr} - \frac{2}{\mathcal{M}^2} \varphi(\varphi,_r)^2 + 2\mathcal{H}\varphi,_r +  2\varphi,_{x^j} \Phi,^{x^j}\bigg) \bigg]\right|_{D_L}.
\nonumber
\\
\end{eqnarray}
Approximating a galaxy as a sphere with homogeneous, constant mass density $\rho$ and radius $R$, and using the shell theorem \cite{Newton}, justifies the treatment of its gravitational potential as the Newtonian potential of a point mass
\begin{eqnarray}
\Phi &=& -\frac{GM}{r_s} ~=~ -\frac{4\pi}{3}\frac{G\rho R^3}{r_s},  
\end{eqnarray}
where $M$ is the total source mass and $r_s$ is the radius originating from the sphere's centre.
Within the lens plane, this radius $r_s$ can be expressed in terms of the coordinates in Eqn.\! (\ref{eqn:Coordiates}) as
\begin{eqnarray}
r_s = r\sqrt{(\theta^1)^2 + (\theta^2)^2}, 
\end{eqnarray}
such that
\begin{eqnarray}
\Phi &=& -\frac{GM}{r\sqrt{(\theta^1)^2 + (\theta^2)^2}} .
\end{eqnarray}
The symmetron field profile will be taken from Eqn.\! (\ref{eqn:SymmetronProfile}) but can be simplified by making the reasonable assumption $\varphi_{0,\text{out}} \gg \varphi_{0,\text{in}}$, and only considering radii for which $r_sm_{\text{out}}\ll 1$ and $Rm_{\text{out}}\ll 1$, which is in agreement with the explicit parameters chosen later.
Here the label ``in" denotes objects inside a galaxy and ``out" labels those objects which are in vacuum outside this galaxy.
The resulting profile around a spherically symmetric galaxy is\footnote{It shall be noted that, at least in the considered case, the disformal coupling term does not modify the field profile. 
This is due to the fact that the considered field source is static and non-relativistic, meaning that $\partial_0\varphi =0$, and  $\forall \mu,\nu \neq 0:~ T^{\mu\nu} =0$. 
Consequently, a term $\partial_\mu\varphi\partial_\nu\varphi T^{\mu\nu}$ in the symmetron Lagrangian cannot contribute to the dynamics of the field.}
\begin{eqnarray}
\varphi(r_s) &=& \pm \frac{v}{mr_s} \left[ \mathcal{S} +m(r_s-R) \right] ,
\end{eqnarray}
where for convenience $v := \varphi_{0,\text{out}} $, $m:=m_{\text{in}}$, and
\begin{eqnarray}
\mathcal{S} &:=& \frac{\sinh(mR)}{\cosh(mR)}
\end{eqnarray}
were defined.
This means in the coordinates originating at the observer's position:
\begin{eqnarray}\label{eqn:ApproxSymmProf}
\varphi(r,\theta^1,\theta^2) = \pm \frac{v}{mr\sqrt{(\theta^1)^2 + (\theta^2)^2}} \left[ \mathcal{S} +m(r\sqrt{(\theta^1)^2 + (\theta^2)^2}-R) \right].
\end{eqnarray}
With all this at hand, the effect of symmetron scalar fields on gravitational lensing of a galaxy can be studied.
Since the galaxy is assumed to be spherically symmetric, it is sufficient to investigate the lensing only in one angular direction, meaning that $\theta^2 \equiv 0$ can be considered, and $\theta^1 =: \omega$ be defined.
In this simplified situation, the derivatives relevant for evaluating Eqn.\! (\ref{eqn:23ForceLAw}) are given by
\begin{eqnarray}
\Phi,_{x} = \frac{GM}{r^2\omega^2}  
\end{eqnarray}
for the Newtonian potential, with $x\approx r\omega$, and
\begin{eqnarray}
\varphi,_r &=& \pm \frac{v}{mr^2\omega} (mR-\mathcal{S}) ,
\\
\nonumber
\\
\varphi,_{rr} &=& \mp \frac{2v}{mr^3\omega} (mR-\mathcal{S}) ,
\\
\nonumber
\\
\varphi,_{x} &=& \pm \frac{v}{mr^2\omega^2} (mR-\mathcal{S}) ,
\\
\nonumber
\\
\varphi,^{z} &=& (1-2\Phi)\varphi,_{z} 
\end{eqnarray}
for the field profile. 
Since no terms higher than first order are considered, and $a_\text{today}=1$ is assumed, $\Phi,_{x} \approx \Phi,^{x}$.
\\\\
Now Eqn.\ (\ref{eqn:23ForceLAw}) shall be expressed in terms of $D_L$ and $\omega$, such that a quantitative estimation is possible. For this, Eqn.\! (\ref{eqn:Cee}) will be considered first.
In the lens plane it becomes
\begin{eqnarray}
C|_{D_L} 
&=& 1+ \frac{v^2}{\mathcal{M}^2m^2D_L^2\omega^2} \left[ \mathcal{S} +m(D_L\omega-R) \right]^2
\nonumber
\\
&\phantom{=}& 
\phantom{1}
 +(1-2\Phi|_{D_L}) \frac{v^2B}{m^2D_L^4\omega^2}(mR-\mathcal{S})^2(1+\omega^{-2}) .
\end{eqnarray}
Next, the following terms are considered:
\begin{eqnarray}
&&
\left. \Big[ \varphi,_{rr} - \frac{2}{\mathcal{M}^2} \varphi(\varphi,_r)^2 + 2\mathcal{H}\varphi,_r +  \varphi,_{x^j} (\Phi-\Psi),^{x^j} \Big] \right|_{D_L}   
\nonumber
\\
&=&
\mp \frac{2v(mR-\mathcal{S})}{mD_L^3\omega}\bigg\{ 1 + \frac{v^2(mR-\mathcal{S})}{\mathcal{M}^2m^2D_L^2\omega^2} [ \mathcal{S} +m(D_L\omega-R) ]\phantom{\mp \frac{2v(mR-\mathcal{S})}{mD_L^3\omega}}
\nonumber
\\
&\phantom{=}&
\phantom{\mp \frac{2v(mR-\mathcal{S})}{mD_L^3\omega}}
- \mathcal{H}D_L- \frac{GM}{D_L\omega^3}\bigg\} .
\end{eqnarray}
With this, Eqn.\! (\ref{eqn:23ForceLAw}) results in
\begin{eqnarray}\label{eqn:VeryLongEQn}
\left. \frac{d^2x^i}{dr^2} \right|_{D_L} &=& \frac{2GM}{D^2_L\omega^2}  
+ \frac{2Bv^2(mR-\mathcal{S})^2}{Cm^2D_L^5\omega^3}\times
\nonumber
\\
&&
\times
\bigg[\left( 1+ \frac{2GM}{D_L\omega} \right) \bigg(1
+ \frac{v^2(mR-\mathcal{S})( \mathcal{S} +m(D_L\omega-R) )}{\mathcal{M}^2m^2D_L^2\omega^2} 
\nonumber
\\
&&
\phantom{\times\bigg\{\left( 1+ \frac{2GM}{D_L\omega} \right) }
- \mathcal{H}D_L\bigg) - \frac{GM}{D_L\omega^3}\bigg],
\end{eqnarray}
where the last term in the square brackets is not multiplied by $\Phi$ because this would otherwise be a term of second order in the Newtonian potential.
\\\\ 
In order to allow for an easy comparison between the GR contribution and the term from the disformal coupling, the derivative of the Newtonian potential is factorised out of Eqn.\! (\ref{eqn:VeryLongEQn}):
\begin{eqnarray}
\left. \frac{d^2x^i}{dr^2} \right|_{D_L} &=& \frac{2GM}{D^2_L\omega^2}  
\bigg\{1+ \frac{Bv^2(mR-\mathcal{S})^2}{CG Mm^2D_L^3\omega }\times
\nonumber
\\
&&
\times
\bigg[\left( 1+ \frac{2GM}{D_L\omega} \right) \bigg(1
+ \frac{v^2(mR-\mathcal{S})( \mathcal{S} +m(D_L\omega-R) )}{\mathcal{M}^2m^2D_L^2\omega^2} 
\nonumber
\\
&&
\phantom{\times\bigg\{\left( 1+ \frac{2GM}{D_L\omega} \right) }
- \mathcal{H}D_L\bigg) - \frac{GM}{D_L\omega^3}\bigg]\bigg\},
\end{eqnarray}
and a function
\begin{eqnarray}\label{eqn:FuncF}
F(B) &:=& 
\frac{Bv^2(mR-\mathcal{S})^2}{CG Mm^2D_L^3\omega }
\bigg[\left( 1+ \frac{2GM}{D_L\omega} \right) \bigg(1
\nonumber
\\
&&
+ \frac{v^2(mR-\mathcal{S})( \mathcal{S} +m(D_L\omega-R) )}{\mathcal{M}^2m^2D_L^2\omega^2} 
- \mathcal{H}D_L\bigg) - \frac{GM}{D_L\omega^3}\bigg]
\nonumber
\\
\end{eqnarray}
defined.
\\\\
It shall now be tested for what value of $B$ the contribution from the disformally coupled symmetron to gravitational lensing represented by the function $F$ equals roughly $5$ since the ratio between baryonic mass and DM in galaxies is approximately $1:5$ \cite{Weinberg}.
For this, concrete values for the parameters appearing in Eqn.\! (\ref{eqn:FuncF}) have to be considered.
As an example, the galaxy parameters from Ref.\! \cite{Copeland2016} will be used, i.e.\ a Milky Way-like galaxy with 
\begin{eqnarray}
M &=& 6 \times 10^{11} M_\odot ~\approx~ 6.67 \times 10^{77} \,\text{eV},
\\
R &=& 5\, \text{kpc} ~\approx~ 2.69 \times 10^{26} \,\text{eV}^{-1},
\end{eqnarray}
where the radius $R$ corresponds to the scale length \cite{Fathi} of the Milky Way.
Lensing with galaxies has been observed, for instance, at redshifts of $z=1$ under an angle of $1\,\text{arcmin}$ \cite{Prat}, leading to
\begin{eqnarray}
D_L &\approx& 6.60\times 10^{32} \,\text{eV}^{-1},
\\
\omega &=& \frac{\pi}{10800},
\end{eqnarray} 
where $D_L\approx zD_H$ \cite{Hogg} was used with $D_H$ being the Hubble length.
The symmetron coupling to matter is given by Ref.\! \cite{Copeland2016} as
\begin{eqnarray}\label{eqn:SymmcotoMa}
\mathcal{M} &=& \frac{M_P}{10} ~\approx~  2.43 \times 10^{26} \,\text{eV},
\end{eqnarray}
and the tachyonic mass as
\begin{eqnarray}
\mu &=& 3 \times10^{-30} \,\text{eV},
\end{eqnarray}
which means that the symmetron is screened within the galaxy but unscreened in the vacuum outside{\footnote{The radiative stability of a model with such a small mass is a potential issue. However, in \cite{Burrage:2016xzz} it was shown that a radiatively stable symmetron model can be constructed.}}.
The symmetron background value is chosen to be
\begin{eqnarray}
v &=& \frac{\mathcal{M}}{150},
\end{eqnarray}
for which the appropriate value of $\lambda$ is used.
For completeness, the gravitational constant is given by
\begin{eqnarray}
G &\approx& 6.71 \times 10^{-57} \,\text{eV}^{-2},
\end{eqnarray}
and today's Hubble parameter by
\begin{eqnarray}
\mathcal{H} &\approx& 1.51 \times 10^{-33} \,\text{eV}.
\end{eqnarray} 
Using all these values, it can be read off from Figure \ref{fig:SingleSymmetron} that for 
\begin{eqnarray} 
B &\approx& -3.16\times 10^{28} \,\text{eV}^{-4}
\end{eqnarray}
the equation $F(B)\approx 5$ is fulfilled.
This means, that for this particular value of $B$ the DM contribution to gravitational lensing by the considered example galaxy could be explained by an unscreened symmetron fifth force. 
A numerical analysis shows that this is indeed the only possible solution.
This solution, however, is already excluded by experiments, including collider, tabletop and astrophysical experiments, from which $B < 5.6 \times 10^{-48} \,\text{eV}^{-4}$ is required \cite{Brax2014Ex,Brax2015Ex}.
Realistic variations of the used galaxy parameters do not significantly change this result, and varying the symmetron parameters too far away from those used in Ref.\! \cite{Copeland2016} would potentially spoil the explanations for the radial acceleration for rotating galaxies and the stabilisation of disk galaxies by a symmetron field.
In conclusion, a disformally coupled symmetron scalar field is not able to explain the modification of gravitational lensing that is otherwise attributed to particle DM while simultaneously giving rise to the galactic dynamics effects described in Ref.\! \cite{Copeland2016}.  
\\\\
Besides disformal couplings, Ref.\! \cite{Kading2} also investigated two other ideas: having a small photon mass, which would allow for a conformally coupling symmetron to modify gravitational lensing, and having an axion-like coupling \cite{BraxCoupling} between symmetrons and photons that is generated from quantum corrections.
However, also these two ideas turned out to not allow for a relevant modification of gravitational lensing due to symmetrons.
\newpage
\subsection{Disformal coupling to extended field content}\label{ssec:CosmoScalar}
\begin{figure}[H]
\centering
\includegraphics[scale=0.6]{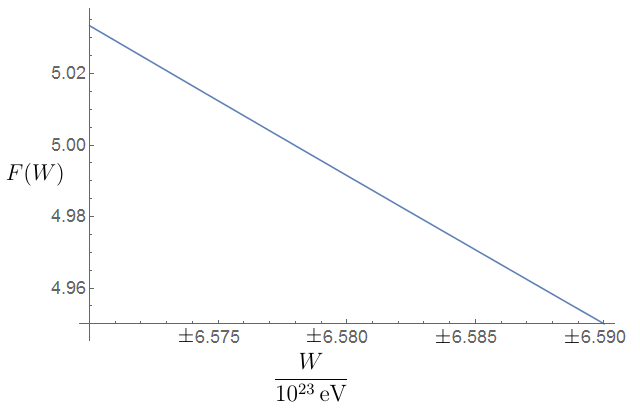} 
%
\caption{$F(W)$ is the lensing contribution by the disformal coupling relative to the Newtonian potential. 
Since $F$ is symmetric in $W$, negative and positive values of $W$ are presented in the same plot with an inverted negative number scale on the abscissa.
The plot shows the range of $W$ for which $F(W)\approx 5$, such that the symmetron fifth force could be an explanation for the modification of gravitational lensing otherwise attributed to particle DM.}
\label{fig:ExtField}
\end{figure}

Since considering only a disformally coupled symmetron was not sufficient to explain the modification of galactic gravitational lensing, it could be explored whether an altered model would be able to deliver a positive result. 
One idea could be to extend the field content, i.e.\ considering an additional scalar field besides a symmetron. 
Introducing an extra field was an attempt made in Ref.\! \cite{KhouryAlter} in order to explain galactic dynamics similarly to Ref.\! \cite{Copeland2016}.
The scalar field that was introduced there has, in contrast to the other fields considered in this thesis, the dimension of a length or inverse mass, and fulfils
\begin{eqnarray}
\pi' &\approx& 1
\end{eqnarray}
with $'$ being a derivative with respect to conformal time. 
\\\\
This field was adapted in Ref.\! \cite{Kading2}, and introduced by the following transformation from the Jordan to the Einstein frame:
\begin{eqnarray}
\tilde{g}_{00}&=& A^2(\varphi)g_{00} + \frac{\varphi^2}{W^2}\partial_0 \pi \partial_0 \pi, \label{eqn:CosmoDisfoTrafo}
\\
\tilde{g}_{ab}&=& A^2(\varphi)g_{ab},
\end{eqnarray}
where $a,b$ are placeholders for spatial coordinates, and the constant $W$ with mass dimension $1$ describes the coupling between the symmetron $\varphi$ and the field $\pi$. 
The conformal coupling $A(\varphi)$ is again given by Eqn.\ (\ref{eqn:CoCoSiSy}).
It is assumed that $\varphi \ll W$, and the Christoffel symbols resulting from Eqn.\! (\ref{eqn:CosmoDisfoTrafo}) can be found in Appendix \ref{app:ChristoCosmo}. 
\\\\
Due to the transformation in Eqn.\ (\ref{eqn:CosmoDisfoTrafo}), the symmetron Lagrangian in Eqn.\! (\ref{eqn:SymmetronLagrangian}) becomes modified to
\begin{eqnarray}
\mathcal{L}_\varphi &=& -\frac{1}{2}\left(\partial\varphi\right)^2 -\frac{1}{2}\left[ \rho_0\bigg(\frac{1}{\mathcal{M}^2}-\frac{2}{W^2}\bigg) - \mu^2 \right]\varphi^2 - \frac{\lambda}{4}\varphi^4,
\end{eqnarray}
which results in a modification of the symmetron mass in Eqn.\! (\ref{eqn:SymmetronMass}):
\begin{eqnarray}\label{eqn:ModSymmetronMass}
m^2 &=& 2\bigg[\mu^2 - \rho_0\left( \frac{1}{\mathcal{M}^2}-\frac{2}{W^2} \right) \bigg]
\end{eqnarray}
where $\rho_0 = \mathcal{H}^2M_P^2$ is the present day cosmological density \cite{Hinterbichler2010}.
The background symmetron $\varphi_0$ keeps the same form as given in Eqn.\! (\ref{eqn:SymmetronBackgroundProf}) but is of course defined in terms of Eqn.\! (\ref{eqn:ModSymmetronMass}).
\\\\
Following the same procedure as in earlier discussions, a force law in the static case can be derived from the geodesic equation (\ref{eqn:Geodesic}):
\begin{eqnarray}\label{eqn:ExFiCoFoLa}
\frac{\partial^2x}{\partial r^2} &=& 2\Phi,_x + \frac{\varphi\varphi,_x}{a^2W^2}. 
\end{eqnarray}
Here the field $\varphi$ has the same profile as in Eqn.\! (\ref{eqn:ApproxSymmProf}), and{,} as in Section \ref{ssec:LensSingleSymm}, the simplification $\theta^2 \equiv 0$ together with the definition $\theta^1 =: \omega$ can be done since the galaxy is assumed to be spherically symmetric: 
\begin{eqnarray}
\varphi(r,\omega) = \pm \frac{v}{mr\omega} \left[ \mathcal{S} +m(r\omega-R) \right].
\end{eqnarray}
Substituting this into Eqn.\! (\ref{eqn:ExFiCoFoLa}), considering it in the lens plane at $r=D_L$, and assuming $a_\text{today}=1$ for the scale factor leaves
\begin{eqnarray}\label{eqn:PuOuNePoSiSy}
\left.\frac{\partial^2x}{\partial r^2}\right|_{D_L}  &=& \frac{2GM}{D_L^2\omega^2} \left[1 + \frac{v^2(mR-\mathcal{S})(\mathcal{S}+m(D_L\omega-R))}{2GMW^2m^2D_L\omega} \right],
\end{eqnarray}
where the factor in front of the square brackets equals $2\Phi,_x$.
\\\\
From Eqn.\! (\ref{eqn:PuOuNePoSiSy}) a function 
\begin{eqnarray}
F(W)&:=& \frac{v^2(mR-\mathcal{S})(\mathcal{S}+m(D_L\omega-R))}{2GMW^2m^2D_L\omega} 
\end{eqnarray}
is defined. 
It describes the ratio between the contributions of the disformal coupling and the Newtonian potential to the lensing by galaxies.
By choosing the same galaxy and symmetron parameters as in Section \ref{ssec:LensSingleSymm}, the value of $W$ can be estimated for which $F(W) \approx 5$, as would be required in order to explain the modification of gravitational lensing otherwise attributed to particle DM.
Due to Eqn.\! (\ref{eqn:ModSymmetronMass}), $m$, $v$ and $\mathcal{S}$ have a $W$-dependence, which has to be taken into account.
\\\\
In Figure \ref{fig:ExtField} it can be seen that 
\begin{eqnarray}\label{eqn:Wvalue}
W &\approx& \pm 6.58 \times 10^{23} \,\text{eV}
\end{eqnarray}
leads to the desired value of $F$. 
\\\\
In conclusion, there are two values for $W$ that would allow the two scalar field model presented here to explain the observed modification of gravitational lensing by a symmetron fifth force.
Future research in the direction presented in Ref.\! \cite{KhouryAlter}, including further developing the theory and making falsifiable predictions that lead to model constraints, might give more insight about the validity of this approach.
How the introduction of $\pi$ affects the galactic dynamics discussed in Ref.\! \cite{Copeland2016} must also be further investigated.

\newpage
\clearpage\mbox{}\clearpage
\newpage
\section{Open quantum dynamics from conformally coupling scalar fields}\label{Sec:OpenQuantumfromCCSF}

Astrophysics and cosmology provide excellent testing grounds for screened scalar field and fifth force models.
However, conducting experiments in those contexts can be quite costly and practically challenging.
Therefore, constraining screened scalar fields in {Earth}-based experiments would, at least in general, be more practicable.
Though, there is a problem with doing this, namely that the effects of these scalar fields should be strongly suppressed, i.e.\ screened, here on {Earth}. 
This complication, however, can at least partially be avoided by performing scalar field tests inside a vacuum chamber, e.g.\ with atom interferometry \cite{Cronin}.
Due to the low environmental density inside a vacuum chamber, some scalar fields are much less screened than they would be on {Earth} outside such a chamber. 
Especially, the chameleon field is an ideal candidate for laboratory experiments in vacuum chambers since its thin-shell effect (see Section \ref{ssec:Chameleons}) allows it to be nearly unscreened if the chamber walls are sufficiently thick. 
This is because the chameleon can become so heavy within the chamber walls that it cannot communicate with the outside world.
Consequently, there have been many successful attempts to constrain chameleon and symmetron fields with atom interferometry \cite{Hinds,Hamilton2015,Burrage:2015lya,Elder:2016yxm,Ben,Jaffe:2016fsh,Sabulsky:2018jma} here on {Earth}. 
See Refs.\! \cite{BurrageSakA,BurrageSakB} for a more comprehensive overview including experiments besides atom interferometry. 
\\\\
To date quantum experiments like atom interferometry are amongst the most powerful tools for constraining scalar fields in laboratory experiments. 
Atom interferometry is an experiment in which the wave-like nature of quantum mechanical atoms is used to create atom interference patterns in a detector, similar to an optical interferometer. 
It allows an experimentalist to perform very precise measurements of changes in the paths on which the involved atoms travel.
This can be used to constrain fifth forces induced by scalar fields since they are expected to cause modifications of the atomic trajectories in the experiment.
However, the change of a path length due to a fifth force is merely a classical effect, meaning that quantum effects caused by the presence of an additional quantum scalar field are not taken into account.
With atom interferometry it is also possible to observe quantum effects like decoherence \cite{Schlosshauer} which have the potential to provide new constraints on screened scalar fields. 
Though, this has not been investigated in any screened scalar field experiment yet.
\\\\
The work in Ref.\! \cite{Kading3}, summarised in Ref.\! \cite{Kading4}, gives the first predictions for quantum effects induced by conformally coupled scalar fields on a test particle like an atom in interferometry. 
This helps to get an understanding of the nature and magnitude of quantum effects that such fields could cause in quantum experiments.
In order to gain knowledge about such effects, it is useful to derive a quantum master equation that describes the evolution of the test particle under influence of a conformally coupled scalar field. 
For this, it is usually assumed that the investigated quantum system is open and interacting with an environment, coining the phrase open quantum system \cite{BreuerPetruccione}.
There are powerful field theoretical tools that allow for the study of open quantum systems (see e.g.\ Ref.\! \cite{CalzettaHuQFT}) but they can be intricate in their usage when actual predictions for experiments are supposed to be made.
The work in Ref.\! \cite{Kading3} addresses this problem, and presents as a main result novel tools based on non-equilibrium QFT that are used to derive a master equation describing the influence of conformally coupled scalar fields on a quantum test particle. 
\\\\
The corresponding derivation and the novel tools that had to be developed for this purpose comprise the main content of this section.
At first, in Section \ref{ssec:OpenQuantum}, it will be explained in more detail what is meant by an open quantum system.
Then a short introduction to atom interferometry will be provided in Section \ref{ssec:AtomInterfero}.
Subsequently, the quantum master equation will be derived in Section \ref{ssec:MasterEqnDerive}. 
While doing so, the used tools and all background needed for their development will be explained on the way. 
Section \ref{ssec:ExpImplications} provides a conclusion with a short discussion of experimental implications that can be expected from the studied quantum effects.

\newpage
\subsection{Open quantum systems}\label{ssec:OpenQuantum}

\begin{figure}[H]
\begin{center}
\includegraphics[scale=0.20]{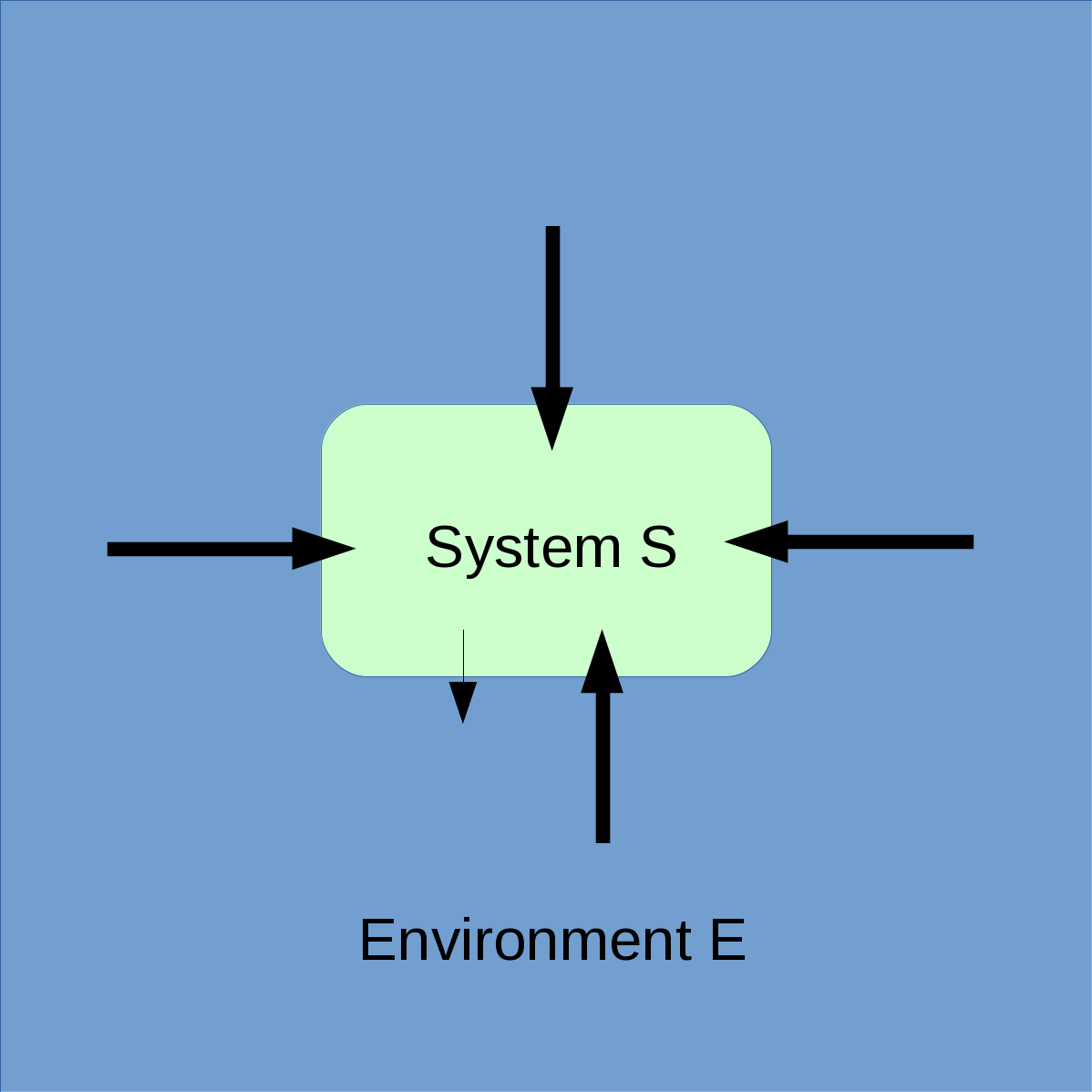}
\caption{Schematic depiction of an open quantum system. The open system S is surrounded by a large environment E. S and E interact with each other and exchange energy, momentum or particles. This interaction affects S much stronger than it affects E. A common assumption (Born-Markov approximation) is that the effect on E is so weak that E relaxes quickly enough for any change to be undone within the time scale of interest, effectively meaning that E is approximately constant in time.\label{fig:OpenQuantumSystem}}
\end{center}
\end{figure}

Similar to a classical system, a realistic quantum system must be open, meaning that it is continuously interacting with one or more uncontrollable environments whose effects on the system cannot be neglected \cite{BreuerPetruccione}\footnote{If not otherwise mentioned, all following arguments in this section refer to those presented in Ref.\! \cite{BreuerPetruccione} as well.}.
Such interactions cause phenomena like energy and momentum diffusion \cite{CalzettaHuQFT}, and ultimately effects like decoherence \cite{Schlosshauer}. 
Typical examples of open quantum systems can be found in quantum optics \cite{Carmichael} but also in fields like early Universe cosmology (see e.g.\ {Refs.} \cite{Lombardo1,Lombardo2,Lombardo3,Boyanovsky1,Boyanovsky2,Boyanovsky3,Boyanovsky4,Burgess2015,Hollowood}) and heavy-ion physics (see e.g.\ {Ref.} \cite{Brambilla1,Brambilla2}).
\\\\
Calculations in open quantum dynamics usually deal with a system $S$ that is surrounded by and interacting with an environment $E$.
An environment could be, for example, an electromagnetic, gravitational or scalar field.
In general, it is assumed that the combined system $C = S + E$ is itself closed. 
The combined system is described by a density operator $\hat{\rho}$ such that the expectation value of an operator $\hat{O}$ acting on the full system is 
\begin{eqnarray}
\langle\hat{O}\rangle_C &=& \text{Tr}(\hat{\rho}\,\hat{O}).
\end{eqnarray}
The so-called Liouville-von Neumann equation
\begin{eqnarray}\label{eqn:LiovilleVonNeumann}
\frac{d}{dt}\hat{\rho}(t) &=& \hat{\mathfrak{L}}(t) \hat{\rho}(t)
\end{eqnarray}
describes the time evolution of this density operator, where $\hat{\mathfrak{L}}(t)$ is the Liouville super-operator (also known as Liouvillian), mapping an operator to an operator, and acting like 
\begin{eqnarray}
\hat{\mathfrak{L}}(t) \hat{\rho}(t) &=& - i[\hat{H}(t),\hat{\rho}(t)]
\end{eqnarray}
with $\hat{H}(t)$ being the Hamiltonian of the closed composite system.
\\\\
A common assumption is that system and environment are only weakly coupled, such that the environment is barely affected by the system, while the system is subject to changes induced by the environment. 
Therefore, the environment is approximately constant over time.
This so-called Born approximation makes it possible to separate $S$ and $E$, such that 
\begin{eqnarray}
\hat{\rho}(t) &\approx& \hat{\rho}_S(t) \otimes \hat{\rho}_E,
\end{eqnarray}
where $\hat{\rho}_{S,E}$ denotes the separated system or environment density operator. 
Furthermore, another commonly used assumption is that any excitations induced in $E$ due to $S$ decay within a time much shorter than a (coarse grained) time scale of interest.
This is implemented by the Markov approximation, which in practical terms means that time integrals are not integrated up to finite times but up to $\infty$. 
Physically this means that memory effects are ignored, meaning that the system does not get feedback from its own effects on the environment and therefore does not depend on any earlier times.
The combination of both approximations is often referred to as Born-Markov approximation.
Figure \ref{fig:OpenQuantumSystem} illustrates the situation when they are applied.
\\ 
\noindent
Since the environment $E$ is often uncontrollable experimentally and difficult to describe, only the subsystem $S$ and its evolution under presence of $E$ are of interest. 
In order to mathematically extract $\hat{\rho}_S$ from the total density operator, a partial trace
\begin{eqnarray}
\text{Tr}_E \hat{\rho} &=& \hat{\rho}_S
\end{eqnarray} 
with respect to the environmental degrees of freedom is taken.
For example, if there are the variables $x_S$ and $x_E$, then 
\begin{eqnarray}
\hat{\rho}_S (x_S) &=& \int dx_E \hat{\rho}(x_S,x_E).
\end{eqnarray}
The time evolution of the so-called reduced density matrix $\hat{\rho}_S$ under presence of the environment is then described by a master equation, which - under the Markov approximation - can have the Lindbladian form, which resembles Eqn.\! (\ref{eqn:LiovilleVonNeumann}) but has a super-operator that not only describes unitary evolution (as in the Liouville-Von Neumann equation) but also non-unitary effects.
However, without the Markov approximation,the master equation, which is then referred to as non-Markovian, does not necessarily have this general form.
A non-Markovian master equation will be derived in Section \ref{ssec:MasterEqnDerive}.
\\\\
Non-unitary effects are characteristic for open quantum systems, and may include, e.g.\ decoherence.
Decoherence is an effect that may explain the emergence of the classical world as a quantum-to-classical transition \cite{Schlosshauer}. 
It means that after a certain time - the decoherence time - a quantum system, represented by a superposition of states, becomes classical, i.e.\ its superposition is destroyed and it remains in a particular state. 
This is mathematically represented by the behaviour of its density matrix $\hat{\rho}_S$, which for a quantum system is idempotent but for a classical {statistical} system is not, i.e.
\begin{eqnarray}
\hat{\rho}_S^2 &=&\hat{\rho}_S,~~~\text{if S is quantum},
\\
\hat{\rho}_S^2 &\neq &\hat{\rho}_S,~~~\text{if S is classical}.
\end{eqnarray}
As an example\footnote{The following discussion is inspired by Ref.\! \cite{Schlosshauer}.}, consider a system given by the superposition of two states:
\begin{eqnarray}\label{eqn:SystemState}
|\Psi_S\rangle &=& \alpha|0_S\rangle + \beta|1_S\rangle,
\end{eqnarray}
where its idempotent density matrix is given by
\begin{eqnarray}
\hat{\rho}_S &=& |\Psi_S\rangle\langle \Psi_S | ~=~ \begin{pmatrix}
                         |\alpha|^2 && \alpha^\ast \beta
										\\
	\alpha\beta^\ast && |\beta|^2
                     \end{pmatrix}.
\end{eqnarray}
An operator $\hat{O}$ acting on this system has then an expectation value
\begin{eqnarray}
\langle \Psi_S | \hat{O} | \Psi_S \rangle &=& \text{tr}(\hat{\rho}_S\,\hat{O})
\nonumber
\\ 
&=& \alpha^2\langle 0_S| \hat{O} | 0_S \rangle + \alpha\beta^\ast\langle 0_S| \hat{O} | 1_S \rangle
\nonumber
\\ 
&&+ \alpha^\ast\beta\langle 1_S| \hat{O} | 0_S \rangle + \beta^2 \langle 1_S| \hat{O} | 1_S \rangle.
\end{eqnarray}
Introducing an interaction with an environment $E$ leads, after a characteristic decoherence time, to an entangled total system of $S$- and $E$-states:
\begin{eqnarray}
|\Psi\rangle = \alpha |0_S\rangle \otimes |{E_0}\rangle + \beta |1_S\rangle \otimes |{E_1} \rangle
\end{eqnarray}
with 
\begin{eqnarray}
\langle {E_m} | {E_n} \rangle = \delta_{mn}, ~~ m,n\in\{0,1\}.
\end{eqnarray}
Using the total density operator 
\begin{eqnarray}
\hat{\rho} &=& |\Psi\rangle \langle \Psi |
\end{eqnarray}
to compute the reduced density matrix 
\begin{eqnarray}
\hat{\rho}_S &=& \text{Tr}_E\hat{\rho} ~=~ \sum_{n\in\{0,1\}} (\mathbb{I}\otimes \langle {E_n}|) |\Psi\rangle \langle \Psi | (\mathbb{I}\otimes | {E_n}\rangle)
\end{eqnarray}
shows that now only diagonal elements remain, and $\hat{\rho}_S$ is not idempotent anymore. 
Consequently, the system $S$ is now classical - it has decohered.

\newpage
\subsection{Atom interferometry}\label{ssec:AtomInterfero}

\begin{figure}[htbp]
\begin{center}
\includegraphics[scale=0.30]{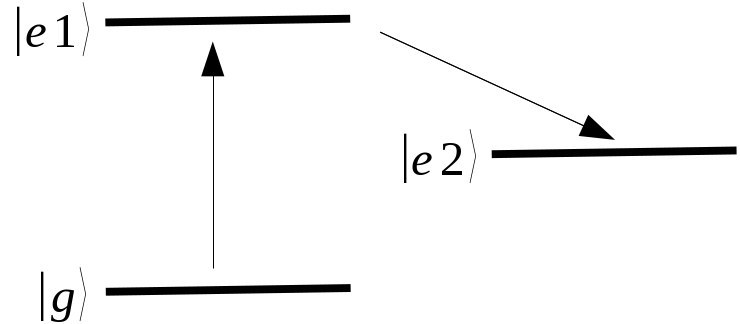}
\caption{\label{fig:AtomicStates} An atom can absorb laser light to go from its ground state $|g\rangle$ to an excited state $|e1\rangle$. Using another laser to induce emission of a photon can cause the excited atom to go to a lower energy level $|e2\rangle$.}
\end{center}
\end{figure}

\begin{figure}[htbp]
\begin{center}
\includegraphics[scale=0.20]{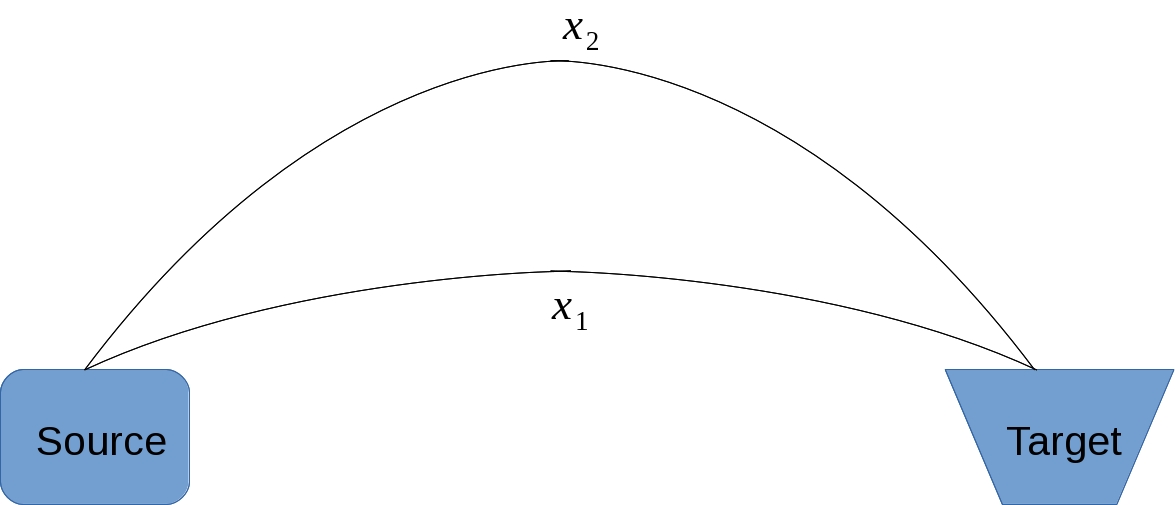}
\caption{\label{fig:AtomInterfero} An atom is produced in a source. Its superposed states travel along different paths but arrive at the same time in a detector where an interference pattern can be observed.}
\end{center}
\end{figure}

\begin{figure}[H]
\begin{center}
\includegraphics[scale=0.40]{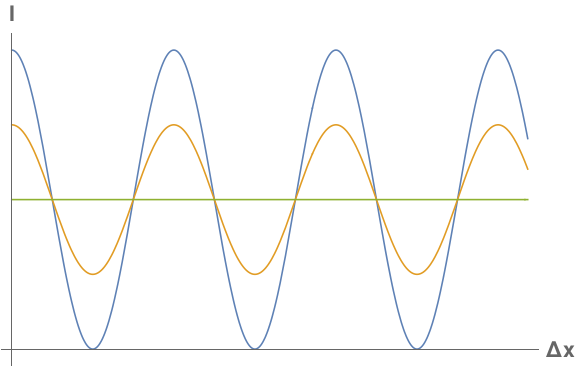}
\caption{\label{fig:Interference}Schematic plot of interference patterns in atom interferometry, in which the intensity $I$ is plotted over the path length difference $\Delta x$. The green line corresponds to the classical intensity $I_\text{cl}$, which appears if every atom in the experiment has decohered. The blue line corresponds to the ideal case in which no decohrence effects are present, and the peak-to-peak amplitude is $2I_\text{cl}$. A case between both extremes is depicted by the orange line: when reaching the detector, some atoms were still superposed, while others were classical, causing the peak-to-peak amplitude to be {larger than $0$ and smaller than} $2I_\text{cl}$.}
\end{center}
\end{figure}

Atom interferometry has successfully been used to constrain light scalar fields like chameleons and symmetrons \cite{Hinds,Hamilton2015,Burrage:2015lya,Elder:2016yxm,Ben,Jaffe:2016fsh,Sabulsky:2018jma}.
Here a short overview of this type of experiment shall be given in order to explain how it can be used to observe decoherence that could, for example, be induced by conformally coupling scalar fields\footnote{Here the open system would be the atoms in the interferometer with an environment formed by the conformally coupling scalar fields.}. 
More details of atom interferometry can be found e.g.\ in Ref.\! \cite{Cronin}.
\\\\
Optical interferometry uses light to create interference patterns in a detector. 
Alterations in these patterns can hint at a distortion or change of the system, which allows light interferometers to be used to detect those.
For example, LIGO makes use of this principle by measuring the change in the length of its two laser interferometer arms during the passage of a gravitational wave \cite{Weiss}.
Atomic interferometry follows the same idea: quantum mechanical atoms are wave-like and can be interfered with each other.
More spectacularly, it is even possible to have two superposed states of the same atom travelling along two different paths and then coming together in a detector in order to form an interference pattern. 
This is actually the same principle as the one behind the famous double slit experiment with electrons.
\\\\
In practice, a superposition of two energy or momentum states of an atom is created using laser light.
Following the schematics in Figure \ref{fig:AtomicStates}, a laser with the right energy to excite an atom from its ground state $|g\rangle$ to a higher energetic state $|e1\rangle$ has only a certain chance $P(g \to e1) <1$ to do so.
Since only a measurement of the atom can tell an observer whether the atom is excited or not, it remains in a superposition 
\begin{eqnarray}
|\Psi\rangle &=& \sqrt{1-P(g \to e1)}|g\rangle + {\sqrt{P(g \to e1)}}|e1\rangle
\end{eqnarray}
after the pulse of laser light has passed.
\\\\
If the atom managed to absorb a photon from the laser, i.e.\ if it got excited, it also absorbed the photon's momentum.
Therefore, the excited state corresponds to a larger momentum than the ground state. 
This means that, starting at an atom source, the two states would travel a defined path towards a detector in different times\footnote{For this, it is assumed that the momentum of a laser pulse photon is aligned with the direction of motion of the atom.}.
Of course this would not lead to the desired interference pattern since for this the two states have to arrive simultaneously. 
Consequently, both states must travel along different paths, i.e.\ on two different interferometer arms.
This idea is illustrated in Figure \ref{fig:AtomInterfero}: one state travels on path $x_1$, the other one on path $x_2$. 
Both then arrive at the same time in the detector and interfere with each other.
In order to control the path length and shape, states can be manipulated via absorption or induced emission \cite{Laser} due to another laser.
In this way, the energies and momenta of the superposed atomic states can be modified during the experiment.
During a realistic atom interferometry experiment, the laser is typically pulsed three times, once to split the wave function apart, once to reflect the two paths back toward each other, and finally to recombine the beam.
This is  analogous to the beam splitter and mirrors used in an optical interferometer. 
\\\\
Depending on the difference 
\begin{eqnarray}
\Delta x &:=& |x_1 - x_2|
\end{eqnarray}
in the lengths of the paths, the measured intensity $I$ in the detector can be between $0$ and $2I_\text{cl}$, where $I_\text{cl}$ is the intensity that would be measured if the atom was merely a classical particle.
A measured intensity $I<I_\text{cl}$ corresponds to destructive and $I>I_\text{cl}$ to constructive interference.
Varying the path length difference $\Delta x$ and plotting the resulting intensities leads to the interference pattern depicted by the blue and orange lines in Figure \ref{fig:Interference}, where the green line corresponds to $I_\text{cl}$.  
\noindent
Such an interference pattern can be used to observe decoherence.
For this, it needs to be understood that decoherence appears as a statistical effect.
This means, that even if an environment is present and interacting with the atoms coming from the source, not every atom superposition will decohere before it reaches the detector.
There is only a certain chance that the superposition decays within a finite time scale, which means that, generally, some atoms will still not be {classical} when being detected.
As a consequence, the appearance of decoherence during an experiment does not necessarily cause the interference pattern to vanish but instead causes the measured intensity amplitudes to decrease since not all observed atoms can interfere but are rather classical (see the orange line in Figure \ref{fig:Interference}).
Hence, measuring intensity peak-to-peak amplitudes that are smaller than $2I_\text{cl}$ is an indication of decoherence.
In the extreme case where all measured atoms have lost their superpositions before reaching the detector, the classical results depicted by the green line in Figure \ref{fig:Interference}) is recovered.

\newpage
\subsection{Derivation of the master equation}\label{ssec:MasterEqnDerive}

Here the quantum master equation describing the evolution of a toy model atom, modelled as a scalar field\footnote{Clearly, this proxy for an atom has some shortcomings. For example, while atoms are complex and generally stable configurations, a scalar field has no complex internal structure, and allows for decay and production processes that are not expected from real atoms. This will shortly be discussed in Section \ref{ssec:OperatorBased}.}, under the influence of a conformally coupled scalar field is finally derived\footnote{This section (including all its subsections) is heavily based on Ref. \cite{Kading3}. }.
The system is described by its density matrix whose time evolution is studied with the master equation.
In Section \ref{ssec:Setup} more details on the physical setup will be given, and subsequently a first expression of the master equation will be derived with the help of the closed time path formalism and the Feynman-Vernon influence functional.
This will be done for a particular model, the $n=-4$ chameleon (see Section \ref{ssec:Chameleons}) but could in principle be done for any other conformally coupled scalar field model as well.
The resulting master equation will be obtained in an abstract field basis whose physical interpretation is intricate.
Consequently, a formalism for projecting into a single particle subspace momentum basis will be developed under guidance of thermo field dynamics.
Afterwards, the projected master equation will be renormalised, and a short quantitative analysis with experimental implications be given. 

\subsubsection{Physical setup}\label{ssec:Setup}

What shall be investigated here is the effect of a scalar field $X$ as an environment on the open system scalar field $\phi$ which is acting as a toy model atom.
A master equation describing the evolution of the system will be derived.
For this, the starting point is the Einstein frame action
\begin{eqnarray}\label{eqn:EiFrAcDe}
S\ &=& \int d^4x  \sqrt{-g} \left[\frac{1}{2}\,M_{P}^2R - \frac{1}{2}g^{\mu\nu}\partial_{\mu}X\partial_{\nu}X - V(X)\right] \nonumber
\\
&& + \int d^4 x \sqrt{-g}A^4(X)\tilde{\mathcal{L}}_m(\tilde{\phi},A^{2}(X)g_{\mu\nu})
\end{eqnarray}
describing system, environment, and their interactions.
Here $R$ is the Ricci scalar, $V(X)$ is the potential of the environmental scalar $X$,
and 
\begin{eqnarray}\label{eqn:JoFrAcDe}
\tilde{\mathcal{L}}_m &=& -\frac{1}{2}\tilde{g}^{\mu\nu}\partial_{\mu}\tilde{\phi}\partial_{\nu}\tilde{\phi}-\frac{1}{2}\tilde{m}^2\tilde{\phi}^2
\end{eqnarray}
is the Jordan frame matter Lagrangian of the single scalar field $\tilde{\phi}$, which will later be expressed in terms of $\phi$.
\\\\
The conformal transformation between Jordan and Einstein frame is given by
\begin{eqnarray}
\tilde{g}_{\mu\nu} &=& A^2(X) g_{\mu\nu},
\end{eqnarray}
where the conformal factor is
\begin{eqnarray}
A^2(X) &=& a + b\frac{X}{\mathcal{M}} + c\frac{X^2}{\mathcal{M}^2} + \mathcal{O}\left( \frac{X^3}{\mathcal{M}^3} \right)
\end{eqnarray}
with $a$, $b$ and $c$ being model-dependent parameters{, and $\mathcal{M}$ being a coupling constant with the dimension of a mass}.
\\\\
The Einstein frame matter Lagrangian resulting from Eqns.\! (\ref{eqn:EiFrAcDe}) and (\ref{eqn:JoFrAcDe}) is then 
\begin{eqnarray}\label{eqn:EiFrLaDe}
\mathcal{L}_m &=& A^4(X)\tilde{\mathcal{L}}_m 
\nonumber
\\
&=& -\frac{1}{2} A^2(X)g^{\mu\nu}\partial_{\mu}\tilde{\phi}\partial_{\nu}\tilde{\phi} 
-\frac{1}{2}A^4(X)\tilde{m}^2 \tilde{\phi}^2.
\end{eqnarray}
Looking at Eqn.\ (\ref{eqn:EiFrLaDe}) it becomes apparent that the atom field $\tilde{\phi}$ is not canonically normalised. 
In order to cure this problem a redefinition
\begin{eqnarray}
\phi &:=& A(X)\tilde{\phi}.
\end{eqnarray}
is applied, leading to
\begin{eqnarray}\label{eqn:ReWrEiFrLaDe}
\mathcal{L}_m  &=& -\frac{1}{2}g^{\mu\nu}\partial_{\mu}\phi\partial_{\nu}\phi - \frac{1}{2}\phi^2g^{\mu\nu}\partial_{\mu}\ln A(X)\partial_{\nu}\ln A(X)
\nonumber
\\
&& + \phi g^{\mu\nu}\partial_\mu\phi \partial_\nu \ln A(X)  
- \frac{1}{2}A^2(X)\tilde{m}^2 \phi^2.
\end{eqnarray}
Clearly Eqn.\! (\ref{eqn:ReWrEiFrLaDe}) is also not canonically normalised but it as canonical as possible, meaning that it contains a canonically normalised kinetic term and some higher order non-canonical kinetic terms. 
It actually turns out that the non-canonical kinetic terms are suppressed compared to the canonical term since they contain factors of $\phi/\mathcal{M} \ll 1$.
Consequently, the non-canonical terms are dropped from here on.
\\\\
For the effective potential of the environmental scalar field $X$, as a sum of $V(X)$ and a coupling to matter, the following form is assumed:
\begin{eqnarray}\label{eqn:PoDecohere}
V^{\rm eff}(X) &=& \pm\frac{1}{2}\mu^2 X^2 + \frac{\lambda}{4!}X^4 + A(X)\rho^{\rm ext},
\end{eqnarray}
where $\rho^{\rm ext}$ is the energy density of an external source that could, for example, be a vacuum chamber in an atom interferometry experiment. 
The potential in Eqn.\! (\ref{eqn:PoDecohere}) could describe different types of scalar field models, e.g.\ $n=-4$ chameleons (see Section \ref{ssec:Chameleons}) or symmetrons (see Section \ref{ssec:Symmetrons}).
In any case it leads to a {non-trivial background configuration} $\langle X \rangle \neq 0$ of the scalar field $X${, which, dropping any corrections of order $\hbar$ or higher, fulfils the classical equation of motion
\begin{eqnarray}
\Box\langle X \rangle \mp\mu^2\langle X \rangle-\frac{\lambda}{3!}\langle X \rangle^3\ &=& \frac{dA(X)}{dX}\bigg|_{\langle X \rangle}\rho^{\rm ext}.
\end{eqnarray}
The field can be expanded around this background into} 
\begin{eqnarray}
X &=& \langle X \rangle + \chi,
\end{eqnarray}
where $\chi$ is a fluctuation.
\\\\
Under the assumption of a constant background, which implies $\langle X \rangle:\text{const.}$ and $\rho^{\rm ext}:\text{const.}$, a constant mass 
\begin{eqnarray}\label{eqn:RescaledMass}
m^2 &:=& \bigg(a+b\frac{\langle X \rangle}{\mathcal{M}} + c\frac{\langle X \rangle^2}{\mathcal{M}^2} \bigg)\tilde{m}^2
\end{eqnarray}
of the scalar field $\phi$ can be defined.
This results in the full Lagrangian of the two scalar system {being} given by
\begin{eqnarray}\label{eqn:FuEiFrLaDeco}
\mathcal{L} &=& -\frac{1}{2}g^{\mu\nu}\partial_{\mu}\phi\partial_{\nu}\phi 
- \frac{1}{2}m^2 \phi^2 
-\frac{1}{2}\alpha_1 m \chi\phi^2-\frac{1}{4}\alpha_2 \chi^2\phi^2
\nonumber
\\
&&
- \frac{1}{2}g^{\mu\nu}\partial_{\mu}\chi\partial_{\nu}\chi - \frac{1}{2}M^2\chi^2 
- \frac{\lambda}{4!}\big(\chi^4 + 4\langle X \rangle \chi^3\big)-V(\langle X \rangle),
\end{eqnarray}
where (up to terms of order $\langle X \rangle^2/\mathcal{M}^2$)
\begin{eqnarray}
\alpha_1\ &:=& \frac{m}{\mathcal{M}}\bigg[\frac{b}{a}\bigg(1-\frac{b}{a}\frac{\langle X \rangle}{\mathcal{M}}\bigg)+2\frac{c}{a}\frac{\langle X \rangle}{\mathcal{M}}\bigg],
\label{eqn:Alpha1}
\\
\alpha_2\ &:=& 2\frac{c}{a}\frac{m^2}{\mathcal{M}^2},
\label{eqn:Alpha2}
\end{eqnarray}
and the squared mass
\begin{eqnarray}
M^2 &:=& \frac{\lambda}{2}\langle X \rangle^2 \pm \mu^2+c\frac{\rho^{\rm ext}}{\mathcal{M}^2} 
\end{eqnarray}
for the $\chi$-field was defined.
\\\\
It will later prove useful to divide the action resulting from the Lagrangian in Eqn.\! (\ref{eqn:FuEiFrLaDeco}) into 
\begin{eqnarray}
S_\phi[\phi]  &{:=}& \int_x \bigg[-\frac{1}{2}g^{\mu\nu}\partial_{\mu}\phi\partial_{\nu}\phi - \frac{1}{2}m^2 \phi^2\bigg],
\label{eqn:FrAcPhi}
\\
S_\chi[\chi] &{:=}& \int_x  \bigg[-\frac{1}{2}g^{\mu\nu}\partial_{\mu}\chi\partial_{\nu}\chi -\frac{1}{2} M^2 \chi^2\bigg],
\label{eqn:FrAcChi}
\\
S_{\chi,{\rm int}}[\chi] &{:=}&  \int_{x\in\Omega_t}\bigg[-\frac{\lambda}{4!}\big(\chi^4 + 4\langle X \rangle\chi^3\big)\bigg],
\label{eqn:SeInAcChi}
\\
S_{\rm int}[\phi,\chi] &{:=}&   \int_{x\in\Omega_t}
\bigg[-\frac{1}{2}\alpha_1 m \chi\phi^2-\frac{1}{4}\alpha_2 \chi^2\phi^2\bigg],
\label{eqn:InAcChiPhi}
\end{eqnarray}
where the constant term arising from $V^{\rm eff}(\langle X \rangle)$ was omitted since it does not contribute to the dynamics, and the shorthand notation
\begin{eqnarray}
\int_x &{:=}& \int d^4x
\end{eqnarray}
introduced.
The actions in Eqns.\! (\ref{eqn:FrAcPhi}) and (\ref{eqn:FrAcChi}) are just free actions of massive scalar fields, while Eqn.\! (\ref{eqn:SeInAcChi}) describes self-interactions of $\chi$, and Eqn.\! (\ref{eqn:InAcChiPhi}) interactions between $\phi$ and $\chi$.
For the subsequent calculation it will be assumed that the interactions of $\phi$ and $\chi$ take place over a finite amount of time, i.e.\ between the preparation of the initial state and the measurement of the system.
Hence, the hypervolume 
\begin{eqnarray}
\Omega_t&:=&[0,t]\times\mathbb{R}^3
\end{eqnarray} 
was introduced.
As an alternative to restricting the time integration to a finite interval, the coupling constants $\lambda$, $\alpha_1$ and $\alpha_2$ could be given an explicit time-dependence, reflecting the switching on and off of the interactions by the experimental apparatus.
More generally, a switching function describing the realistic preparation of the system could be introduced.  

\subsubsection{Functional approach}\label{ssec:FuntionalAppr}

\begin{figure}[htbp]
\begin{center}
\includegraphics[scale=0.18]{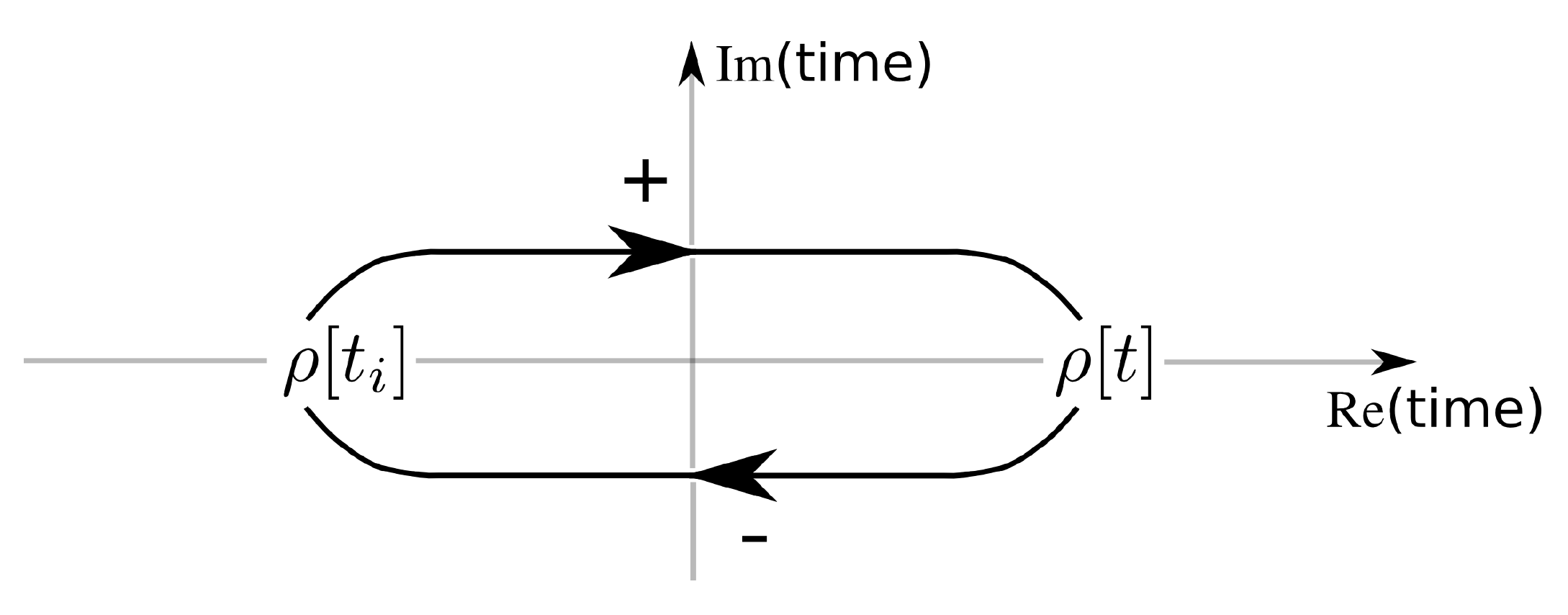}
\caption{\label{fig:ClosedTimePath}(Cf. Ref. \cite{Kading3}) Schematic depiction of a closed time path describing the evolution of a system from an initial time $t_i$ to a final time $t$ and then back to $t_i$.
The system evolves on time branches, labelled with $+$ or $-$, which are slightly shifted on the ordinate.
In this way, they give rise to a small imaginary contribution $\pm i\eta$ to the time variable which is taken in the limit $\eta \to 0$. Operators living on the $+$/$-$ branch are time/anti-time ordered. }
\end{center}
\end{figure}

Next, the master equation for the model set up in Section \ref{ssec:Setup} can be derived.
For this, the example of an $n=-4$ chameleon with $a=1$, $b=c=2$, and $\mu=0$ will be considered.
The coupling constants presented in Eqns.\! (\ref{eqn:Alpha1}) and (\ref{eqn:Alpha2}) therefore become
\begin{eqnarray}
\alpha_1 &=& 2\frac{m}{\mathcal{M}},
\\
\alpha_2 &=& \alpha_1^2.
\end{eqnarray}
Now the aim is to build an open system by tracing out the environmental degrees of freedom from the total density matrix.
More concretely, the combined closed system comprised of $\phi$ and $\chi$ is described by the density matrix $\hat{\rho}(t)$ but for this study, only the evolution of the partial system $\phi$ is of interest.
For this reason, the $\chi$ degrees of freedom have to be traced out, resulting in a reduced density matrix
\begin{eqnarray}
\hat{\rho}_{\phi}(t) &=&  \text{Tr}_{\chi}\hat{\rho}(t),
\end{eqnarray}
where $\text{Tr}_{\chi}$ denotes a {partial} trace with respect to $\chi$.
This trace is evaluated in the following way:
\begin{eqnarray}\label{eqn:ReDeMaTr}
\text{Tr}_{\chi}\hat{\rho}(t) &=& \int d\chi^+_t \langle \chi^+_t | \hat{\rho}(t) |\chi^+_t\rangle 
\nonumber
\\
&=&
\int d\chi^\pm_t \langle \chi^+_t | \hat{\rho}(t) |\chi^-_t\rangle \langle \chi^-_t |\chi^+_t\rangle 
\nonumber
\\
&=&
\int d\chi^\pm_t  \delta(\chi^+_t -\chi^-_t) \hat{\rho}[\chi^\pm;t] ,
\end{eqnarray}
where in the second to last line a complete set of eigenstates of the operator $\hat{x}$ at time $t$ was inserted, $\delta(\chi^+_t -\chi^-_t)$ is a functional delta function, the notations 
\begin{eqnarray}
\int d\chi^\pm_t &:=&  \int d\chi^+_t d\chi^-_t
\end{eqnarray}
and
\begin{eqnarray}\label{eqn:SiCoNots}
\chi_t &:=& \chi(t)
\end{eqnarray}
were introduced,
and the operator
\begin{eqnarray}
\hat{\rho}[\chi^\pm;t] &:=& \langle \chi^+_t | \hat{\rho}(t) |\chi^-_t\rangle
\end{eqnarray}
defined.
In order to simplify the notation, the spatial dependence of $|\chi^\pm_t\rangle$ is not explicitly given.
However, in later sections the notation from Eqn.\! (\ref{eqn:SiCoNots}) will also appear for spatial coordinates.
\\\\
Even though at first glance seemingly superfluous, two copies of the field eigenstates distinguished by labels $\pm$ have been introduced. 
This doubling of degrees of freedom, however, is needed in order to write down a path integral representation of the trace of an operator in terms of the Schwinger-Keldysh closed time path formalism \cite{Schwinger,Keldysh} (see also Figure \ref{fig:ClosedTimePath}).
For the subsequent procedures this will be essential because the objects of interest, the matrix elements of the density matrix, are supposed to be expressed in terms of a path integral formalism.
\\
Now knowing the reduced density matrix as given by the trace in Eqn.\! (\ref{eqn:ReDeMaTr}), the reduced density functional is of interest{:}
\begin{eqnarray}
\rho_{\phi}[\phi^{\pm}_t;t] &:=& \langle\phi^+_t|\hat{\rho}_{\phi}(t)|\phi^-_t\rangle 
\nonumber
\\
&=& \int d\chi_t^{\pm} \delta(\chi_t^+-\chi_t^-)\rho[\phi_t^{\pm},\chi_t^{\pm};t],
\end{eqnarray}
where two copies of the eigenstates of the operator $\hat{\phi}$ at time $t$ were inserted, and
\begin{eqnarray}
\rho[\phi^{\pm}_t,\chi^{\pm}_t;t] &:=& \langle\phi^+_t,\chi^+_t|\hat{\rho}(t)|\phi^-_t,\chi^-_t\rangle.
\end{eqnarray}
Next, it is assumed that the full system described by $\hat{\rho}$ at the initial time $t_i$ is a product state
\begin{eqnarray}
\hat{\rho}(t_i) &=& \hat{\rho}_\phi(t_i) \otimes \hat{\rho}_\chi(t_i), 
\end{eqnarray}
such that, at this time, both subsystems $\phi$ and $\chi$ are independent of each other.
Under this assumption, the reduced density functional at time $t$ is then given by
\begin{eqnarray}\label{eqn:DeFuAtTit}
\rho_\phi[\phi_t^{\pm};t] &=& \int d\phi^{\pm}_i \mathcal{I}[\phi^{\pm}_t,\phi^{\pm}_i;t,t_i] \rho_\phi[\phi^{\pm}_i;t_i],
\end{eqnarray}
where
\begin{eqnarray}\label{eqn:IFPropa}
\mathcal{I}[\phi^{\pm}_t,\phi^{\pm}_i;t,t_i] &=& \int_{\phi_i^{\pm}}^{\phi_t^{\pm}} \mathcal{D}\phi^{\pm} e^{\frac{i}{\hbar} \widehat{S}_{\rm eff}[\phi;t]}
\end{eqnarray}
is the so-called influence functional (IF) propagator (see e.g.\ Ref.\! \cite{CalzettaHuQFT} where this expression can be found).
For obtaining the IF propagator,  
\begin{eqnarray}
\rho_\phi[\phi_t^{\pm};t] &=& \int d\chi^\pm_t  \delta(\chi^+_t -\chi^-_t) \langle \phi^+_t , \chi^+_t| e^{-i(t-t_i)\hat{H}}\hat{\rho}(t_i)e^{i(t-t_i)\hat{H}}|\phi^-_t , \chi^-_t \rangle
\nonumber
\\
\end{eqnarray}
with $\hat{H}$ being the Hamiltonian corresponding to the Lagrangian in Eqn.\! (\ref{eqn:FuEiFrLaDeco}), is used, and insertions of complete sets of field and conjugate momentum eigenstates at all intermediate times between $t_i$ and $t$ on both sides, left and right of $\hat{\rho}(t_i)$, are done.
More precisely, plus/minus type states are inserted left/right of $\hat{\rho}(t_i)$, giving rise to the closed time path formalism \cite{Schwinger,Keldysh}, pictured in Figure \ref{fig:ClosedTimePath}. 
Following e.g.\ Ref.\! \cite{Kapusta}, this then results in the path integral expression in Eqn.\! (\ref{eqn:IFPropa}) with
\begin{eqnarray}
\mathcal{D}\phi^{\pm} &=& \prod\limits_i d\phi^\pm(x_i).
\end{eqnarray}  
The effective action appearing in Eqn.\! (\ref{eqn:IFPropa}) is given by 
\begin{eqnarray}\label{eqn:EffActionAftertrac}
 \widehat{S}_{\rm eff}[\phi;t] &=& \widehat{S}_{\phi}[\phi;t]+\widehat{S}_{\rm IF}[\phi;t],
\end{eqnarray}
where $\widehat{\ }$ indicates functionals that depend on both of the doubled field variables $\phi^+$ and $\phi^-$, e.g.\ the action associated with unitary evolution can be written as
\begin{eqnarray}\label{eqn:UnitEvoEqnAc}
\widehat{S}_{\phi}[\phi;t]&=& \sum_{a=\pm}a S_{\phi}[\phi^a;t] 
\nonumber
\\
&=& S_{\phi}[\phi^+;t]-S_{\phi}[\phi^-;t].
\end{eqnarray}
However, the so-called influence action $\widehat{S}_{\rm IF}$, which generally describes non-unitary evolution, also contains mixes of $+$ and $-$ fields, and is explicitly given by the Feynman-Vernon influence functional \cite{Feynman,CalzettaHuQFT}
\begin{eqnarray}
\widehat{\mathcal{F}}[\phi;t] &=& \exp\left\{\frac{i}{\hbar}\widehat{S}_{\text{IF}}[\phi;t]\right\} 
\nonumber
\\
&=& \int d\chi^{\pm}_t d\chi^{\pm}_i \delta(\chi_t^+-\chi_t^-)\rho_\chi [\chi^{\pm}_i;t_i] 
\times
\nonumber
\\
&&
\times
\int^{\chi^{\pm}_t}_{\chi^{\pm}_i}\mathcal{D}\chi^{\pm}\exp \left\{ \frac{i}{\hbar}\Big(\widehat{S}_{\chi}[\chi;t] +\widehat{S}_{\chi,{\rm int}}[\chi;t]+ \widehat{S}_{\rm int}[\phi,\chi;t]\Big) \right\}.
\nonumber
\\
\end{eqnarray}
The desired quantum master equation can actually be obtained by taking the partial time derivative of Eqn.\! (\ref{eqn:DeFuAtTit}).
In the special case in which only local interactions are induced within the open subsystem this yields
\begin{eqnarray}\label{eqn:EasyMasterEqn}
\partial_t\rho_\phi[\phi_t^{\pm};t] &=& \frac{i}{\hbar}\partial_t\widehat{S}_\text{eff}[\phi;t]\rho_\phi[\phi_t^{\pm};t].
\end{eqnarray}
The influence action does not, in general, include only local interactions, however, and the functional master equation for the model in Eqns. \!(\ref{eqn:FrAcPhi})-(\ref{eqn:InAcChiPhi}) cannot be written as in Eqn.\! (\ref{eqn:EasyMasterEqn}).
\\\\
In order to evaluate the master equation, the influence action is perturbatively computed up to second order in the small parameters 
\\$\lambda, m/\mathcal{M}, X/\mathcal{M} \ll 1$ (not necessary all of the same order of magnitude), which yields
\begin{eqnarray} \label{eqn:PertInflActi}
\widehat{S}_{\text{IF}}[\phi] &=& \sum_{a=\pm}a\Big[\langle S_{\rm int}[\phi^a,\chi^a]\rangle + \langle S_{\chi,{\rm int}}[\chi^a]\rangle\Big] 
\nonumber
\\
&&
+ \frac{i}{2\hbar}\sum_{a,b=\pm}ab\Big[\langle S_{\rm int}[\phi^a,\chi^a]S_{\rm int}[\phi^b,\chi^b]\rangle'
\nonumber
\\
&&
+ \langle S_{\chi,{\rm int}}[\chi^a]S_{\chi,{\rm int}}[\chi^b]\rangle'+2\langle S_{\chi,{\rm int}}[\chi^a]S_{\rm int}[\phi^b,\chi^b]\rangle'\Big],
\end{eqnarray} 
where
\begin{eqnarray} 
\langle A[\chi^{a}]\rangle &:=& \int d\chi^{\pm}_t d\chi^{\pm}_i \delta(\chi_t^+-\chi_t^-)\rho_\chi [\chi^{\pm}_i;t_i]\times
\nonumber
\\
&&
\times
\int^{\chi^{\pm}_t}_{\chi^{\pm}_i} \mathcal{D}\chi^{\pm} A[\chi^{a}]\exp \left\{ \frac{i}{\hbar}\,\widehat{S}_{\chi}[\chi;t] \right\},
\end{eqnarray} 
and
\begin{eqnarray} 
\langle A[\chi^{a}]B[\chi^{b}]\rangle' &=& \langle A[\chi^{a}]B[\chi^{b}]\rangle-\langle A[\chi^{a}]\rangle\langle B[\chi^{b}]\rangle.
\end{eqnarray} 
Here the time arguments of the terms contributing to the influence action in Eqn.\! (\ref{eqn:PertInflActi}) were dropped for convenience.
\\\\
The influence action can be further evaluated by substituting the Eqns.\! (\ref{eqn:SeInAcChi}) and (\ref{eqn:InAcChiPhi}) into Eqn.\ (\ref{eqn:PertInflActi}).
Though, it has to be considered that the chameleon scalar field $\chi$ is actually in thermal contact with the vacuum chamber walls.
It is therefore assumed that chameleon and chamber walls are in thermal equilibrium, such that the initial environmental state is a thermal {Gaussian} state given by 
\begin{eqnarray}
\rho_{\chi}[\chi_i^{\pm};t_i] &=& \frac{1}{\text{Tr}e^{-\beta \hat{H}_{\chi}}}\langle \chi^+_i|e^{-\beta \hat{H}_{\chi}}|\chi^-_i\rangle,
\end{eqnarray}
where $\beta=1/T$ is the inverse thermodynamic temperature and $\hat{H}_{\chi}$ is the free Hamiltonian of the $\chi$ fluctuations. 
The Boltzmann constant is taken to be unity in the used set of units.
Using Wick's theorem \cite{Wick}, the correlation functions resulting from substituting Eqns.\! (\ref{eqn:SeInAcChi}) and (\ref{eqn:InAcChiPhi}) into Eqn.\! (\ref{eqn:PertInflActi}) lead to combinations of scalar $\chi$-propagators of the forms (see e.g.\ Ref.\! \cite{LeBellac})
\begin{eqnarray}
\contraction{}{\chi}{^+_x}{\chi}\chi^+_x\chi^+_y
&=& \langle T\chi_{x} \chi_{y}\rangle  ~=~  \Delta^{++}_{xy}  ~=~ \Delta^{\rm F}_{xy}
\nonumber
\\
&=& - i\hbar \int_k e^{ik\cdot(x-y)}\bigg[\frac{1}{k^2 + M^2 -i\epsilon}+2\pi i f(|k^0|)\delta(k^2+M^2)\bigg],
\nonumber
\\
\\
\contraction{}{\chi}{^-_x}{\chi}\chi^-_x\chi^-_y
 &=& \langle\tilde{T} \chi_{x}\chi_{y}\rangle ~=~  \Delta^{--}_{xy} ~=~ \Delta^{\rm D}_{xy}
\nonumber
\\
&=& + i\hbar \int_k  e^{ik\cdot(x-y)}\bigg[\frac{1}{k^2 + M^2 +i\epsilon}-2\pi i f(|k^0|)\delta(k^2+M^2)\bigg],
\nonumber
\\
\\
\contraction{}{\chi}{^+_x}{\chi}\chi^+_x\chi^-_y
&=& \langle\chi_{y} \chi_{x}\rangle ~=~  \Delta^{+-}_{xy}~=~ \Delta^<_{xy}
\nonumber
\\
&=& \hbar\int_k e^{ik\cdot (x-y)} 2\pi \sgn(k^0)f(k^0)\delta(k^2 + M^2),
\\[0.6em]
\contraction{}{\chi}{^-_x}{\chi}\chi^-_x\chi^+_y
&=&  \langle\chi_{x} \chi_{y}\rangle  ~=~  \Delta^{-+}_{xy}~=~ \Delta^>_{xy}~=~ \Delta^<_{yx}~=~ (\Delta^<)^*_{xy},
\end{eqnarray}
which are the Feynman, Dyson, negative and positive Wightman propagators, respectively. 
The operators $T$ and $\tilde{T}$ denote time- and anti-time-ordering.
Furthermore, the notation 
\begin{eqnarray}
\int_k &:=& \int \frac{d^4k}{(2\pi)^4},
\end{eqnarray}
the signum function 
\begin{eqnarray}
\sgn(k^0) &=& \Theta(k^0) - \Theta(-k^0),
\end{eqnarray}
and the Bose-Einstein distribution \cite{Rammer}
\begin{eqnarray}
f(k^0) &=& \frac{1}{e^{\beta k^0}-1}
\end{eqnarray}
with
\begin{eqnarray}\label{eqn:BoseEisIdenty}
f(-k^0) &=& -\big[1+f(k^0)\big]
\end{eqnarray}
were introduced.
Some relevant properties for the vacuum scalar propagators without thermal corrections can be found in Appendix \ref{app:PropoScaPro}.
\\\\
With these four types of propagators the terms contributing to the influence action in Eqn.\! (\ref{eqn:PertInflActi}) are
\begin{eqnarray}
\langle S_{\rm int}[\phi^a,\chi^a]\rangle &=& - \frac{m^2}{\mathcal{M}^2}\int_x \big(\phi^a_{x}\big)^2 \Delta^{\rm F}_{xx},
\\
\langle S_{\rm int}[\phi^a,\chi^a]S_{\rm int}[\phi^b,\chi^b]\rangle' &=& \langle S_{\rm int}[\phi^a,\chi^a]S_{\rm int}[\phi^b,\chi^b]\rangle 
\nonumber
\\
&=& \frac{m^4}{\mathcal{M}^2}\int_{xy}  \big(\phi^a_{x}\big)^2\big(\phi^b_{y}\big)^2 \Delta^{ab}_{xy},
\\
\langle S_{\chi,{\rm int}}[\chi^a]\rangle &=& -\frac{\lambda}{4!}\int_{x} \Big[3\big(\Delta^{\rm F}_{xx}\big)^2 +\langle X\rangle^4\Big], 
\\
\langle S_{\chi,{\rm int}}[\chi^a]S_{\chi,{\rm int}}[\chi^b]\rangle &=& \frac{\lambda^2}{(4!)^2}\int_{xy} \Big[24\big(\Delta^{ab}_{xy}\big)^4 + 72\big(\Delta^{\rm F}_{xx}\big)^2\big(\Delta^{ab}_{xy}\big)^2     
\nonumber
\\
&&
+ 96\langle X\rangle^2\big(\Delta^{ab}_{xy}\big)^3 + 144\langle X\rangle^2\big(\Delta^{\rm F}_{xx}\big)^2\Delta^{ab}_{xy}
\nonumber
\\
&&
 + 9 \big(\Delta^{\rm F}_{xx}\big)^4 
+ 6\langle X\rangle^4  \big(\Delta^{\rm F}_{xx}\big)^2+ \langle X\rangle^8\Big], 
\\
\langle S_{\chi,{\rm int}}[\chi^a]S_{\chi,{\rm int}}[\chi^b]\rangle' &=& \frac{\lambda^2}{(4!)^2}\int_{xy} \Big[24\big(\Delta^{ab}_{xy}\big)^4 + 72\big(\Delta^{\rm F}_{xx}\big)^2\big(\Delta^{ab}_{xy}\big)^2   
\nonumber
\\
&&
+ 96\langle X\rangle^2\big(\Delta^{ab}_{xy}\big)^3 + 144\langle X\rangle^2\big(\Delta^{\rm F}_{xx}\big)^2\Delta^{ab}_{xy}
 \Big],
 \nonumber
 \\
 \\
\langle S_{\chi,{\rm int}}[\chi^a] S_{\rm int}[\phi^b,\chi^b]\rangle' &=& \langle S_{\chi,{\rm int}}[\chi^a] S_{\rm int}[\phi^b,\chi^b]\rangle 
\nonumber
\\
&=& \frac{\lambda \langle X\rangle m^2}{2\mathcal{M}}\int_{xy} 
 \Delta^{\rm F}_{xx}\Delta^{ab}_{xy}\big(\phi^b_y\big)^2,
\end{eqnarray}
where the factors arising from $V^{\rm eff}(\langle X\rangle)$, originally omitted, are restored in order to illustrate that these do not contribute to $\widehat{S}_{\rm IF}$. Hereafter, it will be left implicit that all time integrals run over the domain $[0,t]$.
\\\\
Altogether, the influence action is given by 
\begin{eqnarray}\label{eqn:FuIFAcPert}
\widehat{S}_{\text{IF}}[\phi;t]  &=&  -\frac{m^2}{\mathcal{M}^2}\sum_{a=\pm}\int_{x} a\big(\phi^a_x\big)^2\Delta^{\rm F}_{xx} +\frac{i}{2\hbar}\int_{xy} \sum\limits_{a,b=\pm}ab\Bigg\{ \frac{m^4}{\mathcal{M}^2}\big(\phi^a_x\big)^2\big(\phi^b_y\big)^2\Delta^{ab}_{xy}   
\nonumber
\\
&&
+\frac{\lambda^2}{24}\Big[\big(\Delta^{ab}_{xy}\big)^4 + 3 \big(\Delta^{\rm F}_{xx}\big)^2\big(\Delta^{ab}_{xy}\big)^2 + 4\langle X\rangle^2 \big(\Delta^{ab}_{xy}\big)^3
\nonumber
\\
&&
\phantom{+\frac{\lambda^2}{24}}
+ 6\langle X\rangle^2\big(\Delta^{\rm F}_{xx}\big)^2\Delta^{ab}_{xy}\Big] +\frac{\lambda \langle X\rangle m^2}{\mathcal{M}}\Delta^{\rm F}_{xx} \Delta^{ab}_{xy}\big(\phi^b_y\big)^2
\Bigg\}.
\end{eqnarray}
From now on $\hbar$ will be set to unity.
\\\\
It can {be} seen that the sum over the terms in squared brackets beginning in the second line of Eqn.\ (\ref{eqn:FuIFAcPert}) has to vanish due to the largest time equation \cite{tHooft, Kobes}
\begin{eqnarray}
\forall \ n\ \in\ \mathbb{N}_0:\qquad  \sum_{a,b=\pm}a b\big(\Delta^{ab}_{xy}\big)^n &=& 0.
\end{eqnarray}
In Appendix \ref{app:Relations} it is shown how this expression can be derived for the vacuum scalar propagators without thermal corrections.
\\\\
Consequently, Eqn.\! (\ref{eqn:FuIFAcPert}) reduces to
\begin{eqnarray}\label{eqn:ReIFAcPertub}
\widehat{S}_{\text{IF}}[\phi;t]  &=& -\frac{m^2}{\mathcal{M}^2}\sum_{a=\pm}\int_{x} a\big(\phi^a_x\big)^2\Delta^{\rm F}_{xx} +\frac{i}{2}\int_{xy} \sum\limits_{a,b=\pm}ab \Bigg\{ \frac{m^4}{\mathcal{M}^2}\big(\phi^a_x\big)^2\big(\phi^b_y\big)^2\Delta^{ab}_{xy}   
\nonumber
\\
&&
+\frac{\lambda \langle X\rangle m^2}{\mathcal{M}}\Delta^{\rm F}_{xx}\Delta^{ab}_{xy}\big(\phi^b_y\big)^2
\Bigg\}.
\end{eqnarray}

\subsubsection{Operator-based approach}\label{ssec:OperatorBased}

After finding Eqn.\! (\ref{eqn:ReIFAcPertub}), a master equation obtained by taking the partial time derivative of Eqn.\! (\ref{eqn:DeFuAtTit}) could in principle be quantitatively evaluated. 
However, its physical interpretation is challenging.
So far, the master equation was only formulated in a basis of the doubled scalar degrees of freedom, see, e.g., Eqn. (\ref{eqn:EasyMasterEqn}).
This abstract field basis makes it intricate to extract physically meaningful information.
Therefore, a formalism shall now be developed that can be used to project such a master equation into a more suitable basis with a clear physical interpretation.
\\\\
Since the resulting equation is supposed to approximate the evolution of relatively heavy cold atoms (e.g.\ Rubidium) as used in atom interferometry experiments, and the production of atom-anti-atom pairs is likely to be strongly suppressed under such conditions, a restriction to a single particle subspace is reasonable.
In addition, momenta are chosen to be the physical variables of interest since momentum superpositions of atoms can be created in atom interferometers.   
Hence, in order to proceed, a projection into a single particle {momentum} subspace shall be derived.
\\\\
Guidance for the derivation of the desired projection formalism is provided by thermo field dynamics (TFD) \cite{Takahasi:1974zn, Arimitsu:1985ez, Arimitsu:1985xm} (see also Ref.\! \cite{Khanna}), an operator-based formulation of the path integral approach with doubled degrees of freedom, which were required for this approach, presented in Section \ref{ssec:FuntionalAppr}. 
The doubling of degrees of freedom is represented by a doubled Hilbert space defined as
\begin{eqnarray}
\widehat{\mathcal{H}} &:=& \mathcal{H}^+\otimes\mathcal{H}^-,
\end{eqnarray}
where $\mathcal{H}^{\pm}$ are Hilbert spaces corresponding to the $\pm$ branch of the closed time path, respectively.
A scalar field operator $\hat{\phi}$ is then embedded by defining the plus- and minus-type field operators
\begin{eqnarray}
\hat{\phi}^+(x) &=&\hat{\phi}(x)\otimes\mathbb{I},
\\
\hat{\phi}^-(x) &=& \mathbb{I}\otimes\hat{\phi}^{\mathcal{T}}(x),
\end{eqnarray}
with analogous expressions for the embeddings of the usual scalar creation and annihilation operators. Here, $\mathcal{T}$ indicates time reversal (see e.g.\ Ref.\! \cite{Srednicki}).
In the interaction picture, the two types of field operators can be written in their usual Fourier decomposition as 
\begin{eqnarray}\label{eqn:PMScaFielFourDeco}
\hat{\phi}^\pm(x)&=& \int d\Pi_{\vec{k}}\Big[\hat{a}^{\pm}_{\vec{k}}e^{\mp i E_{\vec{k}}t\pm i\vec{k}\cdot\vec{x}}+\hat{a}^{\pm\dagger}_{\vec{k}}e^{\pm i E_{\vec{k}}t\mp i\vec{k}\cdot \vec{x}}\Big],
\end{eqnarray}
where the notation introduced in Eqns.\! (\ref{eqn:ScFiNota1}) and (\ref{eqn:ScFiNota2}) was used.
When some operators and states here have no time arguments, then this means that they are evaluated at $t=0$.
\\\\
Plus- and minus-type creation and annihilation operators introduced in Eqn.\! (\ref{eqn:PMScaFielFourDeco}) act in their corresponding Hilbert spaces. 
For this, a doubled vacuum state
\begin{eqnarray}
|0\rangle\rangle &:=& |0\rangle \otimes |0\rangle
\end{eqnarray}  
is defined, such that
\begin{eqnarray}
\hat{a}^{+\dag}_{\vec{k}}|0\rangle\rangle &=& |\vec{k}\rangle \otimes |0\rangle ~:=~ |\vec{k}_+\rangle\rangle,
\\
\hat{a}^{-\dag}_{\vec{k}}|0\rangle\rangle &=& |0\rangle \otimes |\vec{k}\rangle ~:=~ |\vec{k}_-\rangle\rangle,
\end{eqnarray}
and
\begin{eqnarray}
\hat{a}^{+}_{\vec{k}}|\vec{p}_+,\vec{p}_-\rangle\rangle &=& (2\pi)^32E_{\vec{k}}\delta^{(3)}(\vec{p}-\vec{k})|\vec{p}_-\rangle\rangle,
\\
\hat{a}^{-}_{\vec{k}}|\vec{p}_+,\vec{p}_-\rangle\rangle &=& (2\pi)^32E_{\vec{k}}\delta^{(3)}(\vec{p}-\vec{k})|\vec{p}_+\rangle\rangle,
\end{eqnarray}
where $|\vec{p}_+,\vec{p}_-\rangle\rangle := |\vec{p}\rangle \otimes |\vec{p}\rangle$ and so on.
\\\\
The density operator of an isolated system can be embedded as
\begin{eqnarray}
\hat{\rho}^+(t) &:=& \hat{\rho}(t) \otimes \hat{\mathbb{I}},
\end{eqnarray}
where $\hat{\mathbb{I}}$ is the unit operator.
With the help of the state \cite{Arimitsu:1985ez}
\begin{eqnarray}
|1\rangle\rangle &:=& |0\rangle\rangle + \int d\Pi_{\vec{p}_1} |\vec{p}_{1+},\vec{p}_{1-}\rangle\rangle 
\nonumber
\\
&&
+\frac{1}{2!}\int d\Pi_{\vec{p}_1} d\Pi_{\vec{p}_2} |\vec{p}_{1+},\vec{p}_{2+},\vec{p}_{1-},\vec{p}_{2-}\rangle\rangle + \dots
\end{eqnarray}
the expectation value of an operator $\hat{O}(t)$ can be written as 
\begin{eqnarray}
\langle \hat{O}(t) \rangle &=& \text{Tr}\hat{O}(t)\hat{\rho}(t) ~=~  \langle\langle 1|\hat{O}^+(t)\hat{\rho}^+(t)|1\rangle\rangle.
\end{eqnarray}
In addition, this state can, for example, be used to rewrite the Schr\"odinger picture Liouville equation (\ref{eqn:LiovilleVonNeumann}) in a Schr\"odinger-like form\footnote{In order to see that this is true, it has to be shown that $[\hat{H},\hat{\rho}(t)]\otimes\hat{\mathbb{I}}|1\rangle\rangle =\widehat{H}\hat{\rho}^+(t)|1\rangle\rangle$.
After expanding $\widehat{H}$, the first term of the commutator on the left-hand side equals the first term on the right-hand side. All what remains to be shown is that $\hat{\rho}(t)\otimes \hat{H}|1\rangle\rangle = \hat{\rho}(t)\hat{H} \otimes\hat{\mathbb{I}}|1\rangle\rangle$. For simplicity, in what follows, only single particle states will be considered. It can be seen 
\begin{eqnarray*}
\hat{\rho}(t)\otimes \hat{H}|1\rangle\rangle &=& (\hat{\rho}(t)\otimes \hat{H}) \int d\Pi_{\vec{p}} |\vec{p}\,\rangle\otimes |\vec{p}\,\rangle
\\
&=& \int d\Pi_{\vec{k}} d\Pi_{\vec{k}'}d\Pi_{\vec{p}} \rho(\vec{k},\vec{k}';t)|\vec{k}\rangle\langle\vec{k}'|\vec{p}\,\rangle \otimes \hat{H}|\vec{p}\,\rangle
\\
&=& \int d\Pi_{\vec{k}} d\Pi_{\vec{k}'} E_{\vec{k}'} \rho(\vec{k},\vec{k}';t) |\vec{k}\rangle \otimes |\vec{k}'\rangle\,,
\end{eqnarray*}
and similarly
\begin{eqnarray*}
\hat{\rho}(t)\hat{H} \otimes\hat{\mathbb{I}}|1\rangle\rangle &=& 
\int d\Pi_{\vec{k}} d\Pi_{\vec{k}'}d\Pi_{\vec{p}} \rho(\vec{k},\vec{k}';t)|\vec{k}\rangle\langle\vec{k}'|\hat{H}|\vec{p}\,\rangle \otimes |\vec{p}\,\rangle
\\
&=& \int d\Pi_{\vec{k}} d\Pi_{\vec{k}'} E_{\vec{k}'} \rho(\vec{k},\vec{k}';t) |\vec{k}\rangle \otimes |\vec{k}'\rangle\,,
\end{eqnarray*}
showing that both expressions are indeed equal.
}
\begin{eqnarray}
\partial_t\hat{\rho}^+(t)|1\rangle\rangle &=& -i\widehat{H}\hat{\rho}^+(t)|1\rangle\rangle,
\end{eqnarray}
where
\begin{eqnarray}
\widehat{H} &:=& \hat{H} \otimes \hat{\mathbb{I}} - \hat{\mathbb{I}} \otimes \hat{H}
\end{eqnarray}
is the Liouvillian operator.
\\\\
Next, in the interaction picture, a density operator can be given in the form 
\begin{eqnarray}
\hat{\rho}(t) &=& \int d\Pi_{\vec{k}} d\Pi_{\vec{k}'} \rho(\vec{k},\vec{k}';t)|\vec{k};t\rangle\langle\vec{k}';t|,
\end{eqnarray}
where, as described earlier, only single particle states were considered, and the matrix element
\begin{eqnarray}\label{eqn:MatriEleDeMAtr}
\langle\vec{p};t|\hat{\rho}(t)|\vec{p}\, ';t\rangle &=& \rho(\vec{p},\vec{p}\, ' ;t)
\end{eqnarray}
is therefore the object of interest for the subsequent studies. 
It should be stressed that all of the basis states and operators are evaluated at equal times, which therefore makes the matrix element in Eqn.\! (\ref{eqn:MatriEleDeMAtr}) picture-independent (see Refs.\! \cite{Millington:2012pf, Millington:2013isa}).
\\\\
Using the TFD language, Eqn.\! (\ref{eqn:MatriEleDeMAtr}) can be expressed as
\begin{eqnarray}
\text{Tr}|\vec{p}\,';t\rangle\langle\vec{p};t|\hat{\rho}(t) &=& \langle\langle 1(t)|(|\vec{p}\,';t\rangle\langle\vec{p};t|\otimes \hat{\mathbb{I}})(\hat{\rho}(t) \otimes \hat{\mathbb{I}})|1(t)\rangle\rangle,
\end{eqnarray}
where there is now a time-dependent state $|1(t)\rangle\rangle$.
\\\\
Now, using
\begin{eqnarray}
\langle\langle 1(t)|(|\vec{p}\,';t\rangle \langle \vec{p};t| \otimes \hat{\mathbb{I}}) &=& \langle\langle \vec{p}_+,\vec{p}\,'_{\! -} ;t|,
\end{eqnarray}
and
\begin{eqnarray}
(\hat{\rho}(t) \otimes \hat{\mathbb{I}}) |1(t)\rangle\rangle &=& \int d\Pi_{\vec {k}} d\Pi_{\vec{k}'} \rho(\vec{k},\vec{k}';t)|\vec{k}_+,\vec{k}_-';t\rangle\rangle,
\end{eqnarray}
it follows that\footnote{For showing this, it should be considered that, in the interaction picture, the base vectors evolve with the free part of the Schr\"odinger picture Hamiltonian $\hat{H}_{0,\text{S}}$, while the state vectors, and therefore the density operator, evolve with the interaction part $\hat{H}_{\text{I},\text{S}}$ (see e.g.\ Refs.\! \cite{Millington:2012pf,Millington:2013isa}). It can then be seen that 
\begin{eqnarray*}
\rho(\vec{p},\vec{p}\,';t) &=& 
\langle\langle 1|(|\vec{p}\,'\rangle\langle\vec{p}|e^{-i\hat{H}_\text{S}t}\hat{\rho} e^{i\hat{H}_\text{S}t}\otimes \hat{\mathbb{I}})|1\rangle\rangle,
\end{eqnarray*}
and 
\begin{eqnarray*}
\partial_t \rho(\vec{p},\vec{p}\,';t) &=& 
-i \langle\langle 1(t)|(|\vec{p}\,';t\rangle\langle\vec{p}|\hat{H}_\text{S}e^{-i\hat{H}_\text{S}t}\hat{\rho} e^{i\hat{H}_\text{S}t}\otimes \hat{\mathbb{I}})|1\rangle\rangle
 \\
 &&
 +i \langle\langle 1(t)|(|\vec{p}\,';t\rangle\langle\vec{p}|e^{-i\hat{H}_\text{S}t}\hat{\rho}\hat{H}_\text{S} e^{i\hat{H}_\text{S}t}\otimes \hat{\mathbb{I}})|1\rangle\rangle
 \\
&=& 
-i \langle\langle 1(t)|(|\vec{p}\,';t\rangle\langle\vec{p};t|\hat{H}\hat{\rho}(t) \otimes \hat{\mathbb{I}})|1(t)\rangle\rangle
 \\
 &&
 +i \langle\langle 1(t)|(|\vec{p}\,';t\rangle\langle\vec{p};t|\hat{\rho}(t) \otimes \hat{H})|1(t)\rangle\rangle
 \\
&=& - i\int d\Pi_{\vec{k}} d\Pi_{\vec{k}'} \rho(\vec{k},\vec{k}';t) \langle\langle \vec{p}_+,\vec{p}\,'_{\! -};t|\widehat{H}(t)|\vec{k}_+,\vec{k}_-';t\rangle\rangle,
\end{eqnarray*}
where $\hat{H}$ here denotes the Dirac picture Hamiltonian.
}
\begin{eqnarray}
\partial_t\rho(\vec{p},\vec{p}\,';t) &=& - i\int d\Pi_{\vec{k}} d\Pi_{\vec{k}'} \rho(\vec{k},\vec{k}';t) \langle\langle \vec{p}_+,\vec{p}\,'_{\! -};t|\widehat{H}(t)|\vec{k}_+,\vec{k}_-';t\rangle\rangle.
\nonumber
\\
\end{eqnarray}
This can be rewritten as
\begin{eqnarray}\label{eqn:MaEqinOpFoReWr}
\partial_t \rho(\vec{p},\vec{p}\,';t) &=& -i\int d\Pi_{\vec{k}} d\Pi_{\vec{k}'} \rho(\vec{k},\vec{k}';t) \langle\langle 0|\hat{a}^+_{\vec{p}}(t)\hat{a}^-_{\vec{p}\,'}(t)\widehat{H}(t)\hat{a}_{\vec{k}}^{+\dagger}(t)\hat{a}_{\vec{k}'}^{-\dagger}(t)|0\rangle\rangle.
\nonumber
\\
\end{eqnarray}
Here, $\widehat{H}\to \widehat{H}_{\rm eff}=  -\partial_t\widehat{S}_{\rm eff}$, is the effective Liouvillian that comes from tracing out the chameleon degrees of freedom, and $\widehat{S}_{\rm eff}$ results from the effective action in Eqn.\! (\ref{eqn:EffActionAftertrac}).
\\\\
Allowing for the fact that $\widehat{H}_{\rm eff}$ is a non-local\footnote{This means that it depends on past times, e.g.\ via an explicit time integration, as is not unexpected when tracing out the environmental fields.}, but time-ordered operator, and after accounting for the free-phase evolution of the right-most creation operators, the expectation value on the right-hand side of Eqn.\! (\ref{eqn:MaEqinOpFoReWr}) can be time-ordered as
\begin{eqnarray}\label{eqn:TiOrMaEqExpr}
\partial_t\rho(\vec{p},\vec{p}\,';t) &=& -i\int d\Pi_{\vec{k}} d\Pi_{\vec{k}'} e^{i(E_{\vec{k}}-E_{\vec{k}'})t}\rho(\vec{k},\vec{k}';t) 
\times
\nonumber
\\
&&
\times
\langle\langle 0|T[\hat{a}^+_{\vec{p}}(t)\hat{a}^-_{\vec{p}\,'}(t)\widehat{H}_{\rm eff}(t)\hat{a}_{\vec{k}}^{+\dagger}(0)\hat{a}_{\vec{k}'}^{-\dagger}(0)]|0\rangle\rangle.
\end{eqnarray}
Replacing the creation and annihilation operators in Eqn.\! (\ref{eqn:TiOrMaEqExpr}) with field operators via
\begin{eqnarray}
\hat{a}^+_{\vec{p}}(t) &=& +i\int_{\vec{x}} e^{-i\vec{p}\cdot\vec{x}}\partial_{t,E_{\vec{p}}}\hat{\phi}^+(t,\vec{x}) ,
\\ 
\hat{a}^{+\dagger}_{\vec{p}}(t) &=& -i\int_{\vec{x}} e^{+i\vec{p}\cdot\vec{x}} \partial_{t,E_{\vec{p}}}^*\hat{\phi}^+(t,\vec{x}),
\\
\hat{a}^-_{\vec{p}}(t) &=& -i\int_{\vec{x}} e^{+i\vec{p}\cdot\vec{x}}\partial_{t,E_{\vec{p}}}^*\hat{\phi}^-(x),
\\
\hat{a}^{-\dag}_{\vec{p}}(t) &=& +i\int_{\vec{x}} e^{-i\vec{p}\cdot\vec{x}}\partial_{t,E_{\vec{p}}}\hat{\phi}^{-}(x),
\end{eqnarray}
where
\begin{eqnarray}
\partial_{t,E_{\vec{p}}} &:=& \overset{\rightarrow}{\partial}_t - iE_{\vec{p}}~,
\end{eqnarray}
leads to
\begin{eqnarray}\label{eqn:LSZredEqOO}
\partial_t\rho(\vec{p},\vec{p}\,';t) &=& 
-i\lim_{\substack{x^{0(\prime)}\,\to\, t^{+}\\y^{0(\prime)}\,\to\, 0^-}}
\int d\Pi_{\vec{k}} d\Pi_{\vec{k}'} e^{i(E_{\vec{k}}-E_{\vec{k}'})t}\rho(\vec{k},\vec{k}';t) 
\nonumber
\\
&&
\times
\int_{\vec{x}\vec{x}'\vec{y}\vec{y}\,'}e^{-i(\vec{p}\cdot\vec{x}-\vec{p}\,'\cdot\vec{x}')+i(\vec{k}\cdot\vec{y}-\vec{k}'\cdot\vec{y}\,')}
\partial_{x^0,E_{\vec{p}}}\partial_{x^{0\prime},E_{\vec{p}\,'}}^*\partial_{y^0,E_{\vec{k}}}^*\partial_{y^{0\prime},E_{\vec{k}'}}
\nonumber
\\
&&
\times  \langle\langle 0|{\rm T}[\hat{\phi}^+(x)\hat{\phi}^-(x')\widehat{H}_{\rm eff}(t)\hat{\phi}^{+}(y)\hat{\phi}^{-}(y')]|0\rangle\rangle.
\end{eqnarray}
Here $x^{0(\prime)}$ approaches $t$ from above and $y^{0(\prime)}$ approaches $0$ from below to ensure that the time-ordering of the operators is equivalent to the original ordering in Eqn.\! (\ref{eqn:TiOrMaEqExpr}).
\\\\
The role of the differential operators in Eqn.\ (\ref{eqn:LSZredEqOO}) is to remove external propagators and replace them with plane-wave factors. 
Therefore, the procedure outlined here for projecting into the single-particle subspace amounts to an LSZ-like reduction\cite{Lehmann:1954rq} (see also, e.g.\ Ref.\! \cite{PeskinSchroeder}) of the four-point function
\begin{eqnarray}
\langle\langle 0| T[\hat{\phi}^+(x)\hat{\phi}^-(x')\widehat{H}_{\rm eff}(t)\hat{\phi}^{+}(y)\hat{\phi}^{-}(y')]|0\rangle\rangle.
\end{eqnarray}
The master equation in the form of Eqn.\! (\ref{eqn:LSZredEqOO}) can be translated into the path integral formalism similarly to Eqn.\! (\ref{eqn:HigherOrderCorre}) and becomes
\begin{eqnarray}\label{eqn:ProjMasterEqnYo}
\partial_t\rho(\vec{p},\vec{p}\,';t) &=& 
i\lim_{\substack{x^{0(\prime)}\,\to\, t^{+}\\y^{0(\prime)}\,\to\, 0^-}}
\int d\Pi_{\vec{k}} d\Pi_{\vec{k}'} e^{i(E_{\vec{k}}-E_{\vec{k}'})t} \rho(\vec{k},\vec{k}';t) 
\nonumber
\\
&&
\times 
\int_{\vec{x}\vec{x}'\vec{y}\vec{y}\,'} e^{-i(\vec{p}\cdot\vec{x}-\vec{p}\,' \cdot\vec{x}')+i(\vec{k}\cdot\vec{y}-\vec{k}'\cdot\vec{y}\,')}
\partial_{x^0,E_{\vec{p}}} \partial_{x^{0\prime},E_{\vec{p}\,'}}^*\partial_{y^0,E_{\vec{k}}}^*\partial_{y^{0\prime},E_{\vec{k}'}}
\nonumber
\\
&&
\times 
\int\mathcal{D}\phi^{\pm} e^{i\widehat{S}_{\phi}[\phi]}\phi^+(x)\phi^-(x')\Big(\partial_t\widehat{S}_{\rm eff}[\phi;t]\Big)\phi^{+}(y)\phi^{-}(y'),
\nonumber
\\
\end{eqnarray}
where $\widehat{H}_{\rm eff}(\phi^{\pm})$ got replaced by $-\Big(\partial_t\widehat{S}_{\rm eff}[\phi;t]\Big)$.

\subsubsection{Quantum master equation}

\begin{figure}

\centering

\subfloat[][]{\includegraphics[scale=0.2]{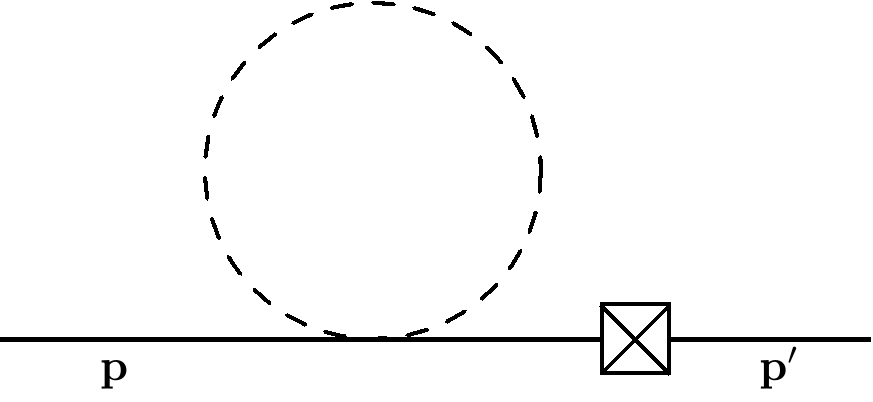}}
\qquad
\subfloat[][]{\includegraphics[scale=0.2]{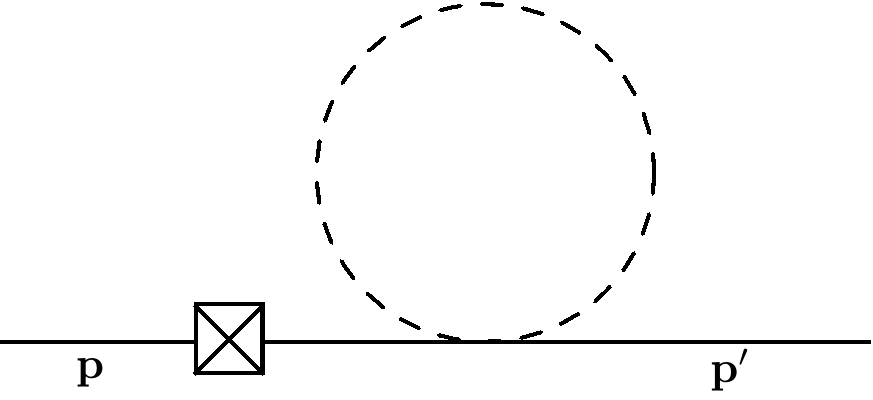}}

\subfloat[][]{\includegraphics[scale=0.2]{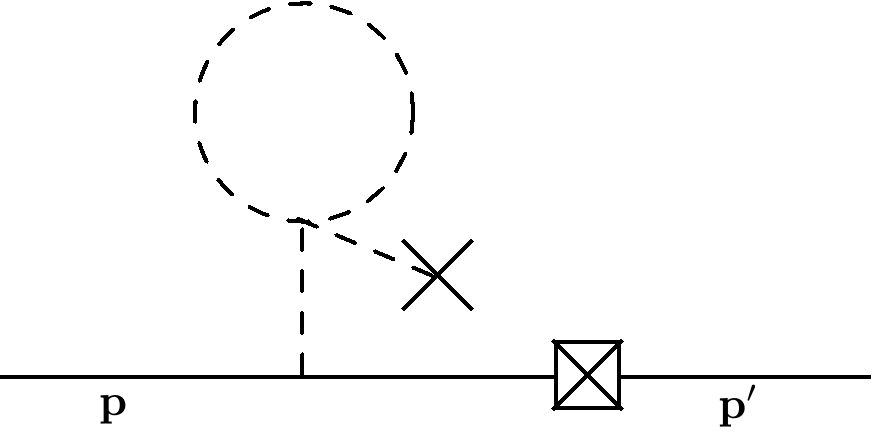}}
\qquad
\subfloat[][]{\includegraphics[scale=0.2]{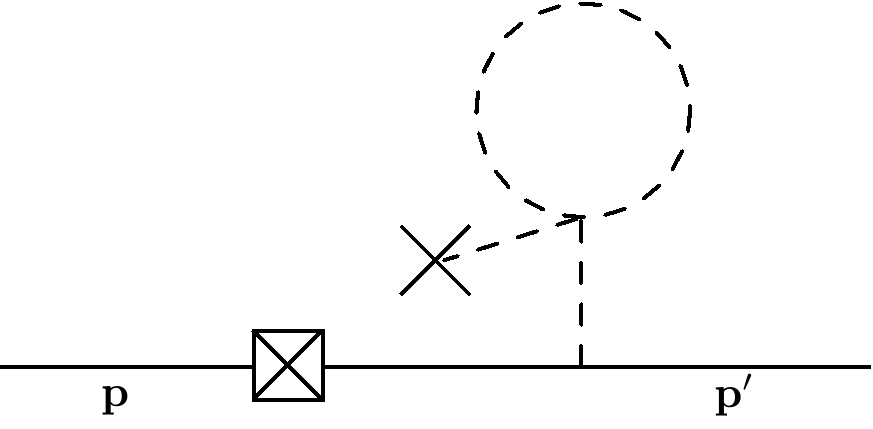}}

\subfloat[][]{\includegraphics[scale=0.2]{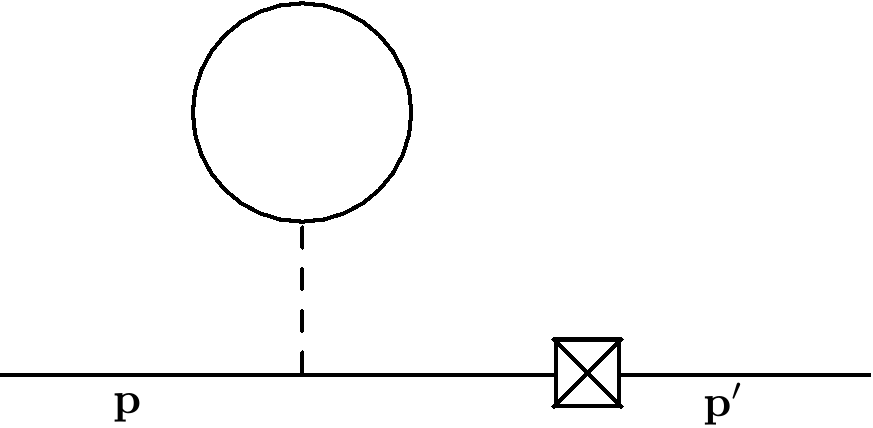}}
\qquad
\subfloat[][]{\includegraphics[scale=0.2]{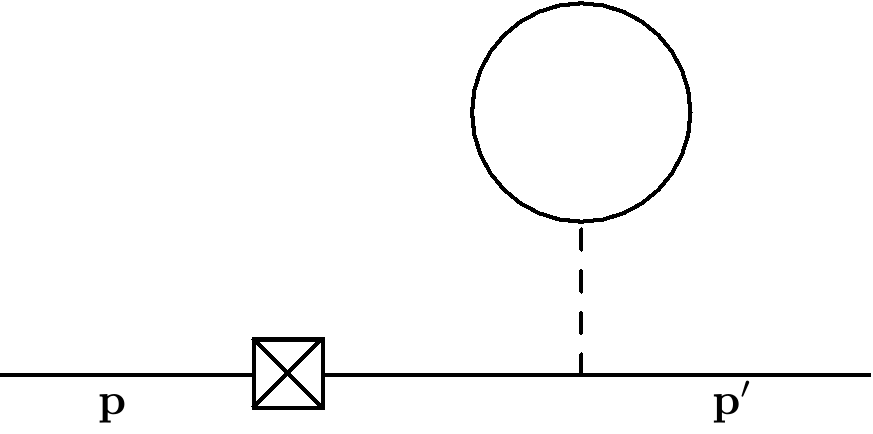}}

\subfloat[][]{\includegraphics[scale=0.2]{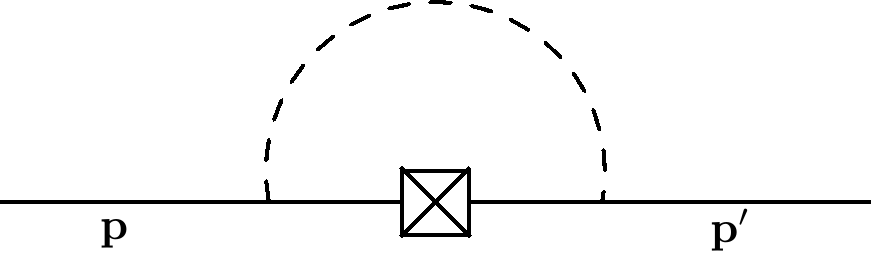}}

\subfloat[][]{\includegraphics[scale=0.2]{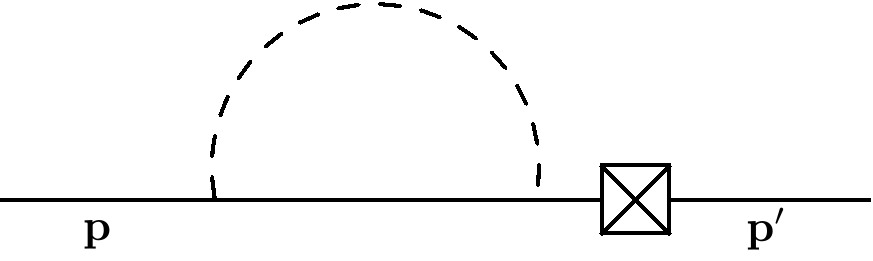}}
\qquad
\subfloat[][]{\includegraphics[scale=0.2]{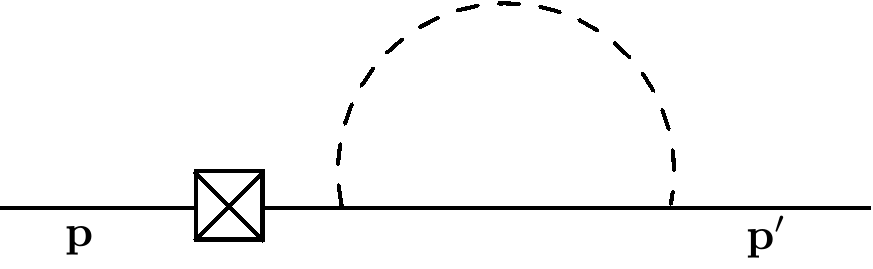}}

\subfloat[][]{\includegraphics[scale=0.175]{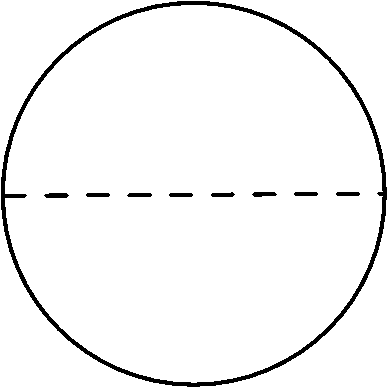}}

\caption{\label{fig:feyns} (Cf. Ref. \cite{Kading3}) Diagrammatic representation of the various terms contributing to the right-hand side of the quantum master equation \ref{eqn:MasterEQua1}. Dashed lines represent chameleons, solid lines atoms, crossed boxes density operator insertions, and crosses insertions of $\langle X \rangle$.}

\end{figure}

Eqn.\! (\ref{eqn:ProjMasterEqnYo}) is the quantum master equation in the desired single particle momentum subspace basis.
Its last line contains contractions of the involved fields that lead to the appearance of scalar propagators, which for the field $\phi$ will generally be denoted as $D$.
Though, it must be remarked that due to the form of Eqn.\! (\ref{eqn:LSZredEqOO}), no contractions between plus- and minus-type fields are possible, i.e.\ only $++$ or $--$ contractions are allowed, leading to the Feynman and Dyson propagators
\begin{eqnarray}\label{eqn:PhiFeynmanProp}
\langle\langle 0|T\hat{\phi}^+_{x} \hat{\phi}^+_{y}|0\rangle\rangle &=& D^{++}_{xy} ~=~ D^{\rm F}_{xy} ~=~ - i\hbar \int_k\frac{e^{ik(x-y)}}{k^2 + m^2 -i\epsilon},
\\
\langle\langle 0| T  \hat{\phi}^-_{x} \hat{\phi}^-_{y} |0\rangle\rangle &=& D^{--}_{xy} ~=~ D^{\rm D}_{xy} ~=~ + i\hbar \int_k\frac{e^{ik (x-y)}}{k^2 + m^2 +i\epsilon}.
\end{eqnarray}
That only these propagators can appear may be seen from 
\begin{eqnarray}
\langle\langle 0|\hat{\phi}^{+(-)}_x\hat{\phi}^{-(+)}_y|0\rangle\rangle &=& 0,
\end{eqnarray}  
which indicates that, at least in the single-particle subspace, mixing the two types of operators leads to a vanishing result.
\\\\
Furthermore, the propagators resulting from contracting two chameleon fields do not receive thermal corrections since it is assumed that the system $\phi$ is at zero temperature and therefore out-of-equilibrium with the surrounding environment $\chi$.
\\\\
Substituting Eqns.\ (\ref{eqn:EffActionAftertrac}), (\ref{eqn:UnitEvoEqnAc}) and (\ref{eqn:ReIFAcPertub}) into Eqn.\ (\ref{eqn:ProjMasterEqnYo}), and performing the contractions of the $\phi$ fields results in
\begin{eqnarray}\label{eqn:MasterEQua1}
\partial_t\rho(\vec{p},\vec{p}\,';t) &=& -i\big(E^{\phi}_{\vec{p}}-E^{\phi}_{\vec{p}\,'})\rho(\vec{p},\vec{p}\,';t)
-
\frac{im^2}{\mathcal{M}}\left(\frac{1}{E^{\phi}_{\vec{p}}}-\frac{1}{E^{\phi}_{\vec{p}\,'}}\right)\rho(\vec{p},\vec{p}\,';t) 
\nonumber
\\
&&
\times \bigg\{
\frac{\Delta^{\rm F}_{xx}}{\mathcal{M}}
+
\bigg[\frac{m^2}{\mathcal{M}}D^{\rm F}_{xx}+
\frac{\lambda}{2}\langle X\rangle\Delta^{\rm F}_{xx}\bigg]\int_{x^0} \frac{\sin[M(x^0-t)]}{M}\bigg\}
\nonumber
\\
&&
-
\frac{4m^4}{\mathcal{M}^2}\int_{x^0}\int_{\vec{k}}\Bigg\{
\Bigg[\rho(\vec{p},\vec{p}\,';t) \frac{\cos\big[E^{\phi}_{\vec{p}}(t-x^0)\big] }{E^{\phi}_{\vec{p}}2E^{\chi}_{\vec{k}}2E^{\phi}_{\vec{p}-\vec{k}}}
\nonumber
\\
&&
\phantom{-\frac{4m^4}{\mathcal{M}^2}\int_{x^0}\int_{\vec{k}}\Bigg\{}
\times 
\exp\big[-iE^{\phi}_{\vec{p}-\vec{k}}(t-x^0)\big]
\nonumber
\\
&&
-\rho(\vec{p}-\vec{k},\vec{p}\,'-\vec{k};t) 
\frac{\exp \big[i\big(E_{\vec{p}-\vec{k}}^{\phi}-E_{\vec{p}}^{\phi}\big)( t-x^0)\big]}{2E^{\chi}_{\vec{k}}2E^{\phi}_{\vec{p}-\vec{k}} 2E^{\phi}_{\vec{p}\,'-\vec{k}} }\Bigg]
\nonumber
\\
&&
\times\Big[\exp \big[-iE_{\vec{k}}^{\chi}(t-x^0)\big]+2\cos\big[E_{\vec{k}}^{\chi}(t-x^0)\big]f\big(E_{\vec{k}}^{\chi}\big)\Big] 
\nonumber
\\
&&
+\big(\vec{p}\leftrightarrow \vec{p}\,'\big)^*\Bigg\}
\nonumber
\\
&&
 -\frac{4m^4}{\mathcal{M}^2} \rho(\vec{p},\vec{p}\,';t)\sum_{s\,=\,\pm}\int_x\int_{\vec{k}_1\vec{k}_2}\frac{1}{2E_{\vec{k}_1}^{\phi}2E_{\vec{k}_1-\vec{k}_2}^{\phi}2E_{\vec{k}_2}^{\chi}}
\nonumber
\\
&&
\times 
\cos\big[\big(E_{\vec{k}_1}^{\phi}+E_{\vec{k}_1-\vec{k}_2}^{\phi}+sE_{\vec{k}_2}^{\chi}\big)(x^0-t)\big]
s\big[1+f\big(sE_{\vec{k}_2}^{\chi}\big)\big],
\nonumber
\\
\end{eqnarray}
where the energies were labelled with the respective field they correspond to.
An example of the evaluated field contractions leading to the terms in Eqn.\! (\ref{eqn:MasterEQua1}) can be found in Appendix \ref{app:Contractions}.
The right-hand side of Eqn.\! (\ref{eqn:MasterEQua1}) is represented diagrammatically in Figure \ref{fig:feyns}. 
Figures \ref{fig:feyns} (a) and (b) correspond to the $\chi$ tadpole insertion arising from the first term in the second line of Eqn.\! (\ref{eqn:MasterEQua1}), and (c) and (d) to the $\chi$ (lollipop) tadpole from the third term in the second line.  
Figures \ref{fig:feyns} (e) and (f) correspond to the $\phi$ tadpoles, arising from the second term in the second line, and the remaining Figures \ref{fig:feyns} (g)-(i) are the $\chi-\phi$ bubble diagrams, appearing in the third to seventh lines of Eqn.\! (\ref{eqn:MasterEQua1}). 
The final two lines of Eqn.\! (\ref{eqn:MasterEQua1}) contain a contribution from the absorptive part of the disconnected vacuum diagram shown in Figure \ref{fig:feyns} (j). 
The disconnected diagrams could be absorbed order-by-order in a redefinition of the matrix element $\rho(\vec{p},\vec{p}\,';t)\to \rho(\vec{p},\vec{p}\,';t)(1+\text{disconnected diagrams})$, taking into account that the time-derivative on the left-hand side of Eqn.\! (\ref{eqn:MasterEQua1}) counts at a finite order in the coupling constants. 
For the present discussions, however, the vacuum diagram is left explicit throughout for completeness.
\\\\
The terms in the third to ninth lines of Eqn.\ (\ref{eqn:MasterEQua1}) include decay (i.e.\ {$\chi\to\phi\phi$}) and production processes (i.e.\ {$\phi\phi\to\chi$}), which would not be expected to be present for realistic atoms, which are complex and stable configurations of many elementary particle fields rather than a simple scalar one. 
These decay and production processes arise here because such processes are permitted in the simple scalar field theory that has been used as a toy proxy for the atom.
\\\\
Next, the remaining $x^0$ integration in Eqn.\! (\ref{eqn:MasterEQua1}) can be done to yield 
\begin{eqnarray}\label{eqn:MaEqAfTiIn}
\partial_t\rho(\vec{p},\vec{p}\,';t) &=& -i\big(E^{\phi}_{\vec{p}}-E^{\phi}_{\vec{p}\,'})\rho(\vec{p},\vec{p}\,';t)
-\frac{im^2}{\mathcal{M}}\left(\frac{1}{E^{\phi}_{\vec{p}}}-\frac{1}{E^{\phi}_{\vec{p}\,'}}\right)\rho(\vec{p},\vec{p}\,';t) 
\nonumber
\\
&&
\times \bigg\{
\frac{\Delta^{\rm F}_{xx}}{\mathcal{M}}
+
\bigg[\frac{m^2}{\mathcal{M}}D^{\rm F}_{xx}+
\frac{\lambda}{2}\langle X\rangle \Delta^{\rm F}_{xx}\bigg]\frac{\cos(Mt)-1}{M^2}\bigg\}
\nonumber
\\
&&
+i
\frac{4m^4}{\mathcal{M}^2}\sum_{s\,=\,\pm}\int_{\vec{k}}\Bigg\{\Bigg[
 \frac{\rho(\vec{p},\vec{p}\,';t)}{E^{\phi}_{\vec{p}}2E^{\chi}_{\vec{k}}2E^{\phi}_{\vec{p}-\vec{k}}} \frac{s}{\big(sE_{\vec{k}}^{\chi}+E^{\phi}_{\vec{p}-\vec{k}}\big)^2-\big(E_{\vec{p}}^{\phi}\big)^2}
\nonumber
\\
&&
\times\Big[\Big(sE_{\vec{k}}^{\chi}+E^{\phi}_{\vec{p}-\vec{k}}\Big)\Big(1-\exp\big[-i\big(s E_{\vec{k}}^{\chi}+E_{\vec{p}-\vec{k}}^{\phi}\big)t\big]\cos\big(E_{\vec{p}}^{\phi}t\big)\Big)
\nonumber
\\
&&
-iE_{\vec{p}}^{\phi}\exp\big[-i\big(s E_{\vec{k}}^{\chi}+E_{\vec{p}-\vec{k}}^{\phi}\big)t\big]\sin\big(E_{\vec{p}}^{\phi}t\big)\Big]\big[1+f\big(sE_{\vec{k}}^{\chi}\big)\big]
\nonumber
\\
&& 
+\rho(\vec{p}-\vec{k},\vec{p}\,'-\vec{k};t) 
\frac{1}{2E^{\chi}_{\vec{k}}2E^{\phi}_{\vec{p}-\vec{k}} 2E^{\phi}_{\vec{p}\,'-\vec{k}}} \frac{s}{sE_{\vec{k}}^{\chi}+E_{\vec{p}-\vec{k}}^{\phi}-E_{\vec{p}}^{\phi}}
\nonumber
\\
&&
\times\Big(1-\exp\big[i(sE_{\vec{k}}^{\chi}+E_{\vec{p}-\vec{k}}^{\phi}-E_{\vec{p}}^{\phi}\big)t\big]\Big)f(sE^{\chi}_{\vec{k}})\Bigg]  - \big(\vec{p}\leftrightarrow \vec{p}\,'\big)^*\Bigg\}
\nonumber
\\
&& 
-\frac{4m^4}{\mathcal{M}^2} \rho(\vec{p},\vec{p}\,';t)\sum_{s\,=\,\pm}\int_{\vec{x}}\int_{\vec{k}_1\vec{k}_2}\frac{\sin\big[\big(E_{\vec{k}_1}^{\phi}+E_{\vec{k}_1-\vec{k}_2}^{\phi}+sE_{\vec{k}_2}^{\chi}\big)t\big]}{E_{\vec{k}_1}^{\phi}+E_{\vec{k}_1-\vec{k}_2}^{\phi}+sE_{\vec{k}_2}^{\chi}} 
\nonumber
\\
&& 
\times
\frac{s\big[1+f\big(sE_{\vec{k}_2}^{\chi}\big)\big]}{2E_{\vec{k}_1}^{\phi}2E_{\vec{k}_1-\vec{k}_2}^{\phi}2E_{\vec{k}_2}^{\chi}},
\end{eqnarray}
where the dummy parameter $s=\pm$ was introduced in order to simplify the sum over the two energy flows in the thermal contributions (see e.g.\ Ref.\! \cite{LeBellac}), and the identity in Eqn.\! (\ref{eqn:BoseEisIdenty}) was made use of.
Specifically, it was used to show that
\begin{eqnarray}
&&\exp\big[-iE_{\vec{k}}^{\chi}(t-x^0)\big]+2\cos\big[E_{\vec{k}}^{\chi}(t-x^0)\big]f\big(E_{\vec{k}}^{\chi}\big) 
\nonumber
\\
&=& \sum_{s\,=\,\pm}s\exp \big[-is E_{\vec{k}}^{\chi}(t-x^0)\big]\big[1+f\big(sE_{\vec{k}}^{\chi}\big)\big].
\end{eqnarray}
It is apparent that all terms on the right-hand side of Eqn.\! (\ref{eqn:MaEqAfTiIn}), except for those in the first line and the first term in the second line, vanish for $t\to 0$.
This is expected since those terms arose from a remaining time integral in Eqn.\! (\ref{eqn:MasterEQua1}) whose measure becomes nil in this limit. 
In addition, it can be observed that the right-hand side of Eqn.\ (\ref{eqn:MaEqAfTiIn}) becomes real under $\vec{p}\,' \to \vec{p}$, and 
is consistent with the usual property of a density matrix element $\rho(\vec{p},\vec{p}\,',t) = \rho^*(\vec{p}\,',\vec{p},t)$.
\subsubsection{Renormalised master equation}

The terms in Eqn.\! (\ref{eqn:MaEqAfTiIn}) yield quadratic and logarithmic ultra-violet divergences.
For the sake of completeness, short comments on the renormalisation of these terms will be given here, and the final expression of the master equation, including counterterms, presented. 
A more detailed discussion of these issues can be found in Ref.\! \cite{Kading3}.
\\\\
At the considered order, there are three relevant counterterms: the mass counterterms for the $\phi$ and $\chi$ fields, and the tadpole counterterm for the $\chi$ field.
They lead to a counterterm action
\begin{eqnarray}
\delta \widehat{S}_{\rm IF} &=& -\sum_{a\,=\,\pm}a\bigg[\int_{x}\delta\alpha \chi^a_x + \frac{1}{2}\int_{xy}\delta m^2_{xy}\phi_x^a\phi_y^a + \frac{1}{2}\int_{xy}\delta M^2_{xy}\chi_x^a\chi_y^a\bigg] 	
\nonumber
\\
\end{eqnarray}
with each counterterm being given by
\begin{eqnarray}
\delta m^2_{xy} &=& -\frac{2m^2}{\mathcal{M}^2} \Delta^{{\rm F}(T=0)}_{xx} \delta^{(4)}_{xy} 
\nonumber
\\
&&
+ \int_{pp'}e^{ipx-ip'y}(2\pi)^4\delta^4(p-p')\mathrm{Re} \Pi^{(T=0)}_{\rm non-loc}(- p^2)\Big|_{\vec{p}\,=\,\vec{\aleph}} ,
\\
\delta M^2_{xy} &=& - \bigg[\frac{2m^2}{\mathcal{M}^2} D^{{\rm F}(T=0)}_{xx}+\frac{\lambda}{2} \Delta^{{\rm F}(T=0)}_{xx}\bigg] \delta^{(4)}_{xy}
\nonumber
\\
&&
+ \int_{pp'}e^{ipx-ip' y}(2\pi)^4\delta^4(p-p')\mathrm{Re} \Sigma^{(T=0)}_{\rm non-loc}(- p^2)\Big|_{\vec{p}\,=\,\vec{\aleph}} ,
\\
\delta \alpha &=& - \frac{m^2}{\mathcal{M}} D^{{\rm F}(T=0)}_{xx} - \frac{\lambda\langle X\rangle}{2} \Delta^{{\rm F}(T=0)}_{xx},
\end{eqnarray}
where the index $T=0$ indicates that this quantity is taken at zero temperature\footnote{The thermal corrections are assumed to be tempered, such that they do not contain any ultra-violet divergences.}, and
\begin{eqnarray}
i\Pi_{\rm non-loc}^{(T=0)}(-p^2) &=& \bigg(-\frac{2im^2}{\mathcal{M}}\bigg)^2\int_k\frac{- i}{k^2+M^2-i\epsilon} \frac{- i}{(k-p)^2+m^2-i\epsilon}
\nonumber
\\
\end{eqnarray}
and
\begin{eqnarray}
i\Sigma_{\rm non-loc}(- p^2) &=& \bigg(-\frac{2im^2}{\mathcal{M}}\bigg)^2\int_k\frac{- i}{k^2+m^2-i\epsilon} \frac{- i}{(k-p)^2+m^2-i\epsilon}
\nonumber
\\
\end{eqnarray}
are the non-local bubble self-energy and the non-local chameleon self-energy, respectively.
\\\\ 
The counterterms are non-local in time.
This must be the case because, as mentioned earlier, some of the terms in Eqn.\! (\ref{eqn:MaEqAfTiIn}), including most of the divergent ones, are vanishing in the limit $t\to0$. 
Introducing $t$-independent counter\-terms, which themselves contain non-finite terms, would therefore amount to artificially creating a divergence at $t=0$, which would also persist for all later finite times.
Only having $t$-dependent counterterms which also vanish in $t\to 0$ and have the same time evolution as the divergences can cure this problem.
\\\\
The non-locality of the counterterms makes sense if it is considered that the external preparation and measurement of the system take place over a finite time, which means that time-translational invariance and Lorentz invariance are broken.
This gives rise to a non-trivial dependence on the zeroth momentum component which also has to be reflected by the counterterms. 
For this reason, the subtraction point for the renormalisation has been taken to only be a fixed three-momentum $\vec{\aleph}$.
\\\\
After renormalisation, the master equation becomes 
\begin{eqnarray}\label{eqn:RenMasEqn}
\partial_t\rho(\vec{p},\vec{p}\,';t) &=& - \big[ i u(\vec{p},\vec{p}\,';t) + \Gamma(\vec{p},\vec{p}\,';t) \big] \rho(\vec{p},\vec{p}\,';t)
\nonumber
\\
&&
+ \int_{\vec{k}} \gamma(\vec{p},\vec{p}\,',\vec{k};t) \rho(\vec{p}-\vec{k},\vec{p}\,'-\vec{k};t),
\end{eqnarray}
where
\begin{eqnarray}
u(\vec{p},\vec{p}\,';t) &:=& E^{\phi}_{\vec{p}} - E^{\phi}_{\vec{p}\,'} \label{eqn:u}
\nonumber
\\
&&
+ \frac{m^2}{\mathcal{M}}\left(\frac{1}{E^{\phi}_{\vec{p}}}-\frac{1}{E^{\phi}_{\vec{p}\,'}}\right)\Delta^{{\rm F}(T\neq 0)}_{xx}\bigg\{
\frac{1}{\mathcal{M}}
+
\frac{\lambda}{2}\langle X\rangle \frac{\cos(Mt)-1}{M^2}\bigg\}
\nonumber
\\
&&
- \Bigg\{\frac{1}{E^{\phi}_{\vec{p}}}\int_{p^0}\frac{\sin\big[\big(p^0-E^{\phi}_{\vec{p}}\big)t\big]}{p^0-E^{\phi}_{\vec{p}}}\Big[{\rm Re} \Pi_{\rm non-loc}(- p^2)
\nonumber
\\
&&
- {\rm Re} \Pi^{(T=0)}_{\rm non-loc}(- p^2)\big|_{\vec{p}\,=\,\bar{\vec{p}}}\Big]  - \big(\vec{p}\leftrightarrow \vec{p}\,'\big)\Bigg\}, \label{eq:coeffs_u} 
\\
\Gamma(\vec{p},\vec{p}\,';t) &:=& \frac{1}{E^{\phi}_{\vec{p}}}\int_{p^0}\frac{\sin\big[\big(p^0-E^{\phi}_{\vec{p}}\big)t\big]}{p^0-E^{\phi}_{\vec{p}}} {\rm Im} \Pi_{\rm non-loc}(- p^2)  + \big(\vec{p}\leftrightarrow \vec{p}\,'\big) , \label{eq:coeffs_Gamma}
\nonumber
\\
\\
\gamma(\vec{p},\vec{p}\,',\vec{k};t) &:=&
i \frac{4m^4}{\mathcal{M}^2}\sum_{s\,=\,\pm}\Bigg\{
\frac{1}{2E^{\chi}_{\vec{k}}2E^{\phi}_{\vec{p}-\vec{k}} 2E^{\phi}_{\vec{p}\,'-\vec{k}}} \frac{s}{sE_{\vec{k}}^{\chi}+E_{\vec{p}-\vec{k}}^{\phi}-E_{\vec{p}}^{\phi}}
\nonumber
\\
&&
\times\Big(1-\exp\big[i(sE_{\vec{k}}^{\chi}+E_{\vec{p}-\vec{k}}^{\phi}-E_{\vec{p}}^{\phi}\big)t\big]\Big)f\big(sE_{\vec{k}}^{\chi}\big) - (\vec{p} \leftrightarrow \vec{p}\,')^*\Bigg\} \label{eq:coeffs_gamma}
\nonumber
\\
\end{eqnarray}
were defined, and the contribution from the disconnected diagram was omitted. 
The self-energy appearing here also contains thermal corrections:
\begin{eqnarray}
i\Pi_{\rm non-loc}(- p^2) &=& \bigg(-\frac{2im^2}{\mathcal{M}}\bigg)^2\int_k\bigg[\frac{- i}{k^2+M^2-i\epsilon} + 2\pi f(|k^0|)\delta(k^2+M^2)\bigg] 
\nonumber
\\
&&
\phantom{\bigg(-\frac{2im^2}{\mathcal{M}}\bigg)^2\int_k}
\times
\frac{- i}{(k-p)^2+m^2-i\epsilon} .
\end{eqnarray}
The coefficients $u$ and $\Gamma$ are real, whereas $\gamma$ is complex. 
Each of the terms in Eqn.\! (\ref{eqn:RenMasEqn}) can be given a clear interpretation: $u$ describes coherent evolution, resulting from the mass shifts, $\Gamma$ corresponds to decays and, together with the real part of $\gamma$, is responsible for decoherence. 
In addition, $\gamma$ also accounts for momentum diffusion due to the coupling between the different momentum states.

\newpage
\subsection{Experimental implications}\label{ssec:ExpImplications}

Now with the renormalised master equation (\ref{eqn:RenMasEqn}) at hand, it shall be estimated whether its predictions could in principle be observed in laboratory experiments in order to constrain conformally coupled scalar fields like the $n=-4$ chameleon\footnote{This section is heavily based on Ref. \cite{Kading3}. }.
However, there are some momentum integrals left that have not been computed. 
They are analytically very challenging and require a numerical calculation which goes beyond the scope of the current analysis. 
Consequently, a full quantitative result cannot be given here for every term appearing in Eqn.\ (\ref{eqn:RenMasEqn}).
Fortunately, the first term in $u$ given in Eqn.\! (\ref{eqn:u}) contains the leading order effect and does not require such an integration.
\\\\
In order to see this, it shall be recalled that the master equation was derived with a rescaled mass defined by Eqn.\ (\ref{eqn:RescaledMass}), which depends on the background value of the chameleon field.
If an experiment was conceived that is sensitive to the absolute mass $\tilde{m}$ of the matter field, which would mean that the phase evolution based on the mass measured in a vanishing ambient value of the chameleon field can be predicted, then the leading effect on the dynamics could be captured by expanding 
\begin{eqnarray}\label{eqn:ExpEnergyinmtild}
E^{\phi}_{\vec{p}} - E^{\phi}_{\vec{p}\,'} &=& \tilde{E}^{\phi}_{\vec{p}} - \tilde{E}^{\phi}_{\vec{p}\,'} + \frac{\tilde{m}^2\langle X\rangle}{\mathcal{M}}\bigg(1+\frac{\langle X\rangle}{\mathcal{M}}\bigg)\Bigg(\frac{1}{\tilde{E}^{\phi}_{\vec{p}}} - \frac{1}{\tilde{E}^{\phi}_{\vec{p}\,'}}\Bigg)
\nonumber
\\
&&
- \frac{\tilde{m}^4\langle X\rangle^2}{2\mathcal{M}^2}\Bigg(\frac{1}{(\tilde{E}^{\phi}_{\vec{p}})^3} - \frac{1}{(\tilde{E}^{\phi}_{\vec{p}\,'})^3}\Bigg) + \mathcal{O}\bigg(\frac{\langle X\rangle^3}{\mathcal{M}^3}\bigg),
\end{eqnarray}
where $\tilde{E}^{\phi}_{\vec{p}}=\sqrt{\vec{p}^{\,2}+\tilde{m}^2}$. 
The first term on the right-hand side of Eqn.\! (\ref{eqn:ExpEnergyinmtild}) corresponds to the ordinary unitary evolution term but all other terms are corrections to it.
Of those, the very first term, i.e.\ the one proportional to $\tilde{m}^2\langle X\rangle/\mathcal{M}$, is of first order in the small parameter $\langle X\rangle/\mathcal{M}$.
Under the assumption that effects caused by the thermal corrections are much smaller than those that arise due to temperature-independent contributions from the chameleon\footnote{This is a standard assumption when people consider the effects of chameleon fields in experiments.}, this particular term is the leading term in the quantum master equation.
All other terms are {either} of second order in small parameters $\langle X\rangle/\mathcal{M}$, $\tilde{m}/\mathcal{M}$ or thermal corrections.
The term of interest is therefore
\begin{eqnarray}\label{eqn:Deltau}
\Delta u &\approx &\left| \frac{\tilde{m}^2\langle X\rangle}{\mathcal{M}})\Bigg(\frac{1}{\tilde{E}^{\phi}_{\vec{p}}} - \frac{1}{\tilde{E}^{\phi}_{\vec{p}\,'}}\Bigg) \right|  .
\end{eqnarray}
\begin{figure}[htbp]
\begin{center}
\includegraphics[scale=0.35]{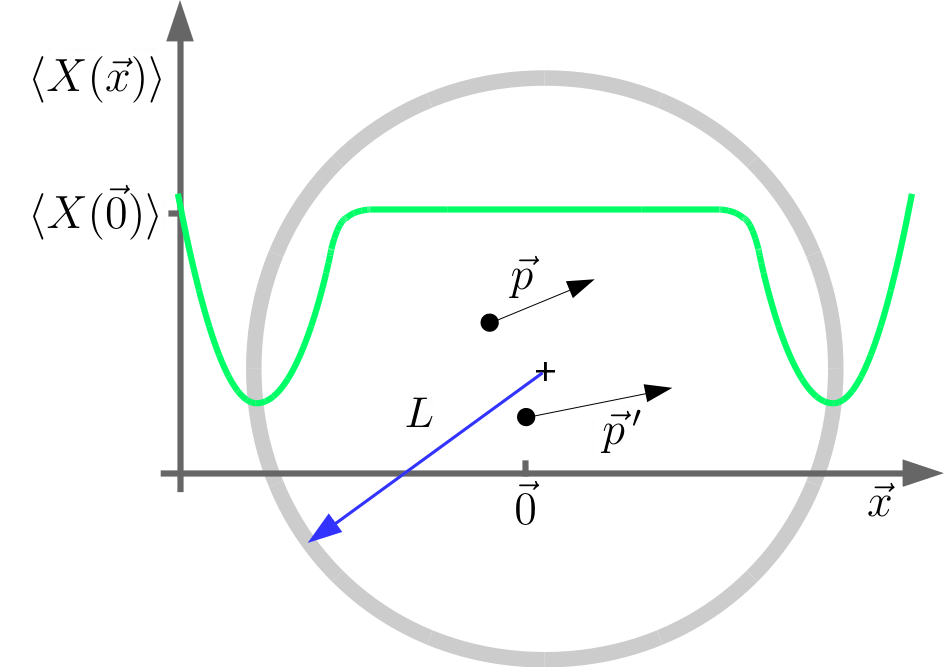}
\caption{(Cf. Ref. \cite{Kading3}) Schematic depiction of a cross-section of the experimental setup considered in this section. $\vec{x}$ is a spatial variable, the grey circle the vacuum chamber wall, $L$ the chamber's radius, and $\langle X(\vec{x}) \rangle$ the chameleon background value at $\vec{x}$. The two black spots represent two superposed atomic states with momenta $\vec{p}$, $\vec{p}\,'$, and the green line the background field profile of the chameleon within the chamber.\label{fig:chamber}}
\end{center}
\end{figure}
The current investigation was started with the hope to make predictions for atom interferometry experiments. 
A typical spherical vacuum chamber has dense walls, such that the chameleon field can be assumed to be constant and heavy within them.
Inside the chamber the chameleon can be much lighter and evolves towards the minimum of its effective potential in Eqn.\! (\ref{eqn:ChamEffPot}). 
However, over a large part of the chameleon parameter space, and for typical vacuum chamber sizes, the chamber radius $L$ is smaller than the Compton wavelength of the chameleon, which means that the field {cannot} reach the actual minimum of its potential.
In this case the chameleon adjusts its value, such that its Compton wavelength becomes equal to chamber radius, which results in modified background and mass values 
\begin{eqnarray}
\langle X \rangle &=& -\frac{q}{\sqrt{\lambda}L},\label{eqn:ModifMiniCham}
\\
M &=&  \frac{q}{\sqrt{2}L},
\end{eqnarray}
where $q = 1.287$ \cite{Khoury2003,Hinds} is a constant that has to be determined numerically by taking into account properties like the chamber geometry.
It can be assumed that the chameleon profile is approximately constant, i.e.\ equal to this modified minimum in Eqn.\! (\ref{eqn:ModifMiniCham}), over the spatial range explored by the atoms.
The form of the chameleon profile inside the chamber is illustrated schematically in Figure \ref{fig:chamber}. 
\\\\
Next, the energies in Eqn.\! (\ref{eqn:Deltau}) are expanded under the non-relativistic assumption $\tilde{m}^2 \gg \vec{p}^{\,2}$, leading to
\begin{eqnarray}\label{eqn:Deltau2}
\Delta u &\approx &\left| \frac{\langle X\rangle}{2\mathcal{M}} \tilde{m}v^2 \right|,
\end{eqnarray}
where the difference in speeds of both {atomic states} is defined by
\begin{eqnarray}
v^2 &=& \frac{\big||\vec{p}|^2 - |\vec{p}\,'|^2\big|}{\tilde{m}^2}.
\end{eqnarray}
For giving an example of a quantitative prediction, the quantum test mass is considered to be a Rubidium-87 atom with $\tilde{m} = 87\,m_u$, with $m_u$ being an atomic mass unit (see e.g.\ Ref.\! \cite{AbramowitzStegun}), and $\lambda = 0.1$, $\mathcal{M} = M_P$, $L = 1\, \text{m}$ and $v =10\, {\rm m}\,{\rm s}^{-1}$ (speeds of up to $6\, {\rm m}\,{\rm s}^{-1}$ were reported in atomic transport experiments \cite{Schmid_2006}) are chosen. 
The conservative value of $\mathcal{M} = M_P$ is motivated by constraints on chameleon theories \cite{BurrageSakA,BurrageSakB}, and the value of $\lambda < 1$ was chosen to remain well within the regime of perturbative validity. 
It should be admitted, however, that the chosen parameters are in tension with the current experimental bounds for the chameleon potential considered here. 
Even so, the aim of this investigation is to provide a conservative estimate of the order of magnitude of the effects. 
\\\\
Substituting the chosen parameter values into Eqn.\! (\ref{eqn:Deltau2}) yields
\begin{eqnarray}
\Delta u &\approx & 10^{-23} \,\text{Hz}.
\end{eqnarray}
The smallest phase shift that is currently measurable in atom interferometry experiments is of order $10^{-8} \,\text{Hz}$ (inferred from Ref.\! \cite{Estimate}). 
This means that the obtained value for $\Delta u$ is far out of reach for current experiments.
The other effects induced by the chameleon field like decoherence and momentum diffusion are even more suppressed.
In conclusion, it can be seen that the effects induced by
the quantum fluctuations of the quartic chameleon considered here are negligibly small and potentially out of reach for any near-future experiment.
This result supports previous work that has treated the chameleon field classically in laboratory experiments, and means that this will still be valid as those experiments continue to improve in sensitivity. 
It is to be expected that similar results would be obtained for other light scalar field models (e.g.\ those presented in Section \ref{ssec:ScreenedScalarModels}) but a possible pathway to experimental detectability is increasing the ratio $\tilde{m}/\mathcal{M}$, which might be an option for models with less constrained parameter spaces.

\newpage
\section{Closing remarks and outlook}\label{sec:Outlook}

To date, gravity is not entirely understood, and in particular the possible nature of its quantised theory is still a great mystery. 
Motivated by attempts to unify gravity with quantum physics, and to explain phenomena like dark matter or dark energy, Einstein's theory of gravity has been modified throughout history.
Some of these modified gravity theories give rise to new scalar degrees of freedom, some of which are expected to cause a gravity-like fifth force. 
Fifth forces, however, are heavily constrained by {Solar System}-based tests \cite{Adelberger}.
So-called screening mechanisms, coining the phrase screened scalar fields, are means to circumvent these constraints by suppressing fifth forces in environments of large mass densities, while allowing for potentially phenomenologically relevant effects to happen in less dense regions of space. 
\\\\
Screened scalar fields arising from modified gravity theories are subjects of ongoing research efforts in theoretical and experimental physics.
Their parameter spaces have successfully been constrained in various observations and experiments. 
However, there is still a lot of work and novel ideas required to {prove or} fully exclude the existence of such fields.
\\\\
In the current thesis, two tests of screened scalar fields have been presented - one from astrophysics, one from quantum physics.
For this, short overviews of general scalar fields, scalar-tensor theories of gravity, and some of the most prominent screened scalar field models were given in Section \ref{Sec:ScreenedScalarFields}. 
\\\\
In Section \ref{Sec:Lensing}, a particular type of scalar field, the symmetron (see Section \ref{ssec:Symmetrons}), was taken as an example to investigate its effect on gravitational lensing by galaxies.
It was shown that, under the assumption of a disformal coupling which had not been considered in previous analyses, the symmetron's contribution cannot explain the modification of lensing by baryonic matter that is otherwise attributed to particle dark matter (see Section \ref{ssec:LensSingleSymm}). 
However, a model that includes an additional scalar field with very particular properties has the potential to do so (confer Section \ref{ssec:CosmoScalar}).
\\\\
Section \ref{Sec:OpenQuantumfromCCSF} dealt with open quantum dynamics induced by conformally coupling scalar fields.
The idea was to investigate whether open quantum dynamical effects like phase shifts and decoherence due to the presence of light scalar fluctuations could be observed in experiments like atom interferometry. 
For this study, methods from non-equilibrium quantum field theory were developed and applied.
In this way, a single particle momentum subspace quantum master equation describing the evolution of a test particle system under influence of a conformally coupling scalar field was derived. 
As an example, the $n=-4$ chameleon was chosen, and a rough estimation of the leading order effect for reasonable experimental parameters showed that, at least for this particular model, it is currently many orders of magnitude away from being detectable in modern atom interferometers (see Section \ref{ssec:ExpImplications}). 
However, there is the possibility that another conformally coupled scalar field model with a larger, less constrained parameter space still left to be explored, could be constrained or detected with such methods.
\\\\
Both presented projects have potential for future work.
For example, it could be possible that there is a parameter regime that allows the symmetron to store enough energy for explaining the modification of gravitational lensing by combining the symmetron fifth force and the additional gravitational pull from the stored energy.
This still needs investigation.
Furthermore, the two-field model presented in Section \ref{ssec:CosmoScalar} deserves further research in order to find constraints on the additional scalar field $\pi$ and the model parameter $W$.
\\\\
The open dynamics project, however, promises a plethora of potential future projects.
For example, instead of using a toy model scalar field, more realistic atom models could be studied, and full numerical solutions for the remaining integrals be obtained.
Applications of the developed projection formalism to gravitational decoherence, particle phenomenology, cosmology, condensed matter, and heavy-ion physics are also at hand.  

\newpage
\begin{appendix}
\section{Fifth forces}\label{app:5Forces}

Fifth forces caused by the presence of conformally coupled scalar fields are a central topic in this thesis. 
Here it will be shown how they appear in the Einstein frame.
At first, in Section \ref{app:CouplingBetwe}, it will be explained how the coupling between a scalar field and matter actually arises.
Subsequently, the general equation for a fifth force experienced by a non-relativistic point particle will be derived in Section \ref{app:DerivOf5Forces}.


\subsection{Coupling between scalar and matter}\label{app:CouplingBetwe}

As previously seen, e.g.\ in Section \ref{Sec:ScreenedScalarFields}, scalar fields appearing in the context of scalar-tensor theories of gravity couple to the trace of the energy-momentum tensor $T_{\mu\nu}$ of other matter.
Here it will be shown how this coupling arises.
For this, an example scalar-tensor theory described by the Einstein frame action (compare with Eqn.\! (\ref{eqn:fRfinal}) in Section \ref{ssec:fR})
\begin{eqnarray}\label{app:ActionAction}
S &=& \int d^4x \sqrt{-g}\left( \frac{M_P^2}{2}R - \frac{1}{2}(\partial\varphi)^2 - V(\varphi) \right) + S_m\left( A^{-1}(\varphi) g^{\mu\nu},\psi \right)
\nonumber
\\
\end{eqnarray}
is considered.
The action $S_m$ is the matter action, $\psi$ represents arbitrary matter fields, $\varphi$ is the modified gravity scalar field, and the conformal transformation between Jordan and Einstein frame is given by
\begin{eqnarray}
\tilde{g}_{\mu\nu} &=& A(\varphi) g_{\mu\nu} .
\end{eqnarray}
Varying the action in Eqn.\! (\ref{app:ActionAction}) with respect to $\varphi$ leads to the equation of motion
\begin{eqnarray}\label{app:EqnOmotion}
0 &=& \sqrt{-g}[\Box\varphi - \partial_\varphi V(\varphi)] + \partial_\varphi \mathcal{L}_m \left( A^{-1}(\varphi) g^{\mu\nu},\psi \right),
\end{eqnarray}
where $\mathcal{L}_m $ is the Lagrangian associated with the matter action $S_m$.
\\\\
Dividing Eqn.\ (\ref{app:EqnOmotion}) by $\sqrt{-g}$, and applying a chain rule to the last term on its right-hand side yields  
\begin{eqnarray}\label{app:EqnBlabla}
0 &=& \Box\varphi - \partial_\varphi V(\varphi) + \frac{1}{\sqrt{-g}}\frac{\partial\mathcal{L}_m \left( A^{-1}(\varphi) g^{\mu\nu},\psi \right)}{\partial A^{-1}(\varphi) g^{\mu\nu}}  \frac{\partial A^{-1}(\varphi) g^{\mu\nu}}{\partial\varphi}
\nonumber
\\
&=& \Box\varphi - \partial_\varphi V(\varphi) + \frac{1}{\sqrt{-g}}\frac{\partial\mathcal{L}_m \left( A^{-1}(\varphi) g^{\mu\nu},\psi \right)}{\partial  g^{\mu\nu}}g^{\mu\nu}  A(\varphi) \partial_\varphi A^{-1}(\varphi)
\nonumber
\\
&=& \Box\varphi - \partial_\varphi V(\varphi) + \frac{1}{2} T^\mu_{~\mu}  \partial_\varphi \ln[A(\varphi)]
,
\end{eqnarray}
where $T^\mu_{~\mu}$ is the trace of the energy-momentum tensor \cite{Fujii}
\begin{eqnarray}
T_{\mu\nu} &=& \frac{-2}{\sqrt{-g}} \frac{\partial\mathcal{L}_m \left( A^{-1}(\varphi) g^{\mu\nu},\psi \right)}{\partial  g^{\mu\nu}}.
\end{eqnarray}
The last line of Eqn.\! (\ref{app:EqnBlabla}) corresponds to the equation of motion that can be derived by using the Euler-Lagrange equations \cite{Zee} with the scalar field Lagrangian
\begin{eqnarray}
\mathcal{L}_\varphi &=& -\frac{1}{2}(\partial\varphi)^2 - V(\varphi) + \frac{1}{2}\ln[A(\varphi)] T^\mu_{~\mu}.
\end{eqnarray} 
Choosing, for example,
\begin{eqnarray}
A(\varphi) &=& e^{2\varphi/\mathcal{M}}
\end{eqnarray}
leads to 
\begin{eqnarray}
\mathcal{L}_\varphi &=& -\frac{1}{2}(\partial\varphi)^2 - V(\varphi) + \frac{\varphi}{\mathcal{M}} T^\mu_{~\mu},
\end{eqnarray} 
which, for the appropriate choice of the potential $V(\varphi)$ and $T^\mu_{~\mu} =-\rho$, is the chameleon Lagrangian in Eqn.\! (\ref{eqn:ChamLagrangian}).

\newpage
\subsection{Derivation of fifth forces}\label{app:DerivOf5Forces}
In this section it will be shown how to derive the fifth force acting on a massive test particle.
The action of a free particle of mass $m$ serves as a starting point, \cite{Zwiebach}
\begin{eqnarray}\label{eqn:FreePartAction}
S &=& -m\int d\tilde{\tau}
\nonumber
\\
&=& -m\int  \sqrt{-\tilde{g}_{\mu\nu} dx^\mu dx^\nu},
\end{eqnarray}
where $\tilde{\tau}$ is the Jordan frame proper time, and $\tilde{g}$ is the Jordan frame metric.
\\\\
Next, the conformal transformation 
\begin{eqnarray}
\tilde{g}_{\mu\nu} &=& A(\varphi)g_{\mu\nu}
\end{eqnarray}
is applied, where the conformal factor is
\begin{eqnarray}\label{eqn:AppConfcoup}
A(\varphi) &\approx & 1+ a\frac{\varphi^\alpha}{\mathcal{M}^\alpha} + \mathcal{O}\left(\frac{\varphi^{2\alpha}}{\mathcal{M}^{2\alpha}} \right)
\end{eqnarray}
with $a$ and $\alpha$ being some positive real numbers, and $\varphi^\alpha \ll \mathcal{M}^\alpha$ is assumed.
Substituting Eqn.\ (\ref{eqn:AppConfcoup}) into Eqn.\ (\ref{eqn:FreePartAction}), and Maclaurin expanding in $\varphi^\alpha/\mathcal{M}^\alpha$ leads to
\begin{eqnarray}\label{eqn:Modifiedaction}
S &=& -m\int  \sqrt{-{g}_{\mu\nu} dx^\mu dx^\nu} \left(1+ \frac{a}{2}\frac{\varphi^\alpha}{\mathcal{M}^\alpha} \right) 
\nonumber
\\
&=& -m\int d\tau \left(1+ \frac{a}{2}\frac{\varphi^\alpha}{\mathcal{M}^\alpha} \right).
\end{eqnarray}
Now it can be seen that Eqn.\! (\ref{eqn:Modifiedaction}) contains the Einstein frame action for a free particle but, in addition, also a potential term depending on the scalar field $\varphi$. 
This means, that the particle was free in the Jordan frame but is interacting in the Einstein frame. 
\\\\
Eqn.\! (\ref{eqn:Modifiedaction}) can be rewritten as \cite{Zwiebach}
\begin{eqnarray}
S &=& -m\int dt \sqrt{1-v^2} \left(1+ \frac{a}{2}\frac{\varphi^\alpha}{\mathcal{M}^\alpha} \right),
\end{eqnarray}
with the velocity $v$.
From this, the Lagrange function
\begin{eqnarray}\label{app:LagrangeFunktion2}
L &=& -m\sqrt{1-v^2} \left(1+ \frac{a}{2}\frac{\varphi^\alpha}{\mathcal{M}^\alpha} \right)
\end{eqnarray}
can be read off. 
In the non-relativistic limit $v \ll 1$ Eqn.\! (\ref{app:LagrangeFunktion2}) becomes 
\begin{eqnarray}\label{app:LagrangeFunktion}
L &\approx& \frac{m}{2}v^2 - m \frac{a}{2}\frac{\varphi^\alpha}{\mathcal{M}^\alpha},
\end{eqnarray}
where the non-dynamic term and terms higher than first order in the small parameters were dropped.
\\\\
The first term on the right-hand side of Eqn.\ (\ref{app:LagrangeFunktion}) is the kinetic term and the second gives the potential
\begin{eqnarray}
V &=& m \frac{a}{2}\frac{\varphi^\alpha}{\mathcal{M}^\alpha}.
\end{eqnarray}
From this the fifth force can be obtained as \cite{Landau}
\begin{eqnarray}
F_\varphi &=& -\nabla V
\nonumber
\\
&=& -\frac{am}{2\mathcal{M}^\alpha} \nabla \varphi^\alpha
\nonumber
\\
&\approx& - \frac{m}{2}\nabla A(\varphi).
\end{eqnarray}
For one of the most common choices, $a=2$ and $\alpha =1$, see e.g.\ Ref.\! \cite{Nicolis2008}, this becomes the familiar
\begin{eqnarray}
F_\varphi &=& -\frac{m}{\mathcal{M}} \nabla \varphi.
\end{eqnarray}
\newpage
\section{Christoffel symbols}
Here the Christoffel symbols that are used in Section \ref{Sec:Lensing} are given. 
The general formula for computing Christoffel symbols from the metric $g_{\mu\nu}$ is \cite{Straumann}
\begin{eqnarray}
\Gamma^\mu_{~\alpha\beta} &=& \frac{1}{2}g^{\mu\rho}(\partial_\alpha g_{\beta\rho} + \partial_\beta g_{\alpha\rho}-\partial_\rho g_{\alpha\beta}).
\end{eqnarray}


\subsection{Perturbed FLRW metric}\label{app:PertubredFLRW}

The Christoffel symbols derived from the perturbed FLRW metric
\begin{eqnarray}
ds^2 &=& a^2(\tau)[-(1+2\Psi)d\tau^2 + (1+2\Phi)\delta_{ij}dx^idx^j] 
\end{eqnarray}
are given by
\begin{eqnarray}
\Gamma^0_{00} & = & \mathcal{H} + \Psi' \nonumber \\ 
\Gamma^0_{i0} & = & \Psi,_i \nonumber \\ 
\Gamma^0_{ij} & = & \delta_{ij}[\mathcal{H} + \Phi' +2\mathcal{H}(\Phi-\Psi)]  \nonumber \\
\Gamma^i_{00} & = & \delta^{ij}\Psi,_j \nonumber \\
\Gamma^i_{j0} & = & \delta^i_j (\mathcal{H} + \Phi') \nonumber \\
\Gamma^k_{ij} & = & 2\delta^k_{(j} \Phi,_{i)} - \delta_{ij}\delta^{kl}\Phi,_l .
\end{eqnarray}
In order to recover the unperturbed Christoffel symbols, setting $\Phi = \Psi=0$ is sufficient.


\subsection{Perturbed Jordan frame FLRW metric with extended field content}\label{app:ChristoCosmo}
In Section \ref{ssec:CosmoScalar} the following transformation between Jordan frame and Einstein frame was introduced:
\begin{eqnarray}
\tilde{g}_{00}&=& A^2(\varphi)g_{00} + \frac{\varphi^2}{W^2}\partial_0 \pi \partial_0 \pi, 
\\
\tilde{g}_{ab}&=& A^2(\varphi)g_{ab},
\end{eqnarray}
where $a,b$ are placeholders for spatial coordinates, and the constant $W$ describes the coupling between the symmetron $\varphi$ and the scalar field $\pi$.
The conformal factor $A^2(\varphi)$ can be ignored for the Christoffel symbols presented here since it was shown in Section \ref{ssec:ConfDisfLens} that it does not have any effect on gravitational lensing.
For completeness, however, it shall be stated that the modification of Christoffel symbols due to a conformal coupling is given by
\begin{eqnarray}
{}_\text{conf.}\Gamma^\alpha_{~\beta\gamma} &=&
\frac12 A(\varphi)^{-1}g^{\alpha\delta}\left(\partial_{\gamma}A g_{\beta\delta} + \partial_{\beta} A g_{\gamma\delta} - \partial_{\delta} A g_{\beta\gamma}\right).
\end{eqnarray}
For the purpose discussed in Section \ref{ssec:CosmoScalar} only the modifications due to the disformal coupling to $\pi$ are of interest.
The explicit expression for the FLRW Christoffel symbols modified in this way from Appendix \ref{app:PertubredFLRW} are  
\begin{eqnarray}
\tilde{\Gamma}^0_{00} &=& \mathcal{H} +\Psi' - \frac{1}{a^2} \frac{\varphi\varphi'}{W^2}
\\
\tilde{\Gamma}^0_{0i} &=& \Psi,_i -\frac{\varphi \varphi,_i}{a^2W^2}
\\
\tilde{\Gamma}^0_{ij} &=& \mathcal{H}(1+a^2\frac{\varphi^2}{W^2}) +\Phi' +2\mathcal{H}(\Phi-\Psi)
\\
\tilde{\Gamma}^i_{00} &=& \delta^{ij} (\Psi,_j - \frac{\varphi \varphi,_j}{a^2W^2}) 
\\
\tilde{\Gamma}^i_{0j} &=& \delta^i_j (\mathcal{H} +\Phi' + \frac{\varphi\varphi'}{a^2W^2}),
\end{eqnarray}
where the left-hand side of each equation corresponds to a Jordan frame Christoffel symbol expressed in terms of the Einstein frame metric and the scalar field $\varphi$ on the right-hand side.

\newpage
\section{Properties of scalar propagators}\label{app:PropoScaPro}
\begin{figure}[H]
\centering

\subfloat[][]{\includegraphics[scale=0.08]{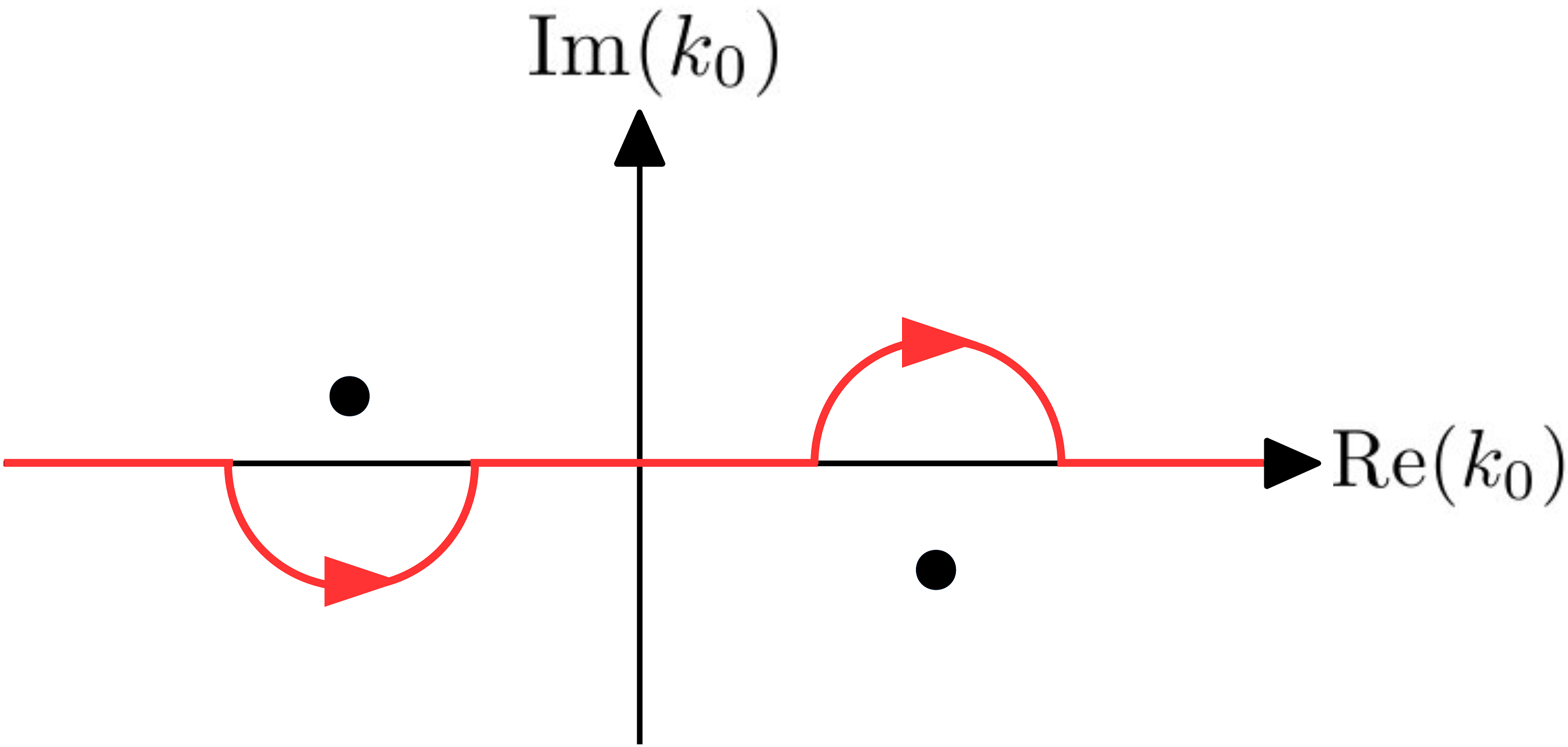}}
~~~\subfloat[][]{\includegraphics[scale=0.08]{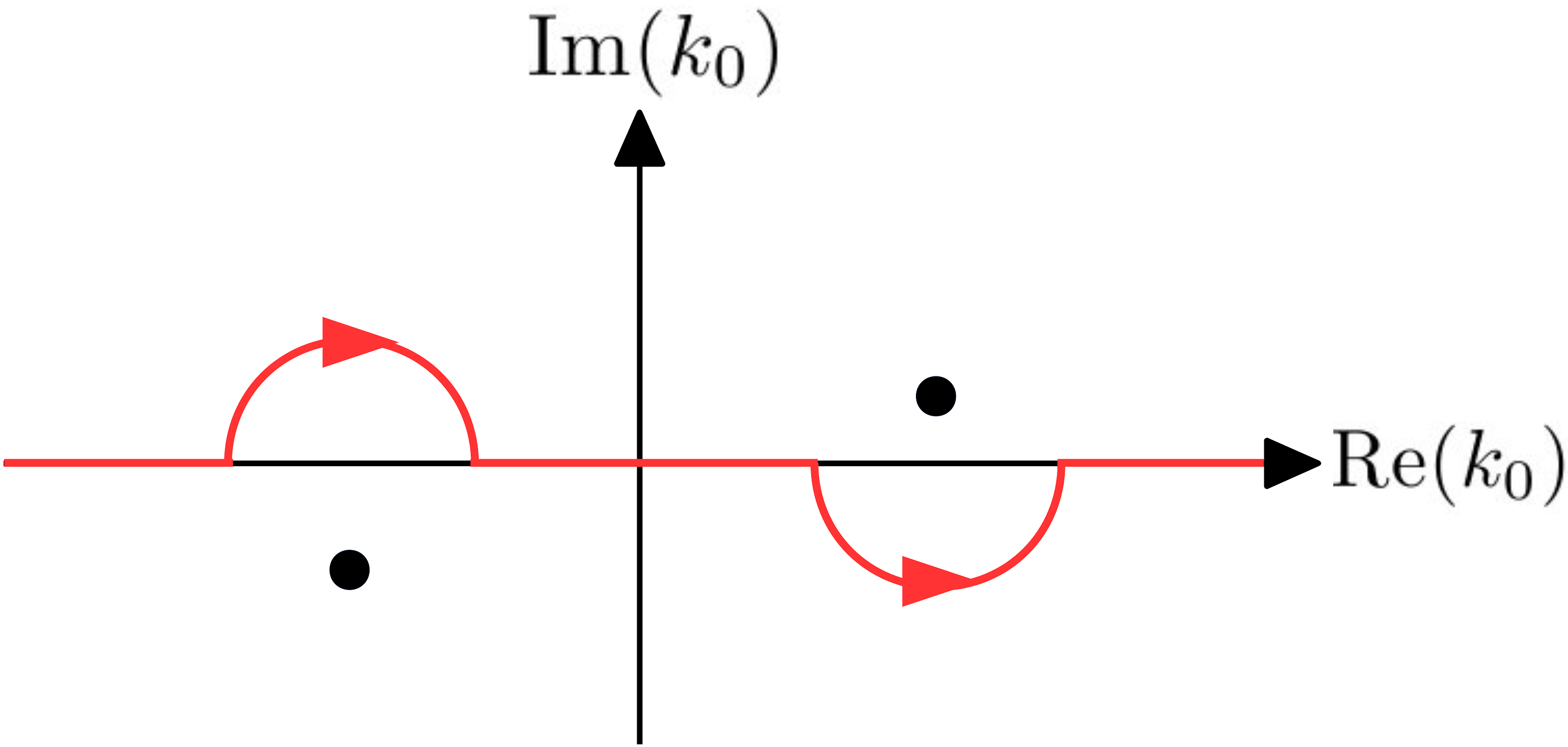}}

\subfloat[][]{\includegraphics[scale=0.08]{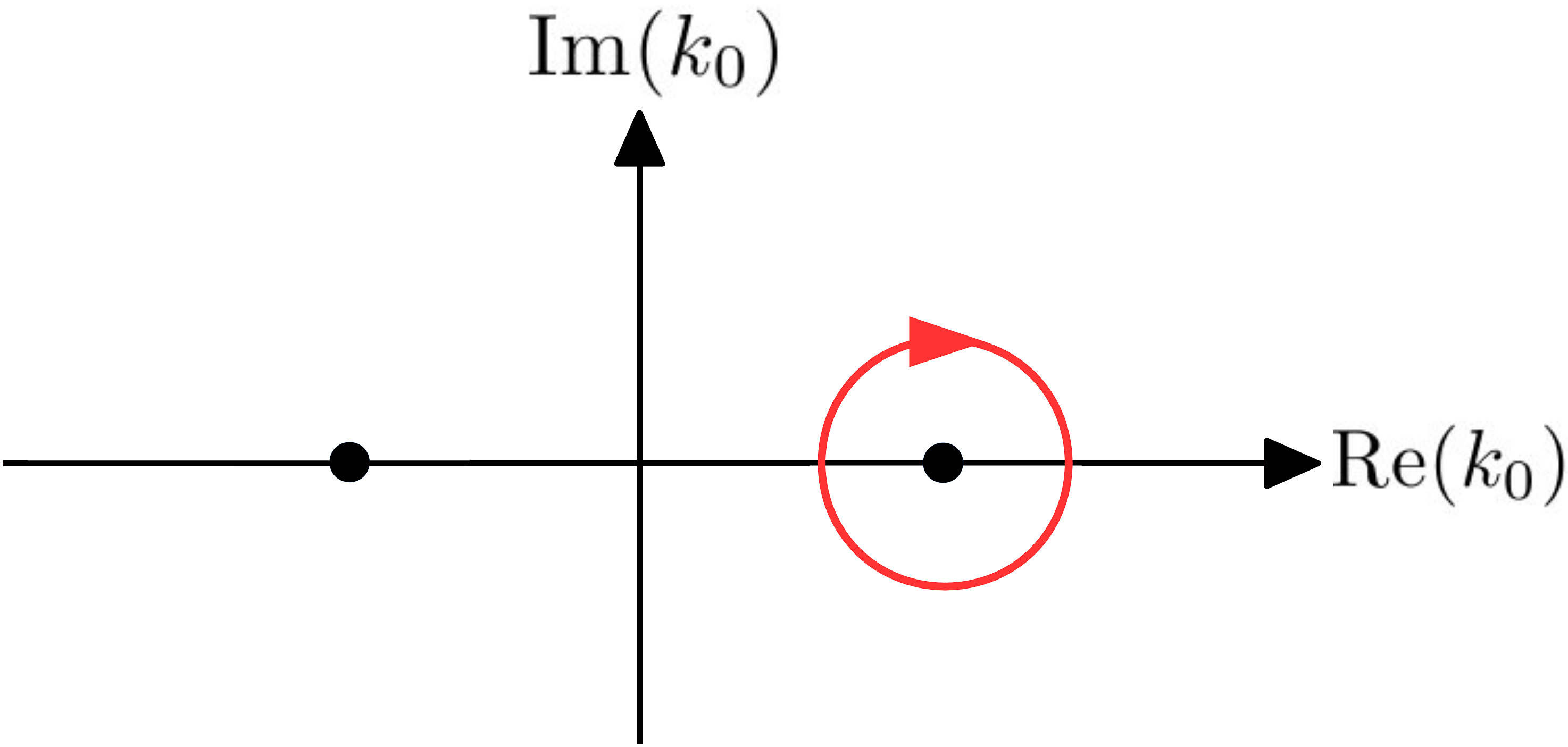}}
~~~\subfloat[][]{\includegraphics[scale=0.08]{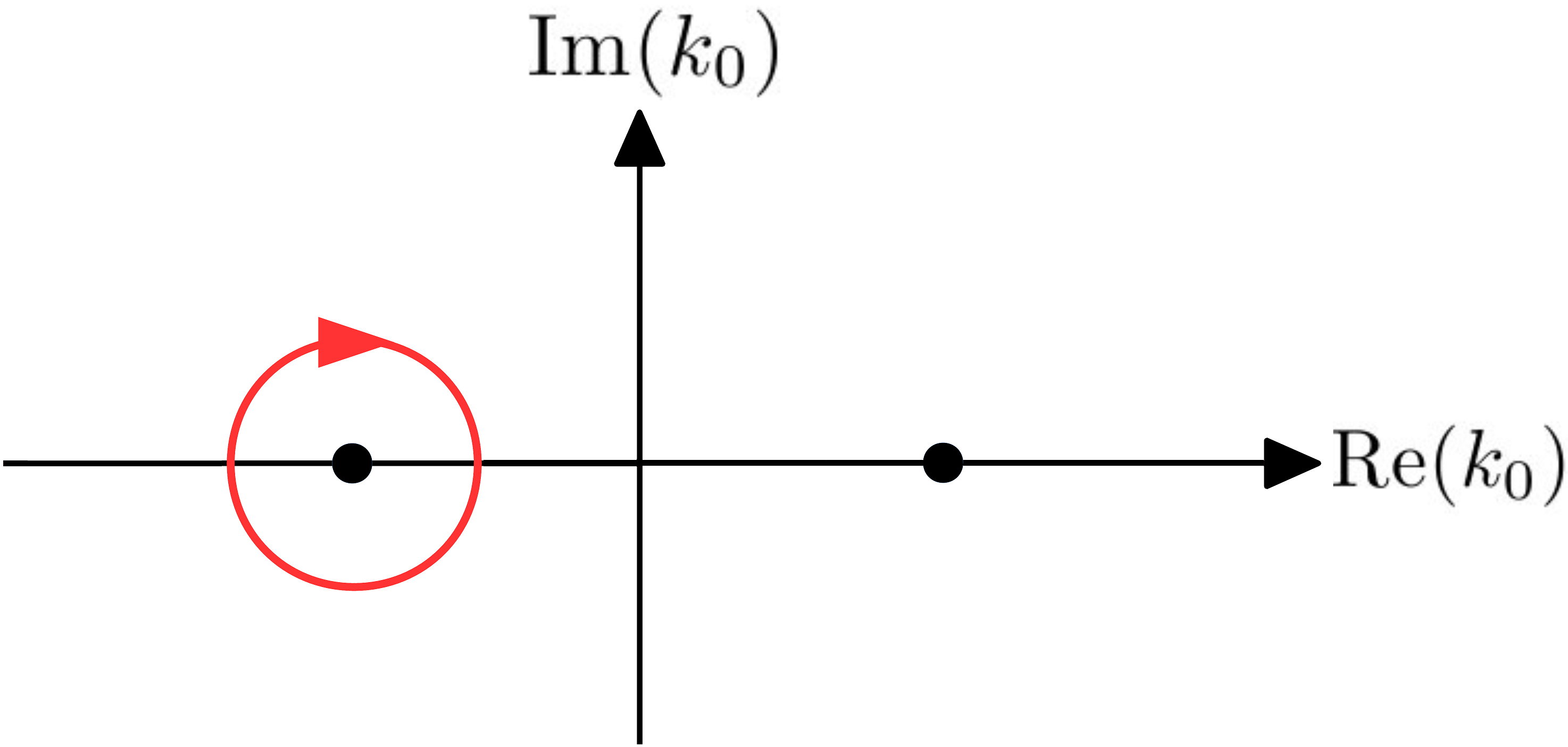}}

\caption{\label{fig:Contours}\cite{GreinerQuantization} Integration contours in the complex plane of (a) Feynman propagator, (b) Dyson propagator, (c) positive frequency Wightman propagator, and (d) negative frequency Wightman propagator.}
\end{figure}

Propagators of scalar fields play a major role in the discussion in Section \ref{Sec:OpenQuantumfromCCSF}. 
Some properties of such propagators, which are vital for this discussion, will be presented and proven here.
In particular, in {Section} \ref{app:CoincidenceLimit} it will be shown that they all coincide in the equal time limit of their arguments, and in {Section} \ref{app:Relations} two important relations between the different types of propagators will be introduced.
\\\\
The used propagators are defined in terms of the doubled scalar degrees of freedom living on a closed time path contour by \cite{CalzettaHuQFT}
\begin{flalign}
\label{app:FeynmanProp}
&\langle T \phi^+ (x) \phi^+ (y) \rangle ~=~ \Delta^{\rm F}_{xy} ~=~ - i\hbar \int \frac{d^4k}{(2\pi)^4} \frac{\exp \{ ik (x-y)\}}{k^2 + m^2 -i\epsilon} 
\\
\label{app:DysonProp}
&\langle T \phi^- (x) \phi^- (y) \rangle ~=~ \Delta^{\rm D}_{xy} ~=~ + i\hbar \int \frac{d^4k}{(2\pi)^4} \frac{\exp \{ ik (x-y)\}}{k^2 + m^2 +i\epsilon}
\\
\label{app:NegWightmanProp}
&\langle \phi^+ (x) \phi^- (y) \rangle = \Delta^<_{xy} ~=~ \int \frac{d^4k}{(2\pi)^4} \exp \{ ik (x-y)\} \Theta(-k^0)2\pi \hbar \delta(k^2 + m^2)
\\
\label{app:PosWightmanProp}
&\langle \phi^- (x) \phi^+ (y) \rangle = \Delta^>_{xy} ~=~ \Delta^<_{yx} ~=~ (\Delta^<_{xy})^*, 
\end{flalign}
where $T$ denotes time-ordering. 
These objects are called Feynman, Dyson and negative/positive frequency Wightman propagators.
Their integration contours in the complex plane are depicted in Figures \ref{fig:Contours}(a)-(d). 
\\\\
A derivation of the explicit expressions for Eqns.\! (\ref{app:FeynmanProp})-(\ref{app:PosWightmanProp}) after integrating over the four-momentum $k$ and a generally nice discussion of propagators can be found in Ref.\! \cite{BogoliubovShirkov}.

\newpage
\subsection{Coincidence limit}\label{app:CoincidenceLimit}
A relevant property of the four propagators in Eqns.\! (\ref{app:FeynmanProp})-(\ref{app:PosWightmanProp}) is that they all coincide in the equal time limit, i.e.
\begin{eqnarray}\label{app:Coincidence}
\lim_{y^0 \to x^0} \Delta^{\rm F}_{xy} &=&  \lim_{y^0 \to x^0} \Delta^{\rm D}_{xy} ~=~ \lim_{y^0 \to x^0} \Delta^<_{xy} ~=~ \lim_{y^0 \to x^0} \Delta^>_{xy}.
\nonumber
\\
\end{eqnarray}
These identities will now be proven, and for this purpose the following notation is introduced:
\begin{eqnarray}
\lim_{y^0 \to x^0} \Delta_{xy} &=:& \Delta_{\vec{x}\vec{y}}.
\end{eqnarray}
At first, it is straightforward to see that at least the last identity in Eqn.\! (\ref{app:Coincidence}) is true:
\begin{eqnarray}\label{app:WightmanPropCoincide}
\Delta^<_{\vec{x}\vec{y}} &=& \int \frac{d^4k}{(2\pi)^3} e^{i\vec{k} (\vec{x}-\vec{y})} \Theta(-k^0) \hbar \delta(k^2 + M^2) 
\nonumber
\\
&=& \int \frac{d^4k}{(2\pi)^3} e^{i\vec{k} (\vec{x}-\vec{y})} \Theta(k^0) \hbar \delta(k^2 + M^2) 
\nonumber
\\
&=& \Delta^>_{\vec{x}\vec{y}},
\end{eqnarray}
where $k^0 \to -k^0$ was used in order to go from one propagator to the other.
\\\\
Showing that also $\Delta^{\rm F,D}_{\vec{x}\vec{y}}$ coincide with the Wightman propagators in the equal time limit is more involved. 
They can be written as 
\begin{eqnarray}\label{app:FDPropPrincipal}
\Delta^{\rm F,D}_{\vec{x}\vec{y}} &=&  \mp i\hbar \int \frac{d^4k}{(2\pi)^4} \frac{\exp \{ i\vec{k} (\vec{x}-\vec{y})\}}{k^2 + m^2 \mp i\epsilon} 
\nonumber
\\
&=&  \mp i\hbar \int \frac{d^4k}{(2\pi)^4} e^{i\vec{k} (\vec{x}-\vec{y})} \left[ \frac{\mathcal{P}}{k^2 + m^2} \pm i\pi\delta(k^2 + m^2)\right],
\end{eqnarray}
where the Sokhotski-Plemelj theorem \cite{Sokhotskii,Plemelj} was made use of, and $\mathcal{P}$ denotes the principle value part \cite{Bronshtein}.
It can already be seen that the latter term in the expression above looks similar to the Wightman propagators.
This means, that the principle value part must be vanishing if Eqn.\! (\ref{app:Coincidence}) is supposed to be correct.
An explicit calculation shows that this is indeed the case:
\begin{eqnarray}
\int\limits_{-\infty}^\infty dk^0 \frac{\mathcal{P}}{-(k^0)^2 + E_{\vec{k}}^2} 
&=& \lim_{\varepsilon \to 0}
\left(\int\limits_{-\infty}^{-E_{\vec{k}}-\varepsilon} + \int\limits_{-E_{\vec{k}}+\varepsilon}^{E_{\vec{k}}-\varepsilon} + \int\limits_{E_{\vec{k}}+\varepsilon}^{\infty}\right)
\frac{dk^0 }{-(k^0)^2 + E_{\vec{k}}^2}
\nonumber
\\
&=& \lim_{\varepsilon \to 0} 
\left[\frac{1}{2E_{\vec{k}}}\log\Big( \frac{E_{\vec{k}}+k^0}{E_{\vec{k}}-k^0} \Big) \right]_{\{-\infty,-E_{\vec{k}}+\varepsilon,E_{\vec{k}}+\varepsilon \}}
^{\{-E_{\vec{k}}-\varepsilon,E_{\vec{k}}-\varepsilon,\infty\}}
\nonumber
\\
&=& \lim_{\varepsilon \to 0} 
\left[\log\Big(\frac{-\varepsilon}{2E_{\vec{k}}}\Big) -\log(-1) + \log\Big(\frac{2E_{\vec{k}}}{\varepsilon}\Big)  \right.\nonumber
\\
&\phantom{=}&\phantom{\lim_{\varepsilon \to 0} }
\left.- \log\Big(\frac{\varepsilon}{2E_{\vec{k}}}\Big) + \log(-1) - \log\Big(\frac{2E_{\vec{k}}}{-\varepsilon}\Big) \right]
\nonumber
\\
&=& 0.
\end{eqnarray}
Therefore, Eqn.\! (\ref{app:FDPropPrincipal}) becomes
\begin{eqnarray}
\Delta^{\rm F,D}_{\vec{x}\vec{y}} &=&   \frac{1}{2}\int \frac{d^4k}{(2\pi)^3} e^{i\vec{k} (\vec{x}-\vec{y})}  \hbar\delta(k^2 + m^2)
\nonumber
\\
&=&   \frac{1}{2}\int \frac{d^4k}{(2\pi)^3} e^{i\vec{k} (\vec{x}-\vec{y})}  (\Theta(k^0) + \Theta(-k^0))\hbar\delta(k^2 + m^2)
\nonumber
\\
&=&   \int \frac{d^4k}{(2\pi)^3} e^{i\vec{k} (\vec{x}-\vec{y})}  \Theta(k^0) \hbar\delta(k^2 + m^2),
\end{eqnarray}
which coincides with the expressions in Eqn.\! (\ref{app:WightmanPropCoincide})


\newpage
\subsection{Relations between propagators}\label{app:Relations}
There are two important relations connecting Feynman, Dyson and Wightman propagators that will be derived now. At first, 
\begin{eqnarray}\label{eqn:PropRel1}
 \Delta^{\rm F,D}_{xy} &=&  \Theta[\pm(x^0-y^0)]\Delta^>_{xy} + \Theta[\pm(y^0-x^0)] \Delta^<_{xy}
\end{eqnarray}
will be proven. For this, Eqns.\! (\ref{app:FeynmanProp}) and (\ref{app:DysonProp}) provide a starting point:
\begin{eqnarray}\label{eqn:PropProve}
\Delta^{\rm F,D}_{X} &=& \pm i \int \frac{d^4k}{(2\pi)^4} e^{i\vec{k}\vec{X}} \frac{e^{-ik^0 X^0}}{(k^0)^2 - E_{\vec{k}}^2 \pm i\epsilon}
\nonumber
\\
&=& \int \frac{d^3k}{(2\pi)^3} \frac{e^{i\vec{k}\vec{X}}}{2E_{\vec{k}}} \left( \Theta(X^0)e^{\mp iE_{\vec{k}} X^0}  +\Theta(-X^0)e^{\pm iE_{\vec{k}} X^0}  \right),
\end{eqnarray}
where in the last line Cauchy's integral formula \cite{Agarwal} was used:
\begin{eqnarray}\label{app:CauchyIntegralFormula}
2\pi i f(a) &=& \omega \oint_\gamma \frac{f(z)}{z-a}dz
\end{eqnarray}
with $\gamma$ being a closed curve in the complex plane and $\omega = +$ or $\omega=-$ for integrating around the curve counterclockwise or clockwise, respectively. 
\\\\
Rewriting Eqn.\! (\ref{eqn:PropProve}) further leads to
\begin{eqnarray}
\Delta^{\rm F,D}_{X} &=& \int \frac{d^4k}{(2\pi)^3} \frac{e^{ikX}}{2E_{\vec{k}}} \left( \Theta(X^0)\delta(k^0 \mp E_{\vec{k}})  +\Theta(-X^0) \delta(k^0 \pm E_{\vec{k}})  \right)
\nonumber
\\
&=& \int \frac{d^4k}{(2\pi)^3} e^{ikX} \delta(k^2 +m^2) \left( \Theta(X^0)\Theta(\pm k^0 )  +\Theta(-X^0) \delta(\mp k^0)  \right)
\nonumber
\\
&=& \Theta(X^0)\Delta^\gtrless_{X} + \Theta(-X^0) \Delta^\lessgtr_X,
\end{eqnarray}
which concludes the proof.
\\\\
The identity in Eqn.\! (\ref{eqn:PropRel1}) can be used for proving the so-called greatest time equation for any positive integer order\footnote{$++$ corresponds to the Feynman propagator, $+-$ to the negative frequency Wightman propagator, and so on.}:
\begin{eqnarray} \label{eqn:PropRel2}
\forall n\in \mathbb{N}: \sum\limits_{a,b=\pm} ab \Delta_{ab}^n &=& 0.
\end{eqnarray}
For convenience, the following proof of this equation will make use of Eqn.\! (\ref{eqn:PropRel1}) in a simplified notation:
\begin{eqnarray}
\Delta_{++,--} &=& \Theta(\pm)\Delta_{-+} + \Theta(\mp)\Delta_{+-}.
\end{eqnarray}
Furthermore, it is noted that for every considered propagator 
\begin{eqnarray}\label{eqn:DoubleTheta}
\Theta(+)\Theta(-)\Delta(x,y) &=& \Theta(+)\Theta(-)\Delta(x_0\equiv y_0, \vec{x},\vec{y}).
\end{eqnarray}
With these ingredients Eqn.\! (\ref{eqn:PropRel2}) is proven as follows: 
\begin{eqnarray}
\sum\limits_{a,b=\pm} ab \Delta_{ab}^n &=&  [\Theta(-)\Delta_{-+} + \Theta(+)\Delta_{+-}]^n + [\Theta(+)\Delta_{-+} + \Theta(-)\Delta_{+-}]^n 
\nonumber
\\
&\phantom{=}&
- \Delta_{-+}^n - \Delta_{+-}^n 
\nonumber
\\
&=& \sum\limits_{k=0}^n {n\choose k}[\Theta^{n-k}(-)\Theta^k(+) + \Theta^{n-k}(+)\Theta^k(-)]\Delta_{-+}^{n-k}\Delta_{+-}^k 
\nonumber
\\
&\phantom{=}&
- \Delta_{-+}^n - \Delta_{+-}^n, 
\end{eqnarray}
where in the third line the binomial theorem \cite{AbramowitzStegun} was used. Extracting from the sum the terms for $k=0$ and $k=n$, these terms sum up to
\begin{eqnarray}\label{eqn:PulledOut}
[\Theta^n(+) + \Theta^n(-)](\Delta^n_{-+} + \Delta^n_{+-})  ,
\end{eqnarray}
while the remaining sum is given by
\begin{eqnarray}\label{eqn:RemainingSum}
\sum\limits_{k=1}^{n-1} {n\choose k}[\Theta^{n-k}(-)\Theta^k(+) + \Theta^{n-k}(+)\Theta^k(-)]\Delta_{-+}^{n-k}\Delta_{+-}^k.
\end{eqnarray}
By making use of Eqns.\! (\ref{eqn:DoubleTheta}) and (\ref{app:Coincidence}), Eqn.\! (\ref{eqn:RemainingSum}) can be rewritten as
\begin{eqnarray}\label{eqn:SumRewritten}
\sum\limits_{k=1}^{n-1} {n\choose k} \Theta^{n-k}(+)\Theta^k(-)(\Delta^n_{-+} + \Delta^n_{+-})
\end{eqnarray}
since Feynman, Dyson and Wightman propagators coincide under this sum. The propagators in Eqn.\! (\ref{eqn:SumRewritten}) can now be extracted from the sum. Recombining Eqn.\! (\ref{eqn:SumRewritten}) with Eqn.\! (\ref{eqn:PulledOut}) and applying the binomial theorem again therefore leads to
\begin{eqnarray}
\sum\limits_{a,b=\pm} ab \Delta_{ab}^n &=& [\Theta(+) + \Theta(-)]^n(\Delta^n_{-+} + \Delta^n_{+-}) - \Delta_{-+}^n - \Delta_{+-}^n 
\nonumber
\\
&=& 0 
\end{eqnarray}
due to 
\begin{eqnarray}
\forall X \in \mathbb{R}: \Theta(+X) + \Theta(-X) &=& 1.
\end{eqnarray}


\newpage
\section{Example: Scalar field contractions}\label{app:Contractions}

In order to give the reader better means to reproduce the master equation (\ref{eqn:MasterEQua1}), an example term from Eqn.\! (\ref{eqn:ReIFAcPertub}) will be evaluated under the projection formalism expressed in Eqn.\ (\ref{eqn:ProjMasterEqnYo}).
For simplicity, the thermal corrections will be ignored and only propagators of the forms given in Eqns.\! (\ref{app:FeynmanProp}) and (\ref{eqn:PhiFeynmanProp}) considered. 
As an example, one of the terms in the last line of Eqn.\! (\ref{eqn:ReIFAcPertub}) is taken, while its constant pre-factor has been dropped for convenience.
Its evaluation under the projection used in Eqn.\ (\ref{eqn:ProjMasterEqnYo}) then goes as follows:
\begin{eqnarray}\label{app:Stern1}
(\ast)&=&
\lim_{\substack{x^{0(\prime)}\,\to\, t^{+}\\y^{0(\prime)}\,\to\, 0^-}}
\int d\Pi_{\vec{k}} d\Pi_{\vec{k}'} e^{i(E_{\vec{k}}-E_{\vec{k}'})t} \rho(\vec{k},\vec{k}';t) 
\nonumber
\\
&&
\times 
\int_{\vec{x}\vec{x}'\vec{y}\vec{y}\,'} e^{-i(\vec{p}\cdot\vec{x}-\vec{p}\,' \cdot\vec{x}')+i(\vec{k}\cdot\vec{y}-\vec{k}'\cdot\vec{y}\,')}
\partial_{x^0,E^\phi_{\vec{p}}} \partial_{x^{0\prime},E^\phi_{\vec{p}\,'}}^*\partial_{y^0,E^\phi_{\vec{k}}}^*\partial_{y^{0\prime},E^\phi_{\vec{k}'}}
\nonumber
\\
&&
\times 
\partial_t\int_{z,z'}  \contraction{}{\phi}{^+_x}{\phi}\phi^+_x\phi^+_{z'} \contraction{}{\phi}{^-_{x'}}{\phi}\phi^-_{x'}\phi^-_{y'} \contraction{}{\phi}{^+_{z'}}{\phi}\phi^+_{z'}\phi^+_{y} \Delta^{\rm F}_{zz'}{.}
\nonumber
\\
\end{eqnarray}
The last line on the right-hand side of Eqn.\! (\ref{app:Stern1})
becomes
\begin{eqnarray}\label{app:Stern2}
&&\partial_t\int_{z,z'}  D^{\rm F}_{xz'} D^{\rm D}_{x'y'} D^{\rm F}_{z'y} \Delta^{\rm F}_{zz'} 
\nonumber
\\
&=&
-\partial_t\int_{z,z'} \int_{q,l,r,n} \frac{e^{iq(x-z')}}{q^2 + m^2 -i\epsilon}  \frac{e^{il(x'-y')}}{l^2 + m^2 +i\epsilon} 
\frac{e^{ir(z'-y)}}{r^2 + m^2 -i\epsilon}  \frac{e^{in(z-z')}}{n^2 + M^2 -i\epsilon} .
\nonumber
\\
\end{eqnarray}
Acting with the Klein-Gordon operators on the right-hand side of Eqn.\! (\ref{app:Stern2}) leads to
\begin{eqnarray}
&&\partial_{x^0,E^\phi_{\vec{p}}} \partial_{x^{0\prime},E^\phi_{\vec{p}\,'}}^*\partial_{y^0,E^\phi_{\vec{k}}}^*\partial_{y^{0\prime},E^\phi_{\vec{k}'}}
\nonumber
\\
&\to &
(q^0 + E^\phi_{\vec{p}})(l^0 - E^\phi_{\vec{p}\,'})(r^0 + E^\phi_{\vec{k}})(l^0 - E^\phi_{\vec{k}'}).
\end{eqnarray}
Next, the relation 
\begin{eqnarray}
\int_{\vec{x}} e^{i\vec{x}(\vec{p}-\vec{k})} &=& (2\pi)^3 \delta^{(3)} (\vec{p}-\vec{k})
\end{eqnarray}
is used to evaluate all spatial integrals in Eqn.\! (\ref{app:Stern1}), and the resulting $\delta$-functions then allow to easily perform the $3$-momentum integrations.
Doing so leaves
\begin{eqnarray}
(\ast) &=& -\lim_{\substack{x^{0(\prime)}\,\to\, t^{+}\\y^{0(\prime)}\,\to\, 0^-}} \frac{\rho(\vec{p},\vec{p}\,';t) }{(2E^\phi_{\vec{p}})(2E^\phi_{\vec{p}\,'})} e^{i(E^\phi_{\vec{p}}-E^\phi_{\vec{p}\,'})t}
\nonumber
\\
&&
\times
\partial_t \int_{z^0,z^{0\prime}} \int_{q^0,l^0,r^0,n^0} (q^0 + E^\phi_{\vec{p}}) (r^0 + E^\phi_{\vec{p}}) (l^0 - E^\phi_{\vec{p}\,'})^2
\nonumber
\\
&&
~\times
\frac{e^{-iq^0 (x^0 - z^{0\prime})}}{-(q^0)^2 + (E^\phi_{\vec{p}})^2 - i\epsilon}
\frac{e^{-il^0 (x^{0\prime} - y^{0\prime})}}{-(l^0)^2 + (E^\phi_{\vec{p}\,'})^2 + i\epsilon}
\nonumber
\\
&&
~\times
\frac{e^{-ir^0 (z^{0\prime} - y^{0})}}{-(r^0)^2 + (E^\phi_{\vec{p}})^2 - i\epsilon}
\frac{e^{-in^0 (z^0 - z^{0\prime})}}{-(n^0)^2 + M^2 - i\epsilon} .
\end{eqnarray}
In order to evaluate the remaining momentum integrals, Cauchy's integral formula (\ref{app:CauchyIntegralFormula}) can be used.
For this, the time derivative is applied, which schematically acts as
\begin{eqnarray}
\partial_t \int_{z^0,z^{0\prime}} F(z^0,z^{0\prime}) &=& \int_{z^0} [F(z^0,t) + F(t,z^{0})],
\end{eqnarray}
and the directions from which the limits of $x^{0(\prime)}$ and $y^{0(\prime)}$ are approached are taken into account.
For example,
\begin{eqnarray}
\lim_{\substack{x^{0\prime}\,\to\, t^{+}\\y^{0\prime}\,\to\, 0^-}}\int_{l^0} 
\frac{e^{-il^0 (x^{0\prime} - y^{0\prime})}}{-(l^0)^2 + (E^\phi_{\vec{p}\,'})^2 + i\epsilon}
&=& -2\pi i \frac{1}{2E^\phi_{\vec{p}\,'}}e^{iE^\phi_{\vec{p}\,'}t},
\end{eqnarray}
where the argument of the exponential function on the left-hand side is always negative, which requires the contour to be closed in the lower half-plane in order to ensure that the contribution at $-i\infty$ vanishes. 
Consequently, the negative pole of the Dyson propagator (see Figure \ref{fig:Contours}(b)) was picked up.
\\\\
Finally, the result is given by
\begin{eqnarray}
(\ast) &=& \frac{\rho(\vec{p},\vec{p}\,';t)}{2E^\phi_{\vec{p}} M} \int_{z^0} e^{iMz^0}.
\end{eqnarray}






\end{appendix}

\newpage
\bibliography{mybib}{}
\bibliographystyle{ieeetr}

\end{document}